\documentclass[a4paper,11pt]{article}
\usepackage{hanging}

\pdfoutput=1

\usepackage{jcappub} 
                     
\usepackage[utf8]{inputenc}
\usepackage[T1]{fontenc}
\usepackage{xcolor}

\usepackage{bm}
\let\vec\bm

\let\ve\bm

\usepackage{soul}
\usepackage{tikz}
\usetikzlibrary{decorations.pathreplacing}
\usetikzlibrary{fadings}
\usepackage{orcidlink}

\newcommand{\Dk}[1]{\frac{d^3#1}{(2\pi)^3}}

\newcommand{\vk}{\ve k}
\newcommand{\vp}{\ve p}
\newcommand{\vq}{\ve q}
\newcommand{\vx}{\ve x}

\newcommand{\vs}{\ve s}

\newcommand{\vr}{\ve r}

\newcommand{\A}{\mathcal{A}}

\newcommand{\Ps}{\mathbf{\Psi}}

\newcommand{\vhn}{\hat{\ve n}}

\def\apjl{ApJL}
\def\apjs{ApJS}

\def\aap{{A\&A}}

\usepackage{booktabs}
\usepackage{float}
\newcommand{\ra}[1]{\renewcommand{\arraystretch}{#1}}

\title{\boldmath Full Modeling and Parameter Compression Methods  in configuration space for DESI 2024 and beyond}

\affiliation{Affiliations are in Appendix \ref{sec:affiliations}}

\author[1]{{S.~Ramirez-Solano}\orcidlink{0009-0006-2096-7974},}
\author[2]{{M.~Icaza-Lizaola}\orcidlink{0000-0002-2547-3184},}
\author[3,1]{{H.~E.~Noriega}\orcidlink{0000-0002-3397-3998},}
\author[1]{{M.~Vargas-Maga\~na}\orcidlink{0000-0003-3841-1836},}
\author[3]{{S.~Fromenteau}\orcidlink{0000-0002-3724-4768},}
\author[3]{{A.~Aviles}\orcidlink{0000-0001-5998-3986},}
\author[1]{{F.~Rodr{\'\i}guez-Mart{\'\i}nez},}
\author[4]{{J.~Aguilar},}
\author[5]{{S.~Ahlen}\orcidlink{0000-0001-6098-7247},}
\author[6]{{O.~Alves},}
\author[7]{{S.~Brieden}\orcidlink{0000-0003-3896-9215},}
\author[8]{{D.~Brooks},}
\author[4]{{T.~Claybaugh},}
\author[9]{{S.~Cole}\orcidlink{0000-0002-5954-7903},}
\author[1]{{A.~de la Macorra}\orcidlink{0000-0002-1769-1640},}
\author[10]{{Arjun~Dey}\orcidlink{0000-0002-4928-4003},}
\author[11]{{B.~Dey}\orcidlink{0000-0002-5665-7912},}
\author[8]{{P.~Doel},}
\author[12,13]{{K.~Fanning}\orcidlink{0000-0003-2371-3356},}
\author[14,15]{{J.~E.~Forero-Romero}\orcidlink{0000-0002-2890-3725},}
\author[16,17,18]{{E.~Gaztañaga},}
\author[19,16,20]{{H.~Gil-Mar\'in}\orcidlink{0000-0003-0265-6217},}
\author[4]{{S.~Gontcho A Gontcho}\orcidlink{0000-0003-3142-233X},}
\author[21,22,23]{{K.~Honscheid},}
\author[24]{{C.~Howlett}\orcidlink{0000-0002-1081-9410},}
\author[10]{{S.~Juneau},}
\author[24]{{Y.~Lai},}
\author[4]{{M.~Landriau}\orcidlink{0000-0003-1838-8528},}
\author[25,26]{{M.~Manera}\orcidlink{0000-0003-4962-8934},}
\author[27]{{M.~Maus},}
\author[28,26]{{R.~Miquel},}
\author[29]{{E.~Mueller},}
\author[1]{{A.~Muñoz-Gutiérrez},}
\author[30]{{A.~D.~Myers},}
\author[17]{{S.~Nadathur}\orcidlink{0000-0001-9070-3102},}
\author[31]{{J.~Nie}\orcidlink{0000-0001-6590-8122},}
\author[32,33,34]{{W.~J.~Percival}\orcidlink{0000-0002-0644-5727},}
\author[4,35,27]{{C.~Poppett},}
\author[36]{{M.~Rezaie}\orcidlink{0000-0001-5589-7116},}
\author[37]{{G.~Rossi},}
\author[38]{{E.~Sanchez}\orcidlink{0000-0002-9646-8198},}
\author[4]{{D.~Schlegel},}
\author[39,6]{{M.~Schubnell},}
\author[40]{{H.~Seo}\orcidlink{0000-0002-6588-3508},}
\author[10]{{D.~Sprayberry},}
\author[6]{{G.~Tarl\'{e}}\orcidlink{0000-0003-1704-0781},}
\author[28,20]{{L.~Verde}\orcidlink{0000-0003-2601-8770},}
\author[10]{{B.~A.~Weaver},}
\author[12,41,13]{{R.~H.~Wechsler}\orcidlink{0000-0003-2229-011X},}
\author[13]{{S.~Yuan}\orcidlink{0000-0002-5992-7586},}
\author[42]{{P.~Zarrouk}\orcidlink{0000-0002-7305-9578},}
\author[31]{{H.~Zou}\orcidlink{0000-0002-6684-3997},}

\emailAdd{sadiramirez@estudiantes.fisica.unam.mx}

\abstract{
In the contemporary era of high-precision spectroscopic surveys, led by projects like DESI, there is an increasing demand for optimizing the extraction of cosmological information from clustering data. This work conducts a thorough comparison of various methodologies for modeling the full shape of the two-point statistics in configuration space. We investigate the performance of both direct fits (Full-Modeling) and the parameter compression approaches (ShapeFit and Standard). We utilize the \texttt{ABACUS-SUMMIT} simulations, tailored to exceed DESI's precision requirements. Particularly, we fit the two-point statistics of three distinct tracers (LRG, ELG, and QSO), by employing a Gaussian Streaming Model in tandem with Convolution Lagrangian Perturbation Theory and Effective Field Theory. We explore methodological setup variations, including the range of scales, the set of galaxy bias parameters, the inclusion of the hexadecapole, as well as model extensions encompassing varying $n_s$ and allowing for $w_0w_a$CDM dark energy model. Throughout these varied explorations, while precision levels fluctuate and certain configurations exhibit tighter parameter constraints, our pipeline consistently recovers the parameter values of the mocks within $1\sigma$ in all cases for a 1-year DESI volume. Additionally, we compare the performance of configuration space analysis with its Fourier space counterpart using three models:  \textsc{PyBird}, \textsc{FOLPS} and \textsc{velocileptors}, presented in companion papers. We find good agreement with the results from all these models. 
}

\begin{document}
\maketitle
\flushbottom
\section{Introduction}
\label{sec:intro}
The study of the large-scale structure (LSS) of the Universe is a fundamental pillar in modern cosmology, and it provides an important observable to investigate its subsequent evolution over time. Observations of the LSS through the distribution of galaxies enable the extraction of valuable information that can be used to compare to theoretical predictions and impose constraints on cosmological parameters. 

Spectroscopic galaxy surveys collect data on the angular position and redshift of millions of galaxies across a wide range of cosmic epochs. Two key phenomena, which have left distinct features in the clustering of galaxies, are extensively utilized for extracting cosmological insights. 

The first phenomenon involves fluctuations in the matter density, arising from acoustic density waves in the primordial plasma of the early universe. These effects are discernible through the peak of the correlation function at comoving scales of $\sim 100\,h^{-1}\,\textrm{Mpc}$, and are commonly referred to as baryon acoustic oscillations (BAO). 

The second key observable arises from apparent anisotropies in the clustering of galaxies along the line-of-sight (LoS), attributed to their peculiar velocities. This phenomenon is known as redshift space distortions (RSD). While RSD introduce complexities in interpreting the observed positions of galaxies as a 3-D map, they also provide a unique opportunity to access information concerning the growth rate of the LSS.

In the last decade, the clustering of galaxies has been analyzed through data collected by spectroscopic galaxy surveys such as the 2dF Galaxy Redshift Survey (2dFGRS) \citep[]{2001MNRAS.328.1039C, 2001MNRAS.327.1297P}, the Baryon Oscillation Spectroscopic Survey (BOSS) \citep[]{2013AJ....145...10D, 2017MNRAS.470.2617A}, the extended Baryon Oscillation Spectroscopic Survey (eBOSS) \citep[]{2016AJ....151...44D,2021PhRvD.103h3533A}, or the Dark Energy Survey (DES) \citep[]{2018MNRAS.481.1149Z,2018PhRvD..98d3526A}. These surveys have significantly increased both precision and volume over the years. At the time of publication, the next generation of surveys is being developed with the EUCLID mission \citep[]{2011arXiv1110.3193L,2016SPIE.9904E..0OR} launching in the summer of 2023, and The Rubin Observatory Large Synoptic Survey Telescope (LSST) \cite{2019ApJ...873..111I} survey expected to have its first light in the summer of 2024. Simultaneously, the Dark Energy Spectroscopic Instrument (DESI) \footnote{\href{https://www.desi.lbl.gov/}{www.desi.lbl.gov}} \citep[]{2016arXiv161100036D, DESI2022.KP1.Instr, DESI2023a.KP1.SV, DESI2023b.KP1.EDR} is currently ongoing and is mapping the universe over an unprecedented range of $14,000\,\textrm{deg}^2$ with an effective volume of $40\, (h^{-1}\,\textrm{Gpc})^3$,  improving the number of observed galaxies by an order of magnitude compared to previous surveys. Plans for a continuing survey, DESI-II \citep[]{2022arXiv220903585S}, that will focus on the $z > 2$ universe are being made, and observations are expected to start in 2026.

DESI targets different tracers of matter at various redshifts, including emission line galaxies (ELG) \citep[]{2020RNAAS...4..180R} at $0.6 < z < 1.6$, luminous red galaxies (LRG) \citep[]{2020RNAAS...4..181Z} at $0.3 < z < 1.0$, and quasars at (QSO) $0.9 < z < 2.1$ \citep[]{2020RNAAS...4..179Y} during the dark-time program. The exceptional accuracy of these surveys will necessitate models capable of predicting clustering statistics with great precision.

Perturbation theory (PT) when applied to data in either the Lagrangian approach \cite{2012MNRAS.426.2719R,2014MNRAS.439.3504S,2017MNRAS.469.1369S,2020MNRAS.499.5527T,2020JCAP...07..062C,2021MNRAS.501.1013Z,2021MNRAS.500..736B} or Eulerian approach \cite{2017JCAP...11..039F,2018JCAP...12..035D,2020PhRvD.102f3533C,2020JCAP...05..005D} has demonstrated to accurately predict galaxy densities, velocities and redshift-space distortions within the quasi-linear regime. Recently, PT has been complemented with effective field theory (EFT) \cite{McDonald:2006mx,McDonald:2009dh,Baumann:2010tm,Carrasco:2012cv,Vlah:2015sea,2021JCAP...03..100C} parameters to account for back-reaction of small scale physics over quasi-linear scales. This has proven effective in confronting theoretical predictions of clustering statistics with observations or simulations. In this work, we follow the Lagrangian PT (LPT)/EFT approach \cite{Matsubara:2007wj,Carlson:2012bu,Vlah:2015sea,Vlah:2016bcl,Aviles:2018thp}.

 In anticipation of the DESI Year 1 data release, the DESI collaboration aims at testing systematic effects of their methodologies using the new \texttt{ABACUS-SUMMIT}  \cite{Abacus} suite of simulations, which are designed to surpass the resolution and volume requirements for cosmological simulations within DESI. Currently, four perturbations approaches for modeling the redshift space clustering statistics are being analyzed and compared within the DESI collaboration. 
 These models include the configuration space \textsc{EFT-GSM} \cite{gsm2023arXiv231017834R} code, which is utilized throughout this work, as well as three Fourier space codes, namely  
\textsc{PyBird} \cite{pybird2020JCAP...05..005D, pybird2020JCAP...06..001C, pybird2021JCAP...01..006D,KP5s4-Lai}, \textsc{FOLPS}$\nu$ \cite{folps2022JCAP...11..038N,KP5s3-Noriega} and \textsc{velocileptors} \cite{velocileptors2021JCAP...03..100C, velocileptors2020JCAP...07..062C,KP5s2-Maus}. A comparison for the three Fourier space codes can be found in the companion paper \cite{KP5s1-Maus} and, in this paper, we compare our configuration space model and the other Fourier space models.

In this work, we assess potential systematics within the pipeline of our EFT-GSM model for the DESI first-year data release. \textsc{EFT-GSM} model by construction is designed to model the redshift space correlation function up to one-loop corrections working on the LPT framework and using the Gaussian Streaming Model \cite{Fisher:1994ks,Scoccimarro:2004tg,Reid:2011ar,Wang:2013hwa,Vlah:2015sea}, in addition to EFT parameters. This framework allows us to model small-scale physics down to scales as low as $20\,h^{-1}\,\textrm{Mpc}$, where non-linear processes appear. Our methodology has been introduced and optimized with neural networks in \cite{gsm2023arXiv231017834R} we test it using the \texttt{Nseries} simulations \cite{2017MNRAS.470.2617A}, a suite of N-body simulations that were designed to analyze the eBOSS data and have a lower resolution than the \texttt{ABACUS-SUMMIT} simulations. The main difference between our EFT-GSM methodology and the other three methodologies is that we fit the correlation function directly, whereas the other methodologies work in Fourier space.\footnote{Reference \cite{KP5s4-Lai} also works with the correlation function, through Fourier transforming the power spectrum obtained with \textsc{PyBird}.}

Two different approaches have been used for extracting cosmological information from the shape of two-point statistics, the compressed methodologies (as the Standard and Shape Fit methodologies), and the direct fits (also called Full-Modelling). The {\it Standard} methodology employed for analyzing the clustering of redshift space tracers throughout SDSS-III BOSS \citep{2011AJ....142...72E,2013AJ....145...10D,2015ApJS..219...12A} and SDSS-IV eBOSS \citep{2021MNRAS.500.3254R,2020MNRAS.498.2354R,2020ApJS..250....8L} requires a one-time calculation of the linear power spectrum within a chosen fiducial cosmology. This power spectrum is then coupled with a PT framework, and the relevant cosmological information are compressed into three free parameters: the linear growth rate of structures $f\sigma_8$ and two scaling parameters, parallel and perpendicular to the line of sight $\alpha_{\perp}$ and $\alpha_{\parallel}$, that parameterize the Alcock-Paczynski effect \cite{AP_1979Natur.281..358A}. 

The {\it ShapeFit}  methodology of \cite{shapefit2021JCAP...12..054B} suggests adding a new free parameter to the {\it Standard} compressed methodology of BOSS and eBOSS, this parameter models the transition of the linear power spectrum from large to small scales, by including a new effective parameter $m$, that modifies the slope of the linear power spectrum. The agnostic nature of compressed methodologies, like the {\it Standard} approach utilized for BOSS and eBOSS, or the more recent {\it ShapeFit} methodology, exhibit a relatively high degree of model independence compared to those derived from {\it Full Modeling} analysis. Moreover, their significantly lower computational costs make them more appealing from a practical standpoint. Although the {\it ShapeFit} methodology has been tested in Fourier space \citep[e.g.][]{shapefit2021JCAP...12..054B}, its performance in configuration space is still an ongoing area of investigation. 

An alternative approach opposed to the compressed nature of the \textit{Standard} and Shape Fit methodologies is to directly explore the posterior distribution of all cosmological parameters. This involves calculating the linear power spectrum at each step of the Markov Chain Monte Carlo (MCMC) exploration. Referred to as the \textit{Full Modeling} or \textit{Direct Fit} approach, it offers the advantage of tighter constraints on cosmological parameters, which are directly constrained. This method has been applied in the analysis of both BOSS \citep{2017MNRAS.464.1640S,2017MNRAS.467.2085G} and eBOSS \citep{2022MNRAS.512.5657S,2022MNRAS.516.1910N}. While these \textit{Full Modeling} approaches can be resource-intensive to construct, advancements in computing capabilities and the evolution of new machine-learning algorithms have accelerated analytical calculations in recent years. Consequently, it has become feasible to model the full shape of the correlation function or power spectrum and simultaneously vary cosmological, bias, and EFT (Effective Field Theory) parameters.

Throughout this work, we evaluate the performance of the three methodologies for extracting cosmological information {\it Standard} and {\it Shape Fit}, and {\it Full Modeling}), and we compare the analysis of these three methods across a range of conditions and across different DESI tracers to have a better understanding of the differences among them. 

This paper gives support to the \textit{DESI 2024: Full-Shape analysis from galaxies and quasars} \cite{DESI2024.V.KP5} which is part of the DESI Year 1 main  results \cite{DESI2024.I.DR1,DESI2024.II.KP3,DESI2024.III.KP4,DESI2024.IV.KP6,DESI2024.VII.KP7B,DESI2024.VIII.KP7C}.

The structure of this paper is as follows. We begin in section \ref{sec:data} by introducing and describing the simulations used throughout this work. Then in Section \ref{sec:model} we describe the theory behind the \textsc{EFT-GSM} model. Section \ref{sec:methodology} describes the implementation of our three different methods: the {\it Standard} methodology, the {\it Shape-Fit} methodology and the {\it Full Modeling} methodology. The material presented up until this point is not new to this work. So, a busy reader can directly skip sections 3 and 4. Section \ref{sec:results_gsm} shows the results of our pipeline for these three distinct methodologies, and explores different configurations of our model. First, we define our baseline analysis and investigate how different variations to it affect our results, these variations include differences in the fitting range explored and different settings for the galaxy bias terms. We continue by exploring the effects on our models constraints under different extensions to our baseline configuration, this extensions include adding the hexadecapole of the correlation function, extending the parameter space explored by adding $n_s$ as a free parameter, and including the parameters of the equation state of dark energy as free parameters. We finalize this section by presenting the results for our three different tracers and presenting a joint analysis where all tracers are analyzed simultaneously. The last subsection presents the comparison between  {\it Shape-Fit} and {\it Full Modeling} methodologies for our baseline analysis. 
Finally, Section \ref{sec:comparison_models} has a comparison between our \texttt{GSM-EFT} model and other EFT models in Fourier space and configuration space. The conclusions of this paper are in section \ref{sec:conclusions}.

\section{Simulations mocks and covariance}
\label{sec:data}

As we have stated, the main goal of our work is to test the performance and analyze the potential systematics of our methodology on a mock representation of the DESI ELG, LRG, and QSO surveys. We begin our work in section \ref{mock_data} by introducing our mock datasets, generated with the \texttt{ABACUS-SUMMIT} simulation. The various analyses presented in this work involve estimating the likelihood that the correlation functions produced by our EFT-GSM model agree with those measured from the \texttt{ABACUS-SUMMIT} simulation across a range of parameters and across different model configurations. The computation of these likelihoods requires estimates of the covariance of our correlation function. The methodology and simulations we utilize to compute this covariance are introduced in Section \ref{cov_mocks}.   

\subsection{Mock Data}
\label{mock_data}
In this study, we test the correlation function model using a set of 25 cubic boxes, with a volume of $\left(2 \, h^{-1} \mathrm{Gpc}\right)^3$, which were generated utilizing the \texttt{ABACUS-SUMMIT} 
high-accuracy $N$-body simulations \cite{Abacus}. Consequently, the combined volume extends to $200 \, h^{-3} \mathrm{Gpc}^3$. 
Each \texttt{ABACUS-SUMMIT} box contains a total of $6912^3$ particles, with each particle carrying a mass of $2 \times 10^9 \, h^{-1} \mathrm{M}_{\odot}$. These simulations adopt the cosmological parameters from Planck 2018 \cite{plank2018_cosmological}: $h=0.6736$, $\omega{\mathrm{cdm}}=0.1200$, $\omega_\mathrm{b}=0.02237$, $\ln(10^{10} A_s)=3.0364$, $n_s=0.9649$, $M_\nu=0.06 \, \mathrm{eV}$, $w_0=-1$.

Our mocks are based on HOD fitted on the \textit{Survey Validation} 3 (SV3) spectroscopic data from DESI and consists of three independent sets of catalogs, each representing a different tracer: LRG at $z=0.8$, ELG at $z=1.1$, and the QSO at $z=1.4$. In order to populate our simulations with galaxies, we utilize the halo occupation distribution (HOD) methodology described in \cite{2023JCAP...05..033R} for ELG and in \cite{2023arXiv230606314Y} for LRG and QSO. The HOD models are calibrated using small-scale wedges (below $5 \, h^{-1} \textrm{Mpc}$) in combination with large-scale bias evolution information, whenever the former is available. These calibrations are done independently for each snapshot of the simulation.

Through this work we utilize the correlation function that we measure from these Abacus mock simulations, these correlation functions are utilized to fit the parameters of our models. Our estimates of the correlation function are defined as the mean of these 25 mock realizations for each tracer, where the $s$-bins in our analysis are spaced at intervals of $\Delta s=4$ $h^{-1} \mathrm{Mpc}$. In Figure \ref{fig:meanABACUS} we show the mean of the \texttt{ABACUS} multipoles for $\ell=0,2,4$ for the three tracers, LRG (left panel), ELG(intermediate panel) and QSO(right panel).

\subsection{Covariance Mocks}
\label{cov_mocks}
One common approach for computing covariance matrices is to utilize the correlation function measured from thousands of different realizations. These simulations are usually computed by methods that are optimized for efficiency rather than accuracy. In this work, we utilize 1000 approximate mocks generated with the Extended Zel'dovich methodology \cite{ezmocks} with the same cosmology of \texttt{ABACUS-SUMMIT}. We compute the covariance matrix $C$ of each tracer between bins $i$ and $j$, using the 1000 EZ mocks, as follows:
\begin{equation}
\begin{aligned}
C_{s}^{(ij)}=\frac{1}{N_\mathrm{mocks}-1}\sum_{m=1}^{N_{\mathrm{mocks}}}\left(\xi_i^m-\bar{\xi}_i\right)\left(\xi_j^m-\bar{\xi}_j\right),
\end{aligned}
\end{equation}
where $N_{\text{mocks}}=1000$ represents the total number of mocks, and $\bar{\xi}_i$ denotes the average of the $i^{th}$ bin in the analysis. The Hartlap factor induces a slight change in the inverse covariance matrix, typically around 0.94 to 0.95, depending on the scales. Hence, its influence in this study is thought to be negligible, as the values vary by less than 0.5\%.

\begin{figure*}
\includegraphics[trim={4cm 0 4cm 0cm},width=150mm]{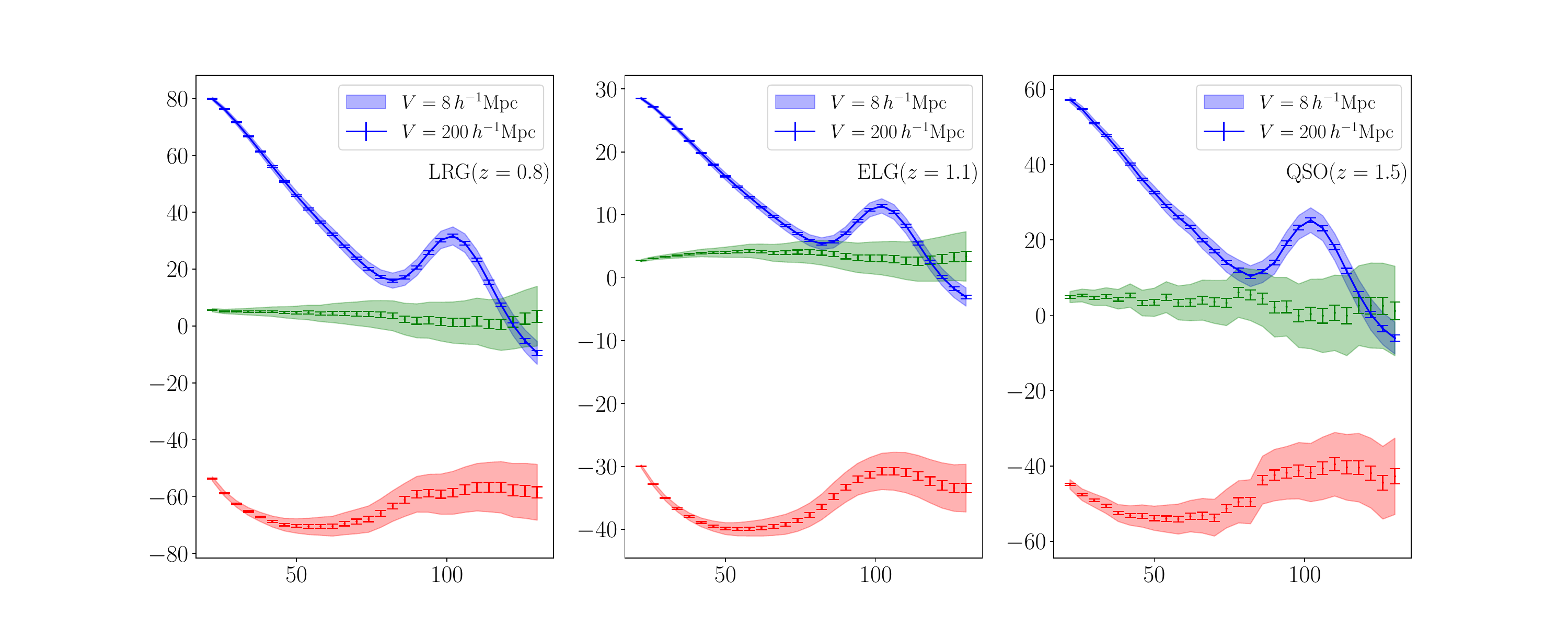}
\caption{\texttt{ABACUS-SUMMIT} simulations for the multipoles of the correlation function for three tracers: LRGs (left), ELGs (middle), and QSOs (right). The shadow region represents the errors of the covariance matrix, while the error bar represents the same error re-scaled by $1/\sqrt{25}$. }   
\label{fig:meanABACUS}
\end{figure*}

In this work, we use three re-scaling for the covariance: without re-scaling that corresponds to a volume of one simulation $8\,h^{-3}\,\textrm{Gpc}^{3}$, re-scaling by a factor 1/5 that corresponds to a volume of $V_{5}=40\,h^{-3}\,\textrm{Gpc}^{3}$  and re-scaling 1/25 corresponding to $V_{25}=200\,h^{-3}\,\textrm{Gpc}^{3}$.
In Figure \ref{fig:meanABACUS} we show as shaded region the square root of the diagonal terms of the \texttt{EZ mocks} covariance without re-scaling for the three tracers, LRG (left panel), ELG (intermediate panel), and QSO (right panel). We also add error bars corresponding to the 1/25 re-scaled covariance.

\section{Modeling Redshift Space Distortions with GSM-EFT }
\label{sec:model}
In this section, we introduce the methodology we follow to compute the theoretical templates of the correlation function using our EFT-GSM model. We begin in section \ref{subsec:gsm} by briefly introducing the theoretical background behind GSM, which in addition to the convolution Lagrangian perturbation theory (CLPT) and EFT, allows us to model the redshift space two-point correlation function. Then, in section \ref{NN_intro}, we summarize the methodology followed to build the machine learning emulators that enhance the efficiency and speed of our analysis.

\subsection{Gaussian streaming model}\label{subsec:gsm}
The GSM-EFT code\footnote{\href{https://github.com/alejandroaviles/gsm}{https://github.com/alejandroaviles/gsm}} \cite{Ramirez2023} is based on the LPT approach, where each cold dark matter fluid element position is parameterized by its initial location  $\vq$ and the Lagrangian displacement field $\Ps(\vq,t)$, as $\vx(\vq,t) = \vq + \Ps(\vq,t)$. Here, the displacement is governed by the Euler equation $\partial_{t}^{2} {\bf \Psi}+2 H \partial_{t} {\bf \Psi}=-\frac{1}{a^2}\nabla {\bf \Phi}({\bf q}+{\bf \Psi})$ and, which we solve perturbatively by expanding the displacement as,
 \begin{equation}
\bf{\Psi}(\bf{q})=\bf{\Psi}^{(1)}(\bf{q})+\bf{\Psi}^{(2)}(\bf{q})+ \cdots.
\end{equation}
Since we work with the observed position $\vec s$, we must consider 
the LoS anisotropies introduced by particles' peculiar velocities. By doing this, the map between Lagrangian and redshift space coordinates is given by,
\begin{equation}
 \vec s = \vq + \mathbf{\Psi} + \vhn \frac{\mathbf{\dot{\Psi}} \cdot \vhn}{H},
\end{equation}\label{eq:qtos}
where $\vhn$ is a unitary vector on the LoS direction.
The correlation function in redshift space is obtained through the conservation of the number of tracers $X$, expressed as $\big[1+\delta_s(\vec s)\big]d^3s = \big[1+\delta_X(\vx)\big]d^3x$.  Following \cite{Scoccimarro:2004tg,Vlah:2018ygt}, one obtains
\begin{equation} \label{xisC}
  1 + \xi_s(\vs) = \int \Dk{k} d^3x \,e^{i\vk\cdot(\vs - \vr)} \exp \left[ \sum_{n=0}^{\infty} \frac{(-i)^n}{n!} k_{i_1}\cdots k_{i_n} \mathcal{C}^{(n)}_{i_1\cdots i_n}(\vr) \right], 
\end{equation}
where the relation between cumulants $\mathcal{C}$ and moments $\Xi$, of  the pairwise density weighted velocity distribution function, up to second order is given by
\begin{align}
 \mathcal{C}^{(0)}(\vr) &=\log[1 + \xi(r)],\\
 \mathcal{C}^{(1)}_i(\vr) &= \frac{\Xi^{(1)}_i(\vr)}{1+\xi(r)}\equiv v_{12,i}, \\
 \mathcal{C}^{(2)}_{ij}(\vr) &=  \frac{\Xi^{(2)}_{ij} (\vr)}{1+\xi(r)} - \mathcal{C}^{(1)}_i(\vr) \mathcal{C}^{(1)}_j(\vr) \equiv  \hat{\sigma}^{2}_{12,ij} - v_{12,i} v_{12,j} = \sigma^{2}_{12,ij},
\end{align}
where $\xi(r)$ is the tracers real space correlation function and the the moments are given by
\begin{equation}
  \Xi^{(n)}_{i_1\cdots i_n}(\vr) = \langle (1+\delta(\vr))(1+\delta(0)) \Delta u_{i_1}\cdots \Delta u_{i_n} \rangle,  
\end{equation}
where  $ \Delta u_{i}=u_{i}(\vr)-u_{i}(0)$, and $u_i = \hat{n}_i \hat{n}_j \dot{\Psi}_j/H  $ is the peculiar velocity field along the LoS in Hubble function units.

We have further introduced the pairwise velocity along the LoS $v_{12,i}$, and the pairwise velocity dispersion moment and cumulant, $\hat{\sigma}^{2}_{12,ij}$ and $\sigma^{2}_{12,ij}$, respectively. Throughout this work we employ CLPT to compute the moments \cite{Matsubara:2007wj,Matsubara:2008wx,Carlson:2012bu,Uhlemann:2015hqa, Vlah:2016bcl,Valogiannis2020}. A detailed description of our implementation of CLPT into the GSM can be found in \cite{gsm2023arXiv231017834R}. The precise equations we use are given below, in \S \ref{sec:shapefit_imp}.

By truncating the sum in the exponential in eq.~\eqref{xisC} up to the second moment, we obtain the standard expression for the {\it Gaussian Streaming Model correlation function}:
\begin{equation}\label{xisGSM}
1 + \xi_s(s_\parallel,s_\perp) = \int_{-\infty}^\infty \frac{dr_\parallel}{\big[2\pi \sigma^2_{12}(r,\mu)\big]^{1/2}}
  \big[1+\xi(r)\big] \exp \left[-\frac{\left(s_\parallel - r_\parallel - \mu v_{12}(r) \right)^2 }{2 \sigma^2_{12}(r,\mu)} \right], 
 \end{equation}
  where the pairwise velocity and pairwise velocity dispersion are projected along the separation vector: $v_{12}=v_{12,i}\hat{r}_i$ and  $\sigma_{12}=\sigma_{12,i}\hat{r}_i\hat{r}_j$. The labels ${}_\parallel$ and ${}_\perp$ are used to indicate components parallel and perpendicular to the LoS direction and $\mu$ is the cosine of the angle between the separation direction of the two galaxies and the LoS. 
  
As noticed in \cite{Vlah:2016bcl}, the second moment of the density distribution is very sensitive to small scale physics, hence the velocity dispersion should be corrected by an EFT parameter, such that
\begin{equation}
    \sigma_{12}^2(r) \rightarrow \sigma_{12}^2(r)  + \sigma^2_\text{EFT} \frac{1+\xi^\text{ZA}(r)}{1+\xi(r)}, 
\end{equation}
where $\xi^\text{ZA}(r)$ is the Zel'dovich Approximation matter correlation function.
One may notice that for sufficiently large $r$, one has $1+\xi(r) \approx 1$, recovering the phenomenological description of the dispersion due to the Fingers of God effect: $\sigma^2_\text{EFT} (1+\xi^\text{ZA}(r)) /(1+\xi(r)) \rightarrow \sigma^2_\text{FoG}$. In our final modeling we will use a second counterterm as shown in section \ref{sec:shapefit_imp}.  

To compute the real space correlation function $\xi$, the pairwise velocity $v_{12}$ and velocity dispersion $\sigma_{12}$ we utilize CLPT/EFT  \cite{Carlson:2012bu,Vlah:2015sea} as implemented in \cite{Vlah:2016bcl}. Specifically, we use eqs.~\eqref{CLPTxiB}, \eqref{v12} and \eqref{s212} displayed below. Those equations also specify how the biases and counter-terms (apart from $\sigma^2_\text{EFT}$ introduced above) enter our correlation function modeling. 

\subsection{Neural Network Accelerators}
\label{NN_intro}

The methodology outlined in the previous section has been implemented into our GSM-EFT code, enabling the prediction of the galaxy correlation function for a given set of cosmological parameters in just a couple of seconds. Despite this relative speed, the sheer volume of evaluations required for a MCMC analysis typically falls on the order of $10^{5}$. Combined with the numerous independent MCMC analyses presented throughout this work, this incentivizes us to accelerate the evaluation time of our models using neural network emulators.

These emulators work as surrogate models that learn to predict the correlation function of the parameters in approximately 0.015 seconds, resulting in a speedup of two orders of magnitude once trained. Throughout this work, we construct a distinct neural network for each redshift and for each multipole of the correlation function. Each of these emulators requires around 30 minutes to train. Our neural network approach is an adaptation of the code presented in \cite{2022JCAP...04..056D}\footnote{Their code is publicly available and can be found at \href{https://github.com/sfschen/EmulateLSS/tree/main}{https://github.com/sfschen/EmulateLSS/tree/main}}, which we modified to work in configuration space instead of Fourier space, as originally designed. The neural network architecture utilizes a Multi-Layer Perceptron with four hidden layers, each containing 128 neurons. In this work, we construct one individual neural network per multipole. 

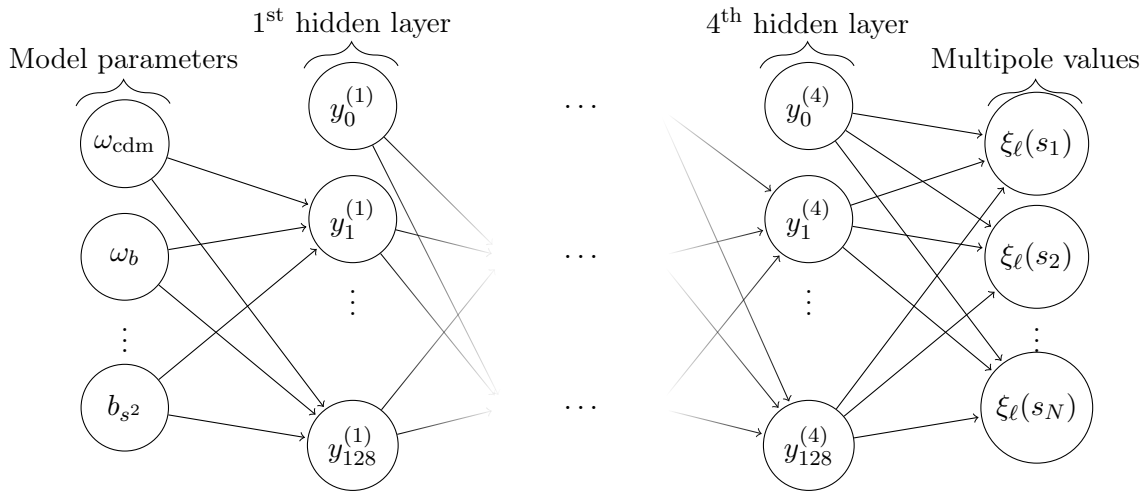
\begin{figure}[t]
	\centering
	\begin{tikzpicture}[shorten >=1pt]
		\tikzstyle{unit}=[draw,shape=circle,minimum size=1.15cm]
		\tikzstyle{hidden}=[draw,shape=circle,minimum size=1.15cm]
 
		\node[unit](x0) at (0,3.5){$\omega_{\textrm{cdm}}$};
		\node[unit](x1) at (0,2){$\omega_b$};
		\node at (0,1){\vdots};
		\node[unit](xd) at (0,0){$b_{s^2}$};
 
		\node[hidden](h10) at (3,4){$y_0^{(1)}$};
		\node[hidden](h11) at (3,2.5){$y_1^{(1)}$};
		\node at (3,1.5){\vdots};
		\node[hidden](h1m) at (3,-0.5){$y_{128}^{(1)}$};
 
		\node(h22) at (5,0){};
		\node(h21) at (5,2){};
		\node(h20) at (5,4){};
		
		\node(d3) at (6,0){$\ldots$};
		\node(d2) at (6,2){$\ldots$};
		\node(d1) at (6,4){$\ldots$};
 
		\node(hL12) at (7,0){};
		\node(hL11) at (7,2){};
		\node(hL10) at (7,4){};
		
		\node[hidden](hL0) at (9,4){$y_0^{(4)}$};
		\node[hidden](hL1) at (9,2.5){$y_1^{(4)}$};
		\node at (9,1.5){\vdots};
		\node[hidden](hLm) at (9,-0.5){$y_{128}^{(4)}$};
 
		\node[unit](y1) at (12,3.5){$\xi_{\ell} (s_1) $};
		\node[unit](y2) at (12,2){$\xi_{\ell} (s_2)$};
		\node at (12,1){\vdots};	
		\node[unit](yc) at (12,0){$\xi_{\ell} (s_N)$};
 
		\draw[->] (x0) -- (h11);
		\draw[->] (x0) -- (h1m);
 
		\draw[->] (x1) -- (h11);
		\draw[->] (x1) -- (h1m);
 
		\draw[->] (xd) -- (h11);
		\draw[->] (xd) -- (h1m);
 
		\draw[->] (hL0) -- (y1);
		\draw[->] (hL0) -- (yc);
		\draw[->] (hL0) -- (y2);
 
		\draw[->] (hL1) -- (y1);
		\draw[->] (hL1) -- (yc);
		\draw[->] (hL1) -- (y2);
 
		\draw[->] (hLm) -- (y1);
		\draw[->] (hLm) -- (y2);
		\draw[->] (hLm) -- (yc);
 
		\draw[->,path fading=east] (h10) -- (h21);
		\draw[->,path fading=east] (h10) -- (h22);
		
		\draw[->,path fading=east] (h11) -- (h21);
		\draw[->,path fading=east] (h11) -- (h22);
		
		\draw[->,path fading=east] (h1m) -- (h21);
		\draw[->,path fading=east] (h1m) -- (h22);
		
		\draw[->,path fading=west] (hL10) -- (hL1);
		\draw[->,path fading=west] (hL11) -- (hL1);
		\draw[->,path fading=west] (hL12) -- (hL1);
		
		\draw[->,path fading=west] (hL10) -- (hLm);
		\draw[->,path fading=west] (hL11) -- (hLm);
		\draw[->,path fading=west] (hL12) -- (hLm);
		
		\draw [decorate,decoration={brace,amplitude=10pt},xshift=-4pt,yshift=0pt] (-0.5,4) -- (0.75,4) node [black,midway,yshift=+0.6cm]{Model parameters};
		\draw [decorate,decoration={brace,amplitude=10pt},xshift=-4pt,yshift=0pt] (2.5,4.5) -- (3.75,4.5) node [black,midway,yshift=+0.6cm]{$1^{\text{st}}$ hidden layer};
		\draw [decorate,decoration={brace,amplitude=10pt},xshift=-4pt,yshift=0pt] (8.5,4.5) -- (9.75,4.5) node [black,midway,yshift=+0.6cm]{$4^{\text{th}}$ hidden layer};
		\draw [decorate,decoration={brace,amplitude=10pt},xshift=-4pt,yshift=0pt] (11.5,4) -- (12.75,4) node [black,midway,yshift=+0.6cm]{Multipole values};
	\end{tikzpicture}
	\caption[Our NN]{Architecture representation of our three Neural Networks, one identical for each multipole ($\ell =0,2,4$) of the correlation function.}
	\label{fig:multilayer-perceptron}
\end{figure}

Figure \ref{fig:multilayer-perceptron} shows a scheme of our methodology. We implemented a decreasing learning rate strategy, starting from $10^{-2}$ and decreasing to $10^{-6}$ in steps of one order of magnitude. We reduce the learning rate with a patience of 1000 epochs, i.e., if there is no improvement for 1000 epochs. At each step, the batch size is doubled. Additionally, we employ the following activation function, as suggested by \cite{2020ApJS..249....5A,2022JCAP...04..056D}:

\begin{equation}
    a(X)=\left[\gamma+(1+e^{-\beta \odot X} )^{-1}(1-\gamma)\right]\odot X.
\end{equation}

These emulators were first introduced and thoroughly tested in \cite{gsm2023arXiv231017834R}. Here, we provide a brief description of the emulator methodology, with a reference to our original work for a more detailed discussion.

To train a neural network, we initially generate 60,000 points in parameter space. This holds true for most results presented in this work, with the exception of certain extensions to our baseline analysis where the models are more complex, which requires a larger training set, as discussed in sections \S\ref{ns_extension} and \S\ref{w0wa_extension} below. Of these points, 50,000 are utilized as the training data, fed into the training algorithm. Additionally, we allocate 5,000 points as the validation set, which the neural network employs to assess the accuracy of the model at each training epoch. Based on the performance of the model, the algorithm may adjust the learning rate or finalize the model. The remaining 5,000 points constitute the test set, which we employ to evaluate the model's accuracy on unseen data. All three sets are independently generated using a Korobov design \citep{korobov1959approximate} to construct uniform samples in a large dimensional space.

\begin{figure}
    \centering
    \includegraphics[width=16cm]{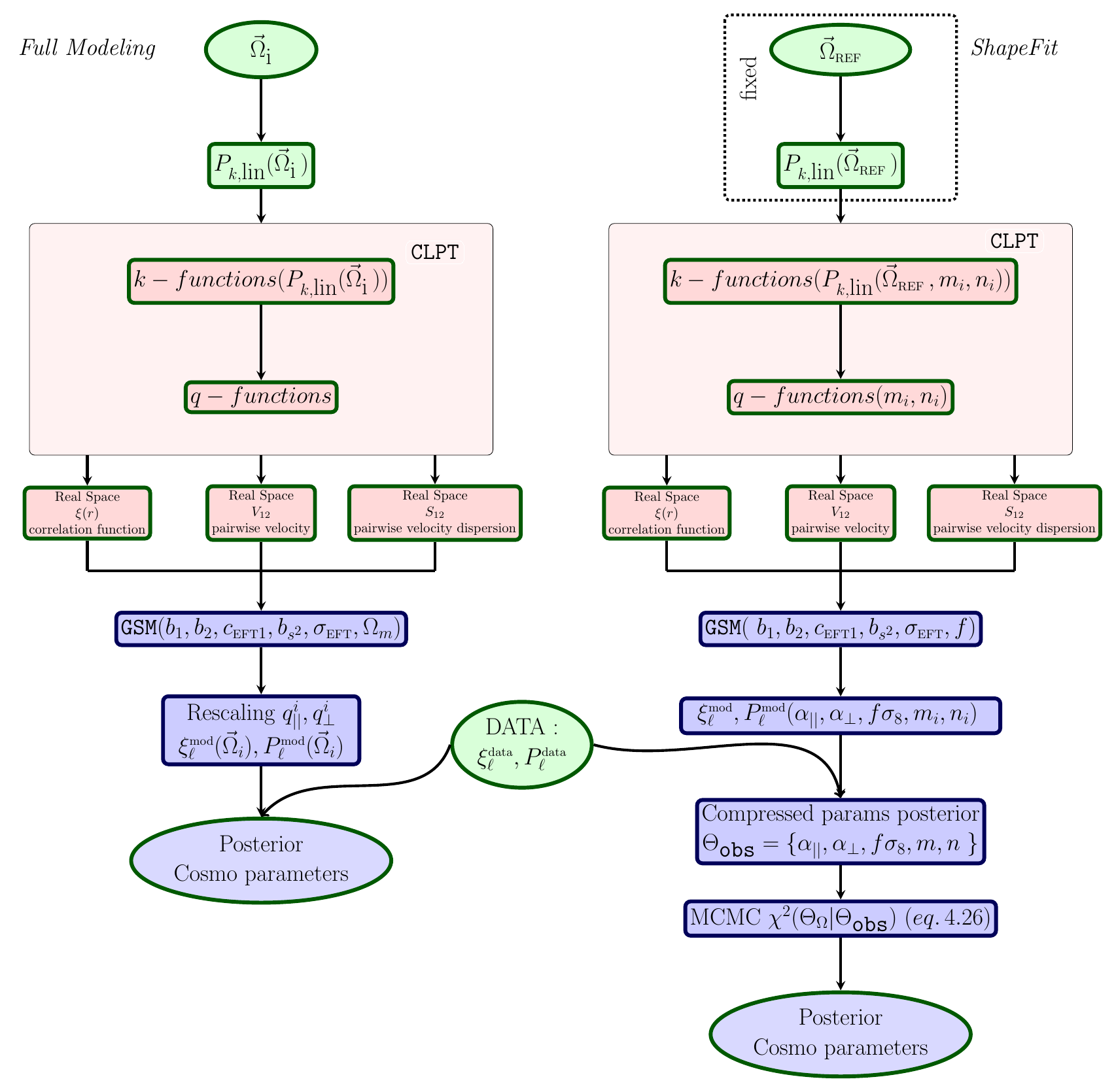}
    \caption{Schemes for the {\it Full Modeling} (left panel) and the {\it ShapeFit} (right panel) analysis, allowing to show the differences in the modeling and the cosmological parameters extraction of the information.}
    \label{fig:methods}
\end{figure}

\section{Extracting Cosmological Information from Clustering}
\label{sec:methodology}

As we illustrate in Figure \ref{fig:methods}, we can distinguish two different approaches for extracting cosmological information from clustering: 

1) \textit{Compressed approaches} \citep[e.g.][]{2011AJ....142...72E,2013AJ....145...10D,2015ApJS..219...12A,2020MNRAS.492.4189I,2021MNRAS.500.3254R,2020MNRAS.498.2354R,2020ApJS..250....8L} shown in the right panel, where the philosophy is to extract the cosmological information from the measured clustering data $\{\xi(s,\mu),p(k,\mu)\} $ through the constraints on compressed variables, $\Theta_{\texttt{obs}}$, that are related with physical observables. 
These compressed variables are further translated into constraints to cosmological parameters using different cosmological models. In this approach, the constraints on the physical observables are proven to be mildly affected by the cosmological model and, for this reason, it is considered agnostic to the cosmological model.  2) 
\textit{Full Modeling} or \textit{Direct fits} \citep[e.g.][]{Ivanov:2019pdj,DAmico:2019fhj, Philcox_2020,Troster_2020,Chen_2022,Semenaite_2023} shown in the left panel, where the extraction of cosmological information from the measured clustering data $\{\xi(s,\mu),p(k,\mu)\}$ is done by comparing them directly with theoretical predictions computed for each cosmological model constraining directly the cosmological (and nuisance) parameters $\theta_{\Omega}$ of the model, without any intermediate step.  
 {\it Full Modeling} has received increasing attention post-Stage-III survey's period, given the optimization of models and codes and the use of sophisticated statistical tools in the last decade, that allows us to compute efficiently the posterior in a highly dimensional space while varying directly the cosmological and nuisance parameters of the model. 

 Within the compressed approaches, we can distinguish two methodologies: 1) \textit{Classical} or \textit{Standard} methodology proposed and developed within spectroscopic survey collaborations (as SDSS-II, BOSS and eBOSS), which is based on compressing the cosmological information into measuring physical observables denoted as BAO and RSD; and 2) \textit{ShapeFit} \citep{shapefit2021JCAP...12..054B} approach that is an extension of the classical approach aiming to achieve more constraining power through the inclusion of two additional parameters that bring information about the early universe. In this section, we briefly describe the three different approaches and their implementation in configuration space for the model used in this work, the \texttt{GSM-EFT}. 

\subsection{Classical methodology}
\label{sec:Standard}
 The {\it Classical} or {\it Standard} methodology 
 
 is based on a \textit{fixed template}, which is generated using a particular perturbative model, computed for a fixed reference cosmology. The {\it fixed template} is modified by three compressed variables associated with late-time dynamics: 
 
 \begin{equation}
 \{ \alpha_{\parallel}, \alpha_{\perp}, f\sigma_8\}, 
 \end{equation}
$\alpha_{\perp}$ and $\alpha_{\parallel}$ parameterize distortions parallel and perpendicular to the line of sight and control the variations along the $r$-axis. The $f\sigma_8$ parameter varies the relative amplitude of our multipoles. The quantity $f\sigma_8$ is the product of the linear growth rate of structures $f$ defined by,

\begin{equation}\label{eq:f}
    f= \frac{d\ln D(a)}{d \ln a},
\end{equation}
where $D$ is the growing mode solution. The amplitude of the matter power spectrum fluctuations on scales up to $8\,h^{-1}\, \text{Mpc}$, $\sigma_{8}$, defined as
\begin{equation}\label{eq:sigma8}
    \sigma_8^2(z)=\int_0^{\infty}  \mathrm{d} (\mathrm{ln} \,  k) \,  k^3P(k, z)W_{TH}^2(k \, 8h^{-1}\mathrm{Mpc})
\end{equation}
where $W_{TH}$ is a spherical top hat filter. In the standard approach, $\sigma_8$ is usually fixed by the reference cosmology and  is completely degenerate with $f$.

The dilation parameters, $\alpha$'s, are defined as follows:
\begin{equation} \label{APequations_SF}
\alpha_{\perp}\left(z_{}\right)=\frac{D_{A}\left(z_{}\right) r_{\mathrm{d}}^{\mathrm{ref}}}{D_{A}^{\mathrm{ref}}\left(z_{}\right) r_{\mathrm{d}}}, 
\quad
\alpha_{\|}\left(z_{}\right)=\frac{H^{\mathrm{ref}}\left(z_{}\right) r_{\mathrm{d}}^{\mathrm{ref}}}{H\left(z_{}\right) r_{\mathrm{d}}},
\end{equation}

where $D_{A}$ denotes the angular diameter distance, $H$ represents the Hubble parameter, and $r_{\mathrm{d}}$ stands for the sound horizon at the drag epoch. The superscript $\texttt{ref}$ means that the estimation is conducted within the reference cosmology\footnote{Here we refer with reference cosmology as a synonym of fiducial cosmology, we are not distinguishing between grid cosmology or template cosmology.}. In our model, we adopt the  alternative parametrization of the distortion parameters:
\begin{equation}\label{alphas}
\alpha=\alpha_{||}^{1/3}\alpha_{\perp}^{2/3}, \quad \epsilon=\left ( \frac{\alpha_{||}}{\alpha_{\perp}} \right)^{1/3}.
\end{equation}

We incorporate distortions in the clustering by substituting \(s\) with \(s\prime(s_{\mathrm{ref}}, \mu_{\mathrm{ref}})\) and \(\mu\) with \(\mu\prime(\mu_{\mathrm{ref}})\). These substitutions are computed using the re-scaling parameters $\alpha$ and $\epsilon$ in the following way:

\begin{eqnarray}
s\prime(s_{\mathrm{ref}},\mu_{\mathrm{ref}})=s_\mathrm{ref} \; \alpha\sqrt{(1+\epsilon)^4 \mu_{\mathrm{ref}}^2 +(1-\mu_{\mathrm{ref}}^2)(1+\epsilon)^{-2}}  , \\
\quad \mu\prime^2(\mu_\mathrm{ref})=\left [  1+ \left (\frac{1}{\mu_{\mathrm{ref}}^2 } -1\right)(1+ \epsilon)^{-6} \right ]^{-1}.
\end{eqnarray}
The multipoles, denoted as $\xi_\ell(s_\mathrm{ref})$, are estimated within the reference cosmology using the transformed coordinates $s\prime(s_\mathrm{ref},\mu_\mathrm{ref})$ and $\mu\prime(\mu_\mathrm{ref})$. 

To incorporate the dilation parameters into our implementation, we employ interpolation for each modeled multipole, $\xi_\ell^\mathrm{model}$, utilizing the transformed coordinates $s\prime(s_\mathrm{ref},\mu_\mathrm{ref})$. Additionally, we calculate the observed Legendre polynomials, $\mathcal{L}^\mathrm{obs}(\mu\prime)$, employing the transformed coordinate $\mu\prime(\mu_\mathrm{ref})$. We formulate $\xi^\mathrm{obs} (s\prime(s_\mathrm{ref},\mu_\mathrm{ref}) , \mu\prime(\mu_\mathrm{ref}))$ by combining the product of each multipole and its corresponding Legendre polynomial. 
The observed multipoles, are expressed as $\xi_\ell (s\prime) = \sum_{ \ell\prime} a_{\ell \ell\prime}\xi_{\ell\prime}(s)$, where the coefficients $a_{\ell \ell\prime}$ are involved in this summation\footnote{As shown in previous work \cite{gsm2023arXiv231017834R}, an excellent approximation is achieved by truncating the summation at \(\ell\prime = 4\).}.

In our {\it Standard} methodology, we fit the data multipoles to the observed multipoles, and then, we get constraints on the compressed parameters $\Theta_{\texttt{obs}}$ and their covariance $C_{\Theta}$. 
The conversion of compressed parameter constraints into cosmological constraints is done separately and is described in \S\ref{compresed_to_cosmo} below.

\subsection{{\it ShapeFit} methodology}
\label{shapefit_mehtod}

The {\it ShapeFit} methodology \citep{brieden_2021,brieden_2021b,Brieden_2022c,Brieden_2022d} emerges as an extension to the {\it Standard} methodology, aiming to preserve its agnostic nature while incorporating additional information to enhance its constraining power, thus making it competitive with {\it Full Modeling}.

This is done by incorporating two additional parameters into the \textit{Standard} methodology variables, compressing the data vector into five parameters ${\alpha_{||}, \alpha_{\perp}, f\sigma_{s8}, m, n}$.  The two new parameters affect the slope of the power spectrum, in two different ways: first, an overall re-scaling of the power spectrum slope, through the parameter $n$, which in principle accounts for a wrong fiducial spectral index $n_s$; and second, a scale-dependent slope, through the parameter $m$, that models the shape of the power spectrum mainly at wave-numbers around and above the turnaround. This re-scaling process is applied to the template power spectrum $P_{\mathrm{ref}}$ to produce a trial spectrum $P_{\mathrm{ref}}^{\prime}$, defined as follows

\begin{equation}
\label{slope}
\ln \left(\frac{P_{\mathrm{ref}}^{\prime}(k)}{P_{\mathrm{ref}}(k)}\right)=\frac{m}{a} \tanh \left[a \ln \left(\frac{k}{k_p}\right)\right]+n \ln \left(\frac{k}{k_p}\right),
\end{equation}
where the pivot scale  $k_p=\pi/r_d$ corresponds to the location where the baryon suppression reaches its maximum. 
 
The two slope re-scaling parameters recover information from the early universe that is imprinted in the broadband shape of the matter power spectrum, this information is omitted in the {\it Standard} methodology.
In order to extract cosmological information, the slope $m$ 
is interpreted in terms of the cosmological parameters through \cite{shapefit2021JCAP...12..054B}
\begin{align}\label{eq:m}
m &=\left.\frac{d}{d \ln k}\left(\ln \left[\frac{T^2_{\text {no-wiggle }}\left(k_p / s, \Theta\right)}{T^2_{\text {no-wiggle }}\left(k_p, \Theta^{\mathrm{ref}}\right)}\right]\right)\right|_{k=k_p} \nonumber\\
&=\left.\frac{d}{d \ln k}\left(\ln \left[\frac{P_{\text {no-wiggle }}^{\text {lin }}\left(k_p / s, \boldsymbol{\Theta}\right)}{P_{\text {no-wiggle }}^{\text {lin }}\left(k_p, \boldsymbol{\Theta^{\mathrm{ref}}}\right)}\frac{\mathcal{P}_{\mathcal{R}}(k, \Theta^{\mathrm{ref}})}{\mathcal{P}_{\mathcal{R}}(k/s, \Theta)}\right]\right)\right|_{k=k_p}
\end{align}
where
$\mathcal{P}_{\mathcal{R}}$ is the primordial power spectrum and $T$ the transfer function. 
To compute the \textit{de-wiggled} linear power spectrum, $P_{\text {no-wiggle }}^{\operatorname{lin}}$ we use the Eisenstein-Hu formulae of \cite{1999ApJ...511....5E}.

In most situations, the parameter $n$ is fixed to zero. However, by letting it free, one can obtain estimations for $n_s$ through:
\begin{equation}
    n = n_{s} - n^\text{ref}_s,
\label{eq:n}
\end{equation}
where $ n^\text{ref}_s$ is the spectral index of the reference cosmology.

Another modification to the {\it Standard} methodology introduced by {\it ShapeFit} involves redefining the amplitude of the clustering. Instead of using the usual $\sigma_8$ parameterization, we introduce $\sigma_{s8}$ as a re-scaling in units of the sound horizon:
\begin{equation}\label{eq:s8}
    \sigma_{s8}^2(z,  \Omega)=\int_0^{\inf}  \mathrm{d} (\mathrm{ln} \,  k) \,  (ks)^3P(ks, z)W_\text{TH}^2(ks \, 8h^{-1}\mathrm{Mpc})
\end{equation}
where $s= r_d/r_d^{\mathrm{ref}}$. By construction, this re-scaling does not vary under the fitting process, and all the compressed variables are expressed in units of the sound horizon. 

 The templates used in {\it ShapeFit}  include non-linearities of the power spectrum, which are commonly computed using Eulerian Perturbation Theory. However, to speed up computations, one can use approximations as,
\begin{align}
P_{22}(k) & =\int \Dk{p} P_{\mathrm{ref}}^{\prime}(p) P_{\mathrm{ref}}^{\prime}(|\vk-\vp|) F_2(\vk, \vk-\vp) \\
& \approx \left(\frac{P_{\mathrm{ref}}^{\prime}(k)}{P_{\mathrm{ref}}(k)}\right)^2 \int \Dk{p} P_{\mathrm{ref}}(p) P_{\mathrm{ref}}(|\vk-\vp|) F_2(\vk, \vk-\vp), \label{P22}
\end{align}
Here, $P_{\mathrm{ref}}(p)$ denotes the linear power spectrum of the reference template, while $P_{\mathrm{ref}}^{\prime}(p)$ represents the linear power spectrum modified by the $m$ \textit{ShapeFit} parameter. Similarly for the rest of the loop integrals. A validity test of such approximations can be found in \cite{KP5s3-Noriega}\footnote{This approximation is important for ShapeFit approach because this computation need to be done at each step of the MCMC chains.}. 

\subsubsection{{\it ShapeFit} in Configuration Space}\label{sec:shapefit_imp}
As we illustrated in Figure \ref{fig:methods}, for the computation of the correlation function, we use the GSM presented in \S \ref{subsec:gsm}  and the Convolution LPT framework \cite{Carlson:2012bu} to obtain the necessary ingredients entering eq.~\eqref{xisGSM}, namely the real space correlation function $\xi(r)$, the pairwise velocity $v_{12}(r)$ and the pairwise velocity dispersion $\sigma_{12}(r)$.

For example, let us consider the simplified version of the one-loop dark matter (unbiased) real space correlation function \cite{Carlson:2012bu} that is given by 
\begin{align}\label{CLPTxi}
&1+\xi(r) = \int  \frac{d^3 q}{(2 \pi)^{3/2} |\mathbf{A}_L|^{1/2}} e^{- \frac{1}{2}(\ve r-\vq)^\mathbf{T}\mathbf{A}_L^{-1}(\ve r-\vq) } 
 \Bigg\{ 1 - \frac{1}{2} A_{ij}^\text{loop}G_{ij} +\frac{1}{6}\Gamma_{ijk}W_{ijk} \Bigg\},
\end{align}
with \textit{$q$-functions} $A_{ij}(\vq) = \langle \Delta_i(\vq)\Delta_j(\vq) \rangle_c$, and $W_{ijk}=  \langle \Delta_i(\vq)\Delta_j(\vq)\Delta_k(\vq) \rangle_c$,  where $\Delta_i = \Psi_i(\vq)-\Psi_i(0)$. In the case of matrix $\mathbf{A}$, which is split into linear and loop pieces
\begin{align}\label{ALij}
 A_{ij}(\vq) &= 2 \int \Dk{k} \big( 1 - e^{i\vp\cdot \vq} \big)\frac{k_i k_j}{k^4} \left( P_L(k) + \frac{3}{7} Q_1(k)  + \frac{10}{21} R_1(k)\right) \nonumber\\
 &=  A^L_{ij}(\vq) +  A^\text{loop}_{ij}(\vq), 
\end{align}
where the \textit{$k$-functions} $Q_1(k)$ and $R_1(k)$ are functions constructed out of  four-point functions of linear density fields, defined in \cite{Matsubara:2007wj}  as,
\begin{align}
    Q_1(k) &=\int \Dk{p} q_1(\vk,\vp) P_L(p) P_L(|\vk-\vp|), \\
    R_1(k) &=\int \Dk{p} r_1(\vk,\vp) P_L(k) P_L(p), 
\end{align}
for kernels $q_1$ and $r_1$. In {\it ShapeFit} we produce the trial matrix  $A\prime_{ij}$, by substituting the trial power spectrum $P\prime_{ref}$ of eq.~\eqref{slope} into eq.~\eqref{ALij}. To speed up the computations, as it is done in the case of the non-linear power spectrum, we evaluate functions $Q_1$ and $R_1$ as 
\begin{align}
    Q_{1}(k;m,n) &= \left(\frac{P_{\mathrm{ref}}^{\prime}(k;m,n)}{P_{\mathrm{ref}}(k)}\right)^2 \int \Dk{p} q_1(\vk, \vp)  P_{\mathrm{ref}}(p) P_{\mathrm{ref}}(|\vk-\vp|), \\
    R_{1}(k;m,n) &=\left(\frac{P_{\mathrm{ref}}^{\prime}(k;m,n)}{P_{\mathrm{ref}}(k)}\right)^2 \int \Dk{p} r_1(\vk, \vp)  P_{\mathrm{ref}}(k) P_{\mathrm{ref}}(p), 
\end{align}
where we have written  explicitly the dependence of the trial functions on the {\it ShapeFit} parameters. 

This method is analogously applied for all $k$-functions, from which the $q$-functions are constructed, and finally for the real space correlation function $\xi(r)$, the pairwise velocity $v_{12}(r)$, and velocity dispersion $\sigma_{12}(r)$, which are necessary to construct the redshift space GSM correlation function.

These ingredients are computed for biased tracers using CLPT, obtaining,
The real space correlation function $\xi_{X}(r)$, which corresponds to the zeroth-order moment for tracer $X$, is obtained within CLPT, \cite{Matsubara:2007wj,Carlson:2012bu,Vlah:2015sea,Uhlemann:2015hqa,Vlah:2018ygt,Aviles:2018thp},
\begin{align}\label{CLPTxiB}
&1+\xi_{X}(r) = \int  \frac{d^3 q}{(2 \pi)^{3/2} |\mathbf{A}_L|^{1/2}} e^{- \frac{1}{2}(\ve r-\vq)^\mathbf{T}\mathbf{A}_L^{-1}(\ve r-\vq) } 
 \Bigg\{ 1 - \frac{1}{2} A_{ij}^{loop}G_{ij} +\frac{1}{6}\Gamma_{ijk}W_{ijk} \nonumber\\
&\quad + b_1 (-2 U_i g_i - A^{10}_{ij}G_{ij}) + b_1^2 (\xi_L - U_iU_jG_{ij}- U_i^{11}g_i) + b_2(\frac{1}{2} \xi_L^2 -U_i^{20}g_i - U_iU_jG_{ij}) \nonumber\\
&\quad - 2 b_1 b_2 \xi_L U_i g_i  -2b_{s^2} g_i V_i^{10}+b_{s^2}^2 \zeta-2 b_1 b_{s^2} g_i V_i^{12}+b_2 b_{s^2} \chi^{12} 
-\frac{1}{2} c_{1,\text{EFT}} \operatorname{tr} G 
\Bigg\},
\end{align}
for bias parameters $b_1$, $b_2$ and $b_{s^2}$ and EFT counterterm $c_{1,\text{EFT}}$. The linear correlation function is
\begin{equation}
\xi_L(q) = \int \Dk{p} e^{i \vp \cdot \vq}   P_L(p),
\end{equation}
and we used the functions
\begin{align}
A_{ij}^{mn} = \langle \delta^{m}(\vq)\delta^{n}(0)\Delta_i \Delta_i \rangle_c, \qquad
U^{mn}_i = \langle \delta^{m}(\vq)\delta^{n}(0)\Delta_i \rangle_c.
\end{align}
The involved $r$ and $q$ dependent tensors are $g_i = (\mathbf{A}_L^{-1})_{ij}(r_j - q_j)$, $G_{ij} = (\mathbf{A}_L^{-1})_{ij} - g_ig_j$, and $\Gamma_{ijk} = (\mathbf{A}_L^{-1})_{\{ij} g_{k\}} - g_ig_jg_k$. 

The pairwise velocity is given by
\begin{align}\label{v12}
&v_{12}(\ve r) = \frac{f\, \hat{r}_i}{1+\xi_{X}(r)}  \int  \frac{d^3 q\, e^{- \frac{1}{2}(\ve r-\vq)^\mathbf{T}\mathbf{A}_L^{-1}(\ve r-\vq) } }{(2 \pi)^{3/2} |\mathbf{A}_L|^{1/2}} 
 \Bigg\{ -g_r \dot{A}_{ri} - \frac{1}{2} G_{rs} \dot{W}_{rsi} \nonumber\\
&\quad + b_1 \left( 2 \dot{U}_i - 2 g_r \dot{A}^{10}_{ri} - 2 G_{rs} U_r \dot{A}_{si} \right) + b_1^2 \left( \dot{U}^{11}_i  - 2 g_r U_r \dot{U}_i  - g_r \dot{A}_{ri} \xi_L  \right)  \nonumber\\
&\quad + b_2 \left( \dot{U}^{20}_i - 2 g_r U_r \dot{U}_i  \right) + 2 b_1 b_2 \xi_L \dot{U}_i  
 \Bigg\},
\end{align}
 and the  pairwise velocity dispersion
\begin{align}\label{s212}
&\sigma^2_{12}(\ve r)= \frac{f^2\, \hat{r}_i\hat{r}_j}{1+\xi_{X}(r)}  \int  \frac{d^3 q \, e^{- \frac{1}{2}(\ve r-\vq)^\mathbf{T}\mathbf{A}_L^{-1}(\ve r-\vq) } }{(2 \pi)^{3/2} |\mathbf{A}_L|^{1/2}}  
 \Bigg\{  \ddot{A}_{ij} - g_r \ddot{W}_{rij} - G_{rs}\dot{A}_{ri}\dot{A}_{sj} \nonumber\\ 
&\quad   + 2 b_1 \left( \ddot{A}^{10}_{ij} -  g_r \dot{A}_{r\{i} \dot{U}_{j\}} - g_r U_r \ddot{A}_{ij} \right) 
+ b_1^2 \left( \xi_L \ddot{A}_{ij} + 2 \dot{U}_i \dot{U}_j \right)  + 2 b_2 \dot{U}_i \dot{U}_j  \,
\Bigg\},
\end{align}
with $q$-coordinate dependent correlators

\begin{align}
&\dot{A}_{ij}^{mn}(\vq) = \frac{1}{f H}\langle \delta^m_1 \delta^n_2 \Delta_i \dot{\Delta}_j \rangle, 
                         \qquad  \ddot{A}_{ij}^{mn}(\vq) = \frac{1}{f^2 H^2}\langle \delta^m_1 \delta^n_2 \dot{\Delta}_i \dot{\Delta}_j \rangle, \nonumber\\
& \dot{W}_{ijk} = \frac{1}{f H}\langle  \Delta_i  \Delta_j \dot{\Delta}_k \rangle,  \qquad \ddot{W}_{ijk} = \frac{1}{f^2 H^2}\langle  \Delta_i  \dot{\Delta}_j \dot{\Delta}_k \rangle,    \nonumber\\
& \dot{U}^{mn}(\vq) = \frac{1}{f H}\langle \delta^m_1\delta^n_2 \dot{\Delta}_i \rangle, 
\end{align}
where $\dot{A}_{ij} \equiv \dot{A}^{00}_{ij}$ and  $\ddot{A}_{ij} \equiv \ddot{A}^{00}_{ij}$. 

Finally, we would like to highlight that our EFT model is not exhaustive, as it does not encompass all PT corrections and counterterms up to 1-loop. Nonetheless, we adopt this simplified approach for two main reasons: firstly, it has been extensively validated in prior research \cite{Ramirez2023}; and secondly, we demonstrate in this work that it satisfactorily meets the requirements of DESI Y1 data. Nevertheless, complete LPT/EFT frameworks are available in the literature \cite{Chen:2020fxs,Chen:2020zjt}.

\subsection{Inferring cosmological parameters from compressed parameters}
\label{compresed_to_cosmo}

As depicted in the bottom row of the schematic in the right panel of Figure \ref{fig:methods}, our compressed methodology incorporates an additional step to derive cosmological parameters from our final compressed parameters. In this section, we briefly summarize the methodology employed to achieve this.

Let $\Theta_{\texttt{obs}}$ represent our final set of compressed parameters. We aim to use them to determine a set of cosmological parameters $\Omega=\{\omega_{m}, A_s, h, ...\}$. For a given value of these parameters we can compute a set of compressed parameters $\Theta_\Omega$ from theory following the equations presented in sections \ref{sec:Standard} and \ref{shapefit_mehtod}.  We use these equations to run an MCMC chain that finds the regions of the parameter space of $\Omega$ for which $\Theta_\Omega$ is a good fit of  $\Theta_{\texttt{obs}}$. Let $C_\Theta$ be the covariance of a given compressed parameter, and then if we assume a Gaussian likelihood\footnote{A Gaussian likelihood is a good approximation when the distributions of $\Theta_{obs}$ are gaussian distributed, what can be verified in the posteriors of the compressed parameters} for our MCMC exploration, the  $\chi^2$ is defined as:

\begin{equation}
    \chi^2(\Theta_\Omega-\Theta_{\texttt{obs}})= (\Theta_\Omega-\Theta_{\texttt{obs}})^{T} C_{\Theta}^{-1}(\Theta_\Omega-\Theta_{\texttt{obs}}),
\label{eq:likelihood_params_AP}
\end{equation}
 
As mentioned earlier, we employ two compression methodologies, namely the \textit{Standard} and \textit{Shape-fit} methods. In our \textit{Standard} approach, $\Theta_\Omega$ is as follows:

\begin{equation}
  \Theta_\Omega=\{ D_A(z,\Omega)/r_d(\Omega), D_H(z)/r_d(\Omega), f\sigma_8(z,\Omega)\}  
\end{equation}

where the equations used for computing the compressed variables are \ref{alphas},  \ref{eq:f} and \ref{eq:sigma8}, and we have explicitly included their dependence on the cosmological models into our notation.

Within our {\it Shape-Fit} methodology, in addition to the equations of 
the {\it Standard} approach, we also need to consider the equations for the additional parameters n and m, thus $\Theta_\Omega$ are given by: 
\begin{equation}
\Theta_\Omega=\{ D_A(z,\Omega)/r_d(\Omega), D_H(z)/r_d(\Omega), f\sigma_{s8}(z,\Omega), m(\Omega), n(\Omega)\},    
\end{equation}

where m, n, and $\sigma_{s8}$ are respectively given by equations \ref{eq:m}, \ref{eq:n}, and   \ref{eq:s8}.
 
Also in the case of {\it ShapeFit}, we have to keep in mind that the $m$ parameter changes the tilt and the amplitude of the power spectrum, which means that we are varying $\sigma_{s 8}$. This deformation of the power spectrum is captured through \citep{shapefit2021JCAP...12..054B}:

\begin{equation} \label{SF_trans}
\begin{split}
f \sigma_{s 8} &= f \frac{\left(f \sigma_{s 8}\right)^{\text {ref }}}{\left(f\left(P_{\operatorname{lin}}\left(k_p\right)\right)^{1 / 2}\right)^{\text {ref }}} \left(\frac{1}{s^3} P_{\text {lin }}\left(k_p / s\right)\right)^{1 / 2} \\
&= \frac{\left(f \sigma_{s 8}\right)^{\mathrm{ref}}}{\left(f A_{s p}^{1 / 2}\right)^{\mathrm{ref}}} (f A_{s p}^{1 / 2} )\times \exp \left(\frac{m}{2a} \tanh \left(a \ln \left(\frac{r_d^{\mathrm{ref}}}{8}\right)\right)\right) \\
&\approx \frac{(f A_{s p}^{1 / 2} )}{\left(f A_{s p}^{1 / 2}\right)^{\mathrm{ref}}}\left(f \sigma_{s 8}\right)^{\mathrm{ref}} 
\end{split}
\end{equation}

where
\begin{equation}
A_{s p}=s^{-3} P_{\text {no-wiggle }}^{\text {lin }}\left(k_p / s, \boldsymbol{\Theta}\right),
\end{equation}
with $\boldsymbol{\Theta}$ the cosmological parameters of the cosmology of reference or the new one.
The last step of equation \ref{SF_trans} can be done thanks to the fact that we only use the amplitude at the pivot scale during the MCMC exploration.

\subsection{{\it Full Modeling} in Configuration space}
\label{fullmodeling_mehtod}
In the {\it Full Modeling} approach, which is illustrated in the left side of Figure \ref{fig:methods}, we use a non-fixed template, which consists of modifying the shape of the linear power spectrum through the variation of the cosmological parameters $\omega_{\mathrm{cdm}}$, $\omega_{b}$, $h$, $A_{s}$, $n_s$, $N_{\text {eff }}$, $\Omega_{\mathrm{ncdm}}$, parameters that also depends on early-time physics, in contrast to the {\it Standard} approach. This methodology involves using a Boltzmann solver code like \textbf{CLASS} or \textbf{CAMB}, and, computing the 1-loop corrections for the perturbative approach used, in each step of the MCMC exploration. Despite its higher computational cost, this allows us to extract cosmological information not only from late-time physics but also from early-time physics (in contrast with {\it Standard} methodology). Hence, the {\it Full Modeling} approach, at least in principle, can impose tighter constraints than the {\it Standard} methodology as it captures more information. As mentioned previously, Shapefit also looks for extract additional information from early-time physics through the additional parameters.

However, the high dimensionality of the parameter space requires the use of new tools to speed up the template generation. In \cite{gsm2023arXiv231017834R}, we presented a neural network approach that enhanced the efficiency of convergence time during MCMC exploration dramatically. 
As clustering measurements employ a cosmology of reference to transform redshift into distance measurements, we need to add two distortion parameters that compensate the Alcock-Paczynski  effect, denoted them as $ q_{||}$ and $q_{\perp}$, and following a similar description of \ref{sec:Standard} with the difference that we do not introduce the BAO scale at the drag epoch, $r_{\text{d}}$, as its effect is already included in the variation of the linear power spectrum.

The geometric distortion parameters, perpendicular, and parallel to the line of sight, in the {\it Full Modeling} approach are defined as
\begin{equation} \label{APequations}
q_{\perp}\left(z_{}\right)=\frac{D_{A}\left(z_{}\right) }{D_{A}^{\mathrm{ref}}\left(z_{}\right) }, 
\quad
q_{\|}\left(z_{}\right)=\frac{H^{\mathrm{ref}}\left(z_{}\right) }{H\left(z_{}\right) },
\end{equation}
here  $D_{A}$ is the angular diameter distance, $H$ is the Hubble parameter, and the $\mathrm{ref}$ superscript indicates that the estimate is done in the reference or fiducial cosmology of the data multipoles. 

The {\it Full Modeling} methodology offers the advantage of capturing a broad range of dynamic-times, as it does not need to fix the early-time dynamics of the universe. However, it requires the adoption of a specific cosmological model, and for exploring alternative models of interest, a complete rerun of the MCMC is needed.

In Table \ref{tab:gsm_priors} we show the varied parameters and their priors for our baseline {\it Full Modeling} analyses. We keep fixed the slope $n_s=0.97$, the effective number of relativistic degrees of freedom $N_\mathrm{eff}=3.046$, the massive neutrino abundance $\omega_\mathrm{ncdm} = 0.00064$ and we impose a big-bang nucleosynthesis (BBN) prior $\mathcal{N}[\mu=0.02237, \sigma=0.00037]$ to $\omega_{b}$ \citep{2018ApJ...855..102C}.

\subsection{\texttt{GSM-EFT} code for compressed and direct fits: Users guide} 

We present our \texttt{GSM-EFT} C-code that computes the multipoles of the one-loop GSM two-point correlation function in redshift space\footnote{\href{https://github.com/alejandroaviles/gsm}{https://github.com/alejandroaviles/gsm}.}.
The \texttt{GSM-EFT} code receives as an input, the linear power spectrum, obtained from a Boltzmann Solver like \texttt{CAMB} or \texttt{CLASS}, as well as a set of nuisance parameters: including biases $\{b_1$, $b_2$, $b_{s^2}$
$\}   $, the EFT parameters $\{\sigma_\text{EFT}^2$, $c_1^\text{EFT} \}$
and the cosmological parameter $\Omega_m$\footnote{$\Omega_m$ is necessary to calculate the growth rate $f$ at the output redshift}. The code produces as  \texttt{GSM-EFT} outputs the correlation function multipoles $\ell=0,2,4$ in redshift space.

\begin{itemize}
\item {\bf \texttt{GSM-CosmoExtractor.}} The repository includes a python wrapper and has two distinct approaches implemented for extracting the cosmological information explored in this work (developed in section \ref{sec:methodology}).
The wrapper also includes a function to compute the  \texttt{CAMB}/\texttt{CLASS} linear power spectrum required as an input by our \texttt{GSM-EFT} code. The methodology for modeling the effects of the re-scaling parameters $\{\alpha_{||},\alpha_\perp\}$ (Eqs. \ref{APequations_SF}) required for compressed parameters, as well as the re-scaling parameters in {\it Full Modeling}, the $\{ q_{||},q_{\perp}\}$ are also implemented in the python wrapper.

\item {\texttt{GSM-EFT} modules }
The \texttt{GSM-EFT} code is composed of two separate modules: \texttt{CLPT} and \texttt{GSM}. The \texttt{CLPT} module computes the real-space correlation function, pairwise velocity, and velocity dispersion. Then, the \texttt{GSM} module receives the \texttt{CLPT} output statistics as inputs along with values of the nuisance parameters (bias and EFT terms), and outputs the redshift-space two-point correlation function.  
 In Section \ref{sec:shapefit_imp} below, we describe the implementation of the compressed approaches in the wrapper as well as in the \texttt{GSM-EFT} code. The \texttt{GSM-EFT} code takes only $\sim1$ second in computing the redshift multipoles, using either the {\it Full Modeling} or {\it ShapeFit } approaches. While this is a relatively small time, given the sheer number of evaluations required for our MCMC analysis and the large number of analyses we require, there is an incentive to accelerate the model evaluation time even further. As introduced in section \ref{NN_intro}, we generated an emulator to reduce the processing time to around $0.015$ seconds per evaluation.

\item {\bf {\it ShapeFit} implementation in \texttt{GSM-EFT}.}
As indicated in Figure \ref{fig:methods}, the implementation of {\it ShapeFit} in \texttt{GSM-EFT} code implies a modification of the \texttt{CLPT} module. This is done through two modifications to the original pipeline:
    First, the k-funtions that depend on the $P_L(k)$ and loop corrections, are pre-computed using the modified linear power spectrum $P_\texttt{ref}(k)$.
    Then, the q-functions are computed from k-functions, this is done trough the
    re-scaling factor  $\left(\frac{P_{\mathrm{ref}}^{\prime}(k)}{P_{\mathrm{ref}}(k)}\right)^2$.
        
\end{itemize}

\subsection{Likelihood and Priors}
\label{Likelihood_and_Priors}

Our exploration of the parameter space is done through the Bayes law, with the Bayesian evidence neglected. For a given model, the posterior probability of a point in the parameter space, denoted as $\mathcal{P}(\boldsymbol{\theta})$, is defined as follows:

\begin{equation}
\mathcal{P}(\boldsymbol{\theta} \mid \boldsymbol{D})\propto \mathcal{L} (\boldsymbol{D} \mid \boldsymbol{\theta}) \times \mathcal{\pi}(\boldsymbol{\theta}), 
\end{equation}
where, $\mathcal{L} (\boldsymbol{D} \mid \boldsymbol{\theta})$  is the likelihood and $\mathcal{\pi}(\boldsymbol{\theta})$ is the prior distributions. If we assume that the parameter space follows a normal distribution, we can express the likelihood as,

\begin{equation}
\label{eq:posterior}
\mathcal{P}(\boldsymbol{D} \mid \boldsymbol{\theta}) \propto  
(\chi^{2})^{\frac{\nu-2}{2}} \exp\left(-\frac{\chi^{2}}{2}\right),
\end{equation}
where $\nu$ is the number of degrees of freedom, and $\chi^{2}$ is defined as $\chi^{2}= (\vec{m}-\vec{d})^{T} C^{-1}(\vec{m}-\vec{d}),$
and the vectors \(\vec{m}\) and \(\vec{d}\) represent the model and data respectively. The covariance matrix of the data is denoted as \(C\).

The exploration of parameter space is done using the publicly available MCMC code \texttt{emcee} \cite{Foreman-Mackey:2012any} which is an ensemble sampler with affine invariance defined in \cite{emceetheory}. This code has some advantages, like enabling parallelization, allowing us to take advantage of the computational power of large clusters. In our work, we employ the priors outlined in Table \ref{tab:gsm_priors} for all of our three independent methodologies. In all the cases we fix the spectral index $n_s=0.97$ unless stated otherwise, the effective number of relativistic degrees of freedom $N_\mathrm{eff}=3.046$, and the massive neutrino abundance $\omega_\mathrm{ncdm} = 0.00064$. 

To guarantee convergence, we employ an integrated auto-correlation time, which is monitored every 100 steps \cite{emceetheory}. Convergence is considered achieved when two conditions are simultaneously satisfied: the chain's length surpasses 100 times the estimated auto-correlation time, and the alteration in this estimation remains below $1\%$.

\begin{table}
\ra{1.4}
\centering                          
\begin{tabular}{lc c c}       
\hline
\multicolumn{4}{c} {Free parameters and priors} 
\\
\hline
\underline{\texttt{Cosmological}}&&\underline{\texttt{Compressed}}& \\
$\, h$ &
$\mathcal{U}[0.55, 0.91]$ & $f \sigma_{8}$&$\mathcal{U}[0, 1]$\\
$\, \omega_{b}$&
$\quad \mathcal{N}[0.02237,0.00037]\quad$ & $\alpha_{\perp}$ & $\mathcal{U}[0.5, 1.5]$ \\
$\, \omega_{\mathrm{cdm}}$ &
$\mathcal{U}[0.08,0.16]$ &$\alpha_{\parallel}$ & $\mathcal{U}[0.5, 1.5]$\\
$\, \log(10^{10} A_s)$ &
$\mathcal{U}[2.0, 4.0]$ &$m$ & $\mathcal{U}[-1, 1]$ \\[0.2cm]
\hline
\multicolumn{4}{c} {\underline{\texttt{Nuisances}}} 
\\
$b_1$ & $\mathcal{U}[0,2.0]$&$b_2$ &$\mathcal{U}[-5,10]$\\
$\sigma^2_\text{EFT}$& $\mathcal{U}[-20,100]$&$c_{1,\text{EFT}}$ &$\mathcal{U}[-100,100]$ \\
$b_{s^2}$ &$\mathcal{U}[-10,10]$& \\[0.2cm]
\hline

\end{tabular}
\caption{Free parameters and priors we use for the baseline {\it Full Modeling} analyses of \texttt{ABACUS-SUMMIT} simulations. The nuisance parameters with their priors are the same for all three methods. We fix the spectral index of the power spectrum to the Planck 2019 value $n_s=0.9649$. 
}    
\label{tab:gsm_priors} 
\end{table}

\section{Cosmological constraints from configuration-space clustering statistics}

\label{sec:results_gsm}
In this section, we discuss and compare the results obtained with our pipeline following three different methodologies: {\it Standard}, {\it ShapeFit}, and {\it Full Modeling}. However, our attention will primarily be on {\it ShapeFit} and {\it Full Modeling}, given their stronger constraining potential. In Section \ref{sec:baseline}, we explore the settings for defining our baseline analysis, in particular the range of the fits, and the bias settings. In the Section \ref{sec:extensions_base}, we explore extensions to the baseline analysis, as including the hexadecapole, the extension of the parameter space including $n_s$ and the $\omega_0-\omega_a$. Section \ref{sec:tracers}  is devoted to exploring the performance of the analysis for the different DESI tracers and their combination of them. Finally, in the last section, we compare the constraining power of the two methodologies, {\it ShapeFit}, and {\it Full Modeling}, in the cosmological parameter space. 

\subsection{Baseline Analysis}\label{sec:baseline}
\begin{figure*}
\centering
\includegraphics[width=75mm,height=75mm]{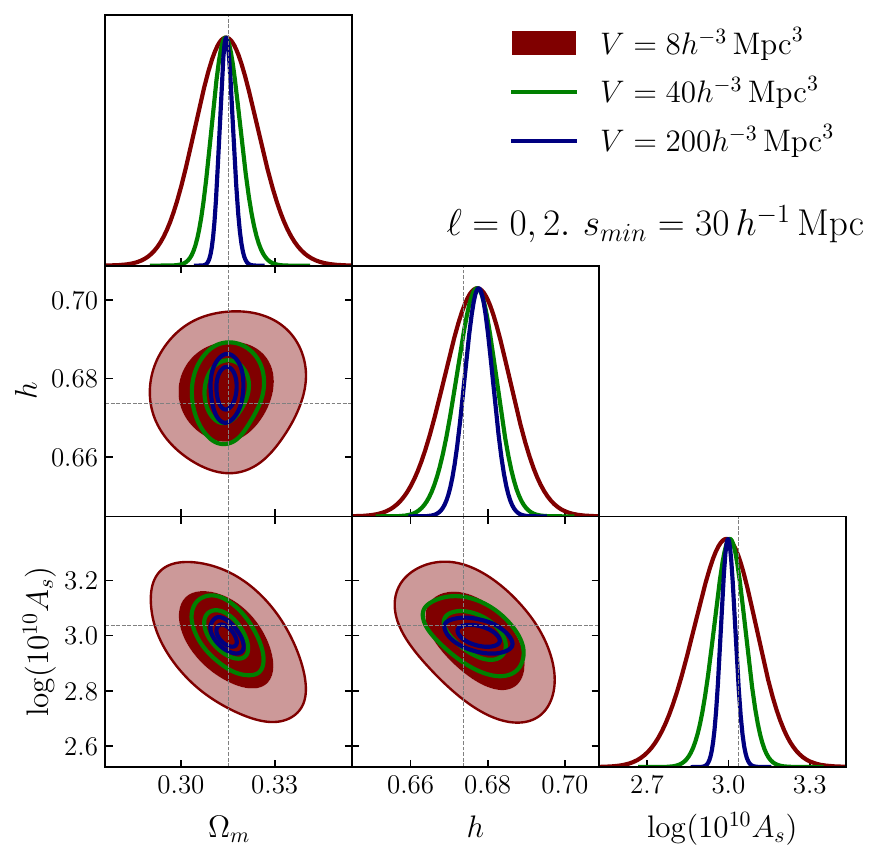}
\includegraphics[width=75mm,height=75mm]{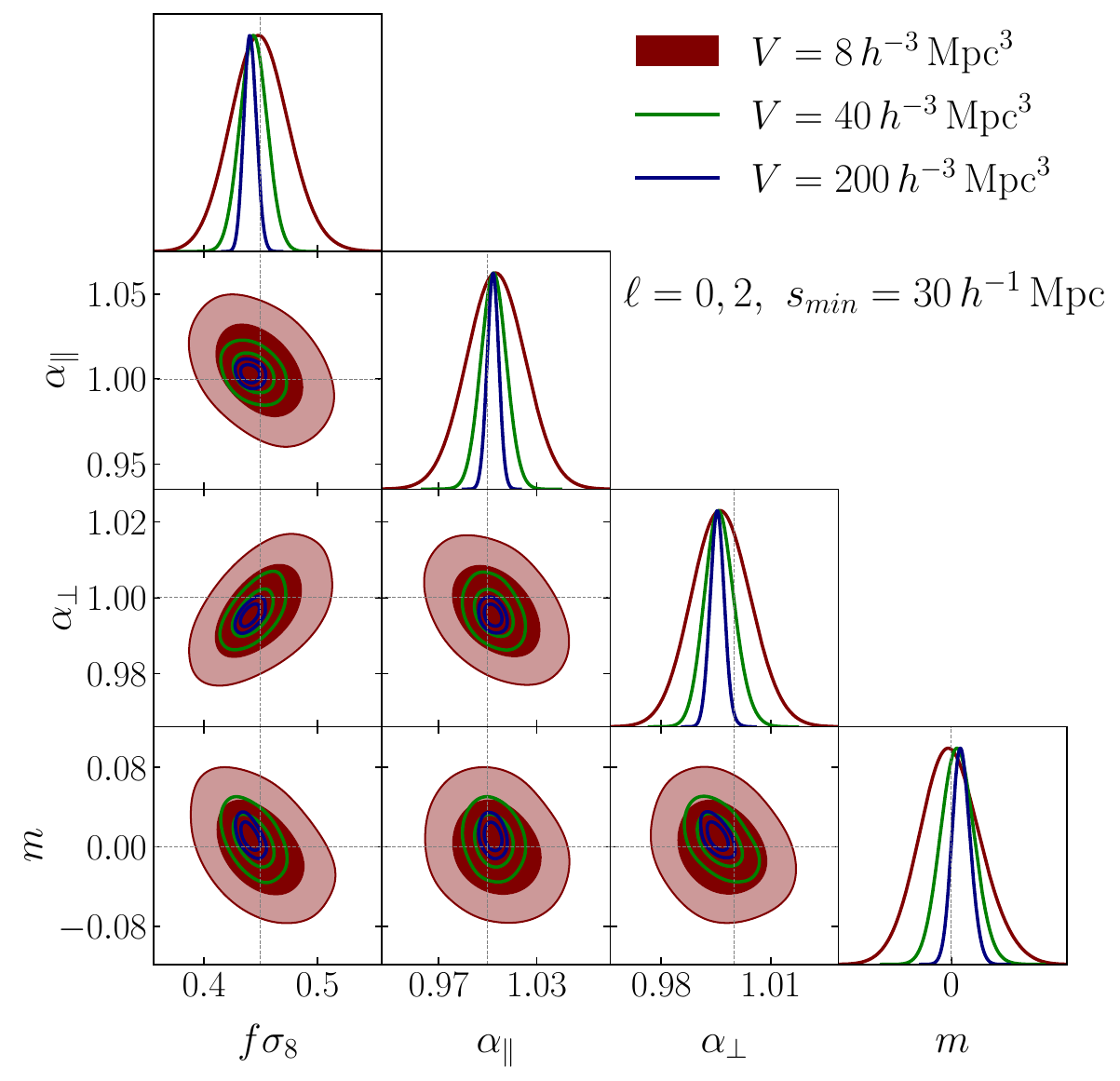}
\caption{The triangular plot shows the posterior distribution of our EFT-GSM templates from section \ref{sec:model}, obtained when we fit the mean of 25 \texttt{ABACUS-SUMMIT} simulations using a different rescaling number for the covariance matrix. On the left, we have the fit resulting from {\it Full-Modeling}, while on the right, we have the {\it Shape Fit}. The non-rescaled covariance matrix is in maroon, the covariance matrix rescaled by 5 in green, and the covariance matrix rescaled by 25 in blue. All the plots are obtained for the LRG tracer with $s_{min}=30 h^{-1} \mathrm{Gpc}$.}
\label{FM_traingular_plot}
\end{figure*}

We start by exploring the behavior of the GSM model for different volumes with \textit{Full  Modeling} and \textit{Shape  Fit}. 
Figure \ref{FM_traingular_plot} shows the posteriors for the LRG tracer with a minimal scale of $s_min= 30\,h^{-1}\,\textrm{Gpc}$ for three different volumes: $8\,h^{-3}\,\textrm{Gpc}^{3}$, which represents approximately the volume for DESI first year, $40\,h^{-3}\,\textrm{Gpc}^{3}$ which represents the volume of DESI by the end of the fifth year, and  $200\,h^{-3}\,\textrm{Gpc}^{3}$ with the total combined volume of the 25 simulations and which represents an unrealistic volume for a survey but allows us to test models with very small error bars. In Table \ref{tab:volumes} we report the central values and 1$\sigma$ errors for the cosmological parameters. As expected, when we extend to larger volumes, our constraints become tighter, maintaining approximately the same central values of the posteriors. In \textit{Full  Modeling} all our constraints tend to fall within the range of $1 \sigma$ for $h$ (0.42$\sigma$ for $V_{1}$, 0.61$\sigma$ for $V_{5}$ and 1.14$\sigma$ for $V_{25}$) and within the range of $1-2 \sigma$  for $\mathrm{ln}(10^{10}A_s)$ (0.46$\sigma$ for $V_1$, 0.11$\sigma$ for $V_5$ and 1.4$\sigma$ for $V_{25}$), while $\Omega_m$ consistently remains below $1\sigma$ (0.05$\sigma$ for $V_{1}$, 0.12$\sigma$ for $V_{5}$ and 0.28$\sigma$ for $V_{25}$). On the other hand, in \textit{Shape  Fit}, we obtain 
constraints within the range of $1-2 \sigma$ $f\sigma_8$ (0.04$\sigma$ for $V_{1}$, 0.50$\sigma$ for $V_{5}$ and 1.7$\sigma$ for $V_{25}$) and below $1 \sigma$  for $\alpha_{\parallel}$ (0.28$\sigma$ for $V_1$, 0.48$\sigma$ for $V_5$ and 0.94$\sigma$ for $V_{25}$), while $\alpha_{\perp}$ remains around $1-2\sigma$ (0.44$\sigma$ for $V_{1}$, 0.97$\sigma$ for $V_{5}$ and 2.42$\sigma$ for $V_{25}$) and lastly, for $m$ we obtain constrains around $1\sigma$ (0.03$\sigma$ for $V_{1}$, 0.41$\sigma$ for $V_{5}$ and 1.14$\sigma$ for $V_{25}$).

The increasing shift for the largest volume can be understood by considering the limitations of our 1-loop corrections, however even for the largest volume we do not detect a significant shift in any parameter. Considering the limitations in accuracy of N-body simulations that we start being sensitive to at these big volumes \citep[e.g.][]{2022MNRAS.515.1854G}, whose systematics potentially could be indistinguishable from inaccuracies in the modeling, and given the interest of comparing the errors with the current DESI year one analysis,
we decided throughout the rest of this work, to report the results for a volume of $8\,h^{-3}\,\textrm{Gpc}^{3}$ (unless otherwise stated).

\begin{center}
\begin{table*}
\ra{1.7}
\begin{center}

\begin{tabular} { c c c c c}

\multicolumn{5}{c}{\text{ Volume (Max. Freedom).}}\\
\hline
\multicolumn{5}{c}{\text{ Full Modeling}}\\
\hline
Volume &$\Omega_{m}$ &$h$ & $\mathrm{ln}(10^{10}A_s) $& \\
\hline
 $V_1 = 8 h^{-3} \mathrm{Mpc^{3}}$  & $0.3147\pm 0.0099 $   & $0.6771\pm 0.0083$ &  $2.98\pm 0.12$ &   \\
 $V_5 = 40 h^{-3} \mathrm{Mpc^{3}}$  & $0.3146\pm 0.0046$   & $0.6768\pm 0.0052$  &  $3.003\pm 0.058$   & \\
$V_{25} = 200 h^{-3} \mathrm{Mpc^{3}}$   &$0.3146\pm 0.0021$     &$0.6776\pm 0.0035$  &$2.999\pm 0.026$  &    \\
\hline 
\multicolumn{5}{c}{\text{ Shape Fit}}\\
\hline
Volume &$f\sigma_{8}$ &$\alpha_{\parallel}$ & $\alpha_{\perp}$& $m$ \\
\hline
 $V_1 = 8 h^{-3} \mathrm{Mpc^{3}}$  & $0.449\pm 0.025 $   & $1.005\pm 0.018$ &  $0.9965\pm 0.0079$ & $-0.001\pm 0.031$  \\
 $V_5 = 40 h^{-3} \mathrm{Mpc^{3}}$  & $0.444\pm 0.012$   & $1.0037\pm 0.0077$  &  $0.9960^{+0.0039}_{-0.0043}$ & $0.007\pm 0.017$  \\
$V_{25} = 200 h^{-3} \mathrm{Mpc^{3}}$   &$0.4408\pm 0.0054$     &$1.0033\pm 0.0035$  &$0.9954\pm 0.0019 $ & $0.0107^{+0.0088}_{-0.0099}$    \\
\hline
\end{tabular}
\caption{Summary of the constraints obtained when fitting the ABACUS LRG sample with our {\it Full Modeling} and {\it Shape Fit} methodology for  $s_{\text{min}} = 30 h^{-1} \mathrm{Mpc}$ using different volumes.}.\label{tab:volumes}

\end{center}
\end{table*}
\end{center}

\subsubsection{Optimizing the fitting range}
\label{sec:baseline_lrg}

In what follows, we present the results of our fits to the mean of the 25 \texttt{ABACUS-SUMMIT}  simulations, employing the methodologies detailed in section \ref{sec:methodology}.
We define our baseline methodology as using the monopole and quadrupole of the correlation function, using the covariance for a volume of $8\,h^{-3}\,\textrm{Gpc}^{3}$, with a maximum scale of $s_{max}=130 h^{-1}\,\textrm{Mpc}$ for the three methodologies. 
For the tests of the baseline settings, we focus on the LRG sample. We explore in this section the performance of the GSM-EFT model for different minimum scales for all methodologies {\it Standard} analysis,  {\it ShapeFit} and {\it Full Modeling}.

\begin{figure*}
\centering
\includegraphics[width=75mm,height=75mm]{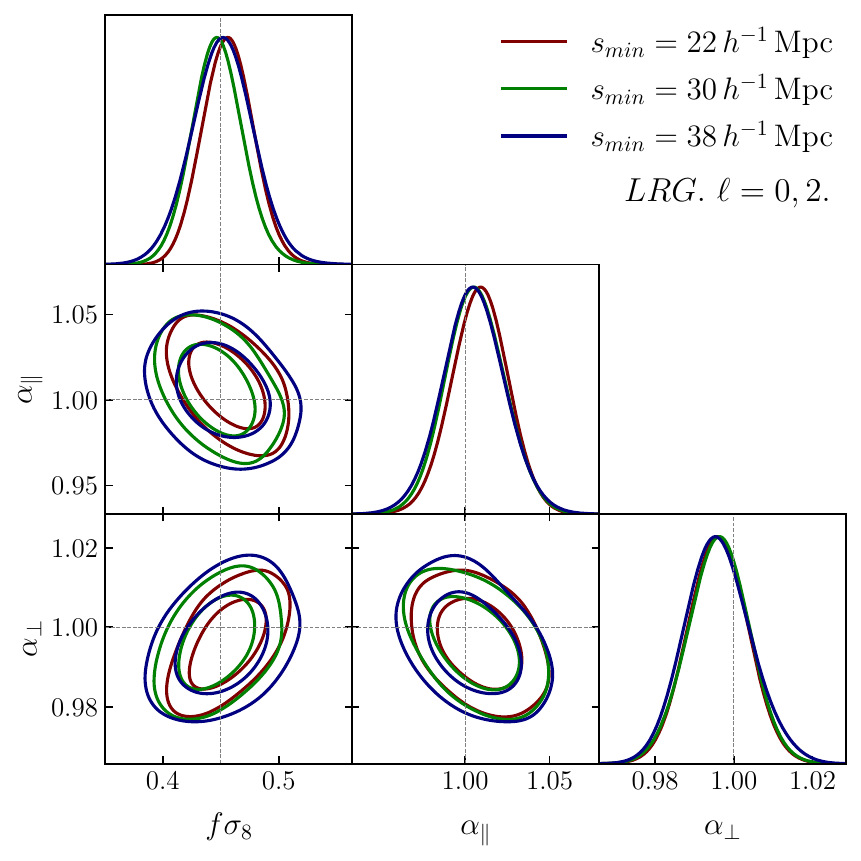}
\includegraphics[width=75mm,height=75mm]{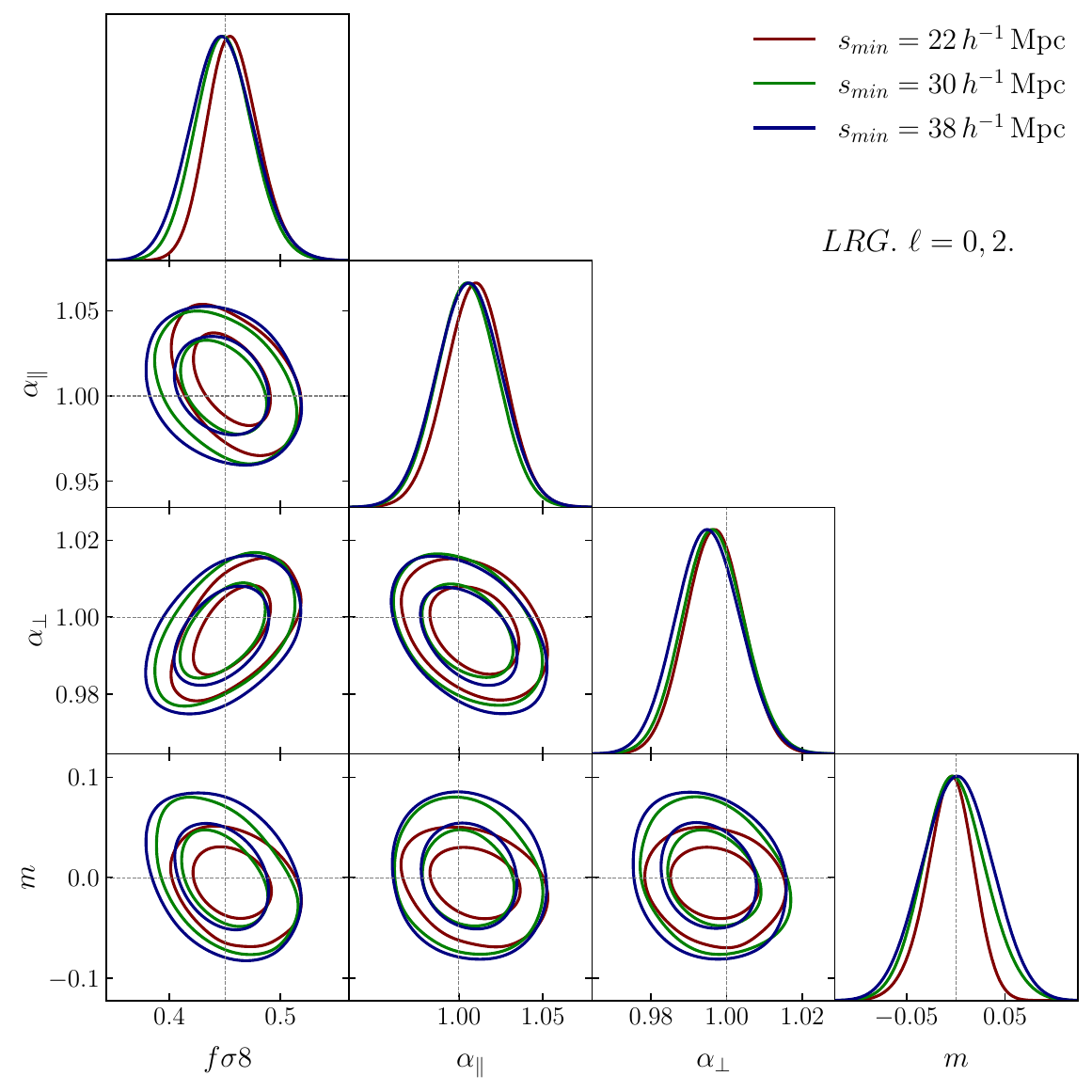}
\caption{The triangular plot shows the posterior distribution obtained from the mean of 25 \texttt{ABACUS-SUMMIT} simulations using a non-rescaled covariance matrix for the LRG tracer. On the left side, we have constraints for the {\it Standard} methodology, while on the right side, we have constraints for the {\it ShapeFit} methodology. We plot three different ranges for the LRG tracers, a minimum scale  $s_{min}=22\, h^{-1} \mathrm{Mpc}$ in red, $s_{min}=30 \,h^{-1} \mathrm{Mpc}$ in green, and $s_{min}=38 \,h^{-1} \mathrm{Mpc}$ in blue.}
\label{FM_traingular_plot_LRG_non_rescaled_compressed_sf}
\end{figure*}

In the {\it Standard} analysis, we constrain $f\sigma_8$, $\alpha_{\perp}$ and $\alpha_{\parallel}$ parameters, introduced in the section \ref{sec:Standard}, and, given that the reference template is computed based on the \texttt{ABACUS-SUMMIT}  cosmology, we expect $\alpha_{\perp}=1$ and $\alpha_{\parallel}=1$. In the {\it ShapeFit} methodology, we also include the $m$ parameter, which, we expect to be $0$.  In Figure \ref{FM_traingular_plot_LRG_non_rescaled_compressed_sf} we show the constraints for both methodologies in our baseline analysis. We compare the results of the {\it Standard} methodology (left) and of the {\it ShapeFit} (right) for three different minimum scale $s_{min}$ equal to: $22\, h^{-1}\,\textrm{Mpc}$, $30\, h^{-1}\,\textrm{Mpc}$, and $38\, h^{-1}\,\textrm{Mpc}$. In Table \ref{tab:smin_compress} we present the quantitative results. We observe that all the results are in agreement with the cosmology of the \texttt{ABACUS-SUMMIT} within $1\sigma$. We observe a similar performance among these two methodologies for the parameters in common, which is something we expect, given the parameter $m$ is intended to provide independent information. This is the extra information that $m$ captures about the slope of the linear power spectrum, which makes {\it ShapeFit} as constraining as {\it Full-Modeling} when interpreting the set of compressed parameters from {\it ShapeFit} into the cosmological parameters of $\Lambda$CDM, as shown in \citep{brieden_2021,brieden_2021b}.

\begin{center}
\begin{table*}
\ra{1.7}
\begin{center}
\begin{tabular} {ccccc}
\multicolumn{5}{c}{Baseline} \\
\hline
\multicolumn{5}{c}{\textit{ ShapeFit }for Maximal Freedom} \\
\hline
$s_{\text{min}}$ &$f\sigma_{8}$ &$\alpha_{\parallel} $ & $\alpha_{\perp} $ & $m$  \\  
\hline
 $22 \, h^{-1} \mathrm{Mpc}$  & $0.457^{+0.021}_{-0.025}$   & $1.010\pm 0.017$ &  $0.9968\pm 0.0074$  &$-0.006^{+0.025}_{-0.022}  $\\
 $30 \, h^{-1} \mathrm{Mpc}$  & $0.449\pm 0.025$   & $1.005\pm 0.018$  & $0.9965\pm 0.0079$ & $-0.001\pm 0.031$\\
$38 \, h^{-1} \mathrm{Mpc}$ &$0.448\pm 0.028$      &$1.006\pm 0.019$   &$0.9951\pm 0.0083$ & $0.002\pm 0.034 $\\
\hline
\multicolumn{5}{c}{ Standard  for Maximal Freedom}\\
\hline
$22 \, h^{-1} \mathrm{Mpc}$ &$0.456\pm 0.021$& $1.009\pm 0.016$&$0.9960\pm 0.0074$          &-\\
$30 \, h^{-1} \mathrm{Mpc}$ &$0.447\pm 0.022$&$ 1.006\pm 0.017$ & $0.9961\pm 0.0076$          &-\\
$38 \, h^{-1} \mathrm{Mpc}$ &$0.452\pm 0.026$&$ 1.005\pm 0.018$ & $0.9960^{+0.0079}_{-0.0087}$ & -\\
\hline
\end{tabular}
\caption{Constraints on our compressed parameters when fitting our ABACUS LRG sample for three minimum scales when using {\it Standard} and {\it ShapeFit} methodology.}\label{tab:smin_compress} 
\end{center}
\end{table*}
\end{center}

\begin{figure*}
\centering
\includegraphics[width=140mm,height=110mm]{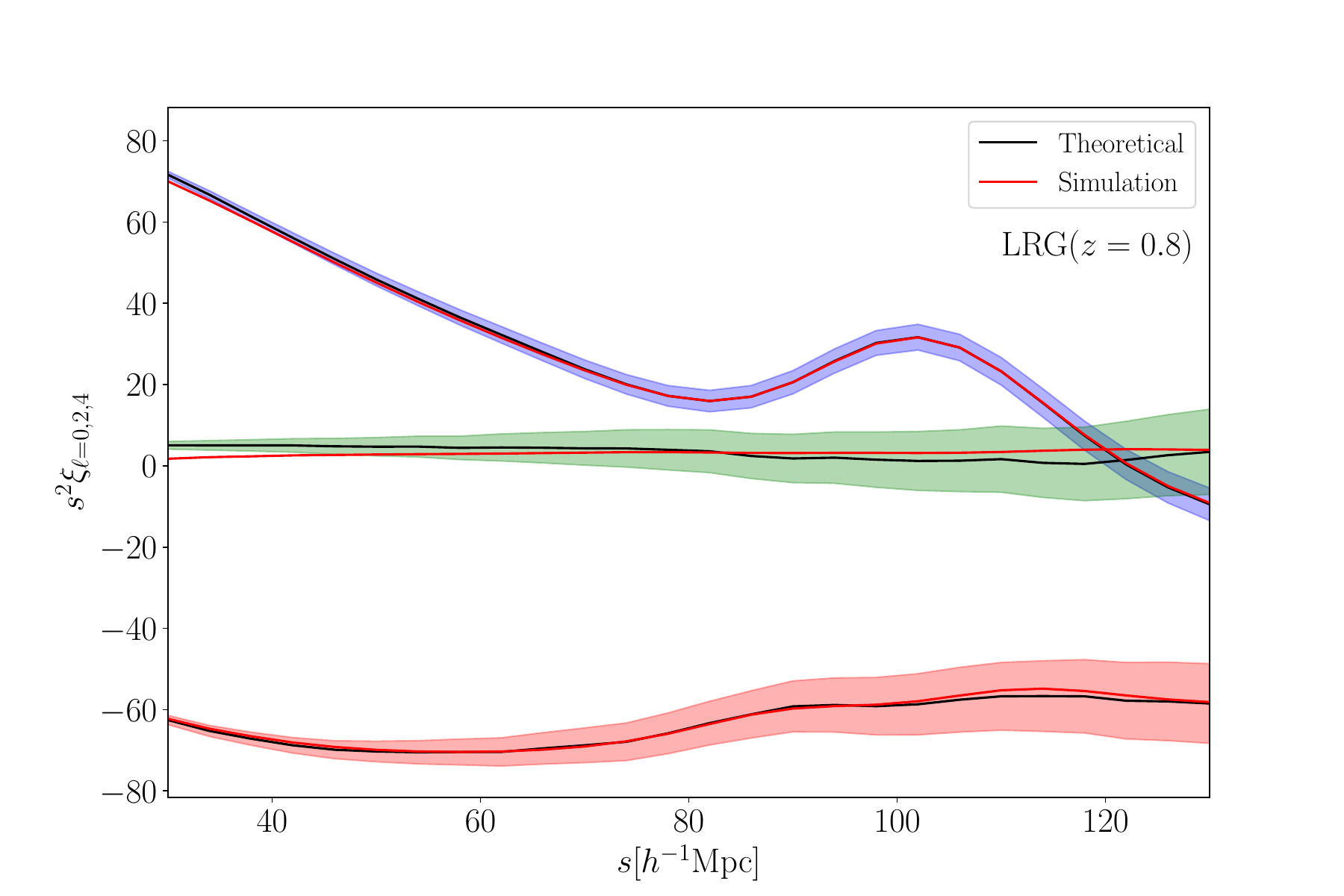}
\caption{This plot shows the correlation function obtained from the mean of 25 \texttt{ABACUS-SUMMIT}  simulations and the mean of the posterior distribution in the range $s_{min}=30 \,h^{-1} \mathrm{Mpc}$ and $s_{max}=130 \,h^{-1} \mathrm{Mpc}$. The shadow regions show the error bars from the non-rescaled covariance matrix for the LRG tracer.}
\label{plot_corr_function}
\end{figure*}

\begin{figure*}
\centering
\includegraphics[width=70mm,height=70mm]{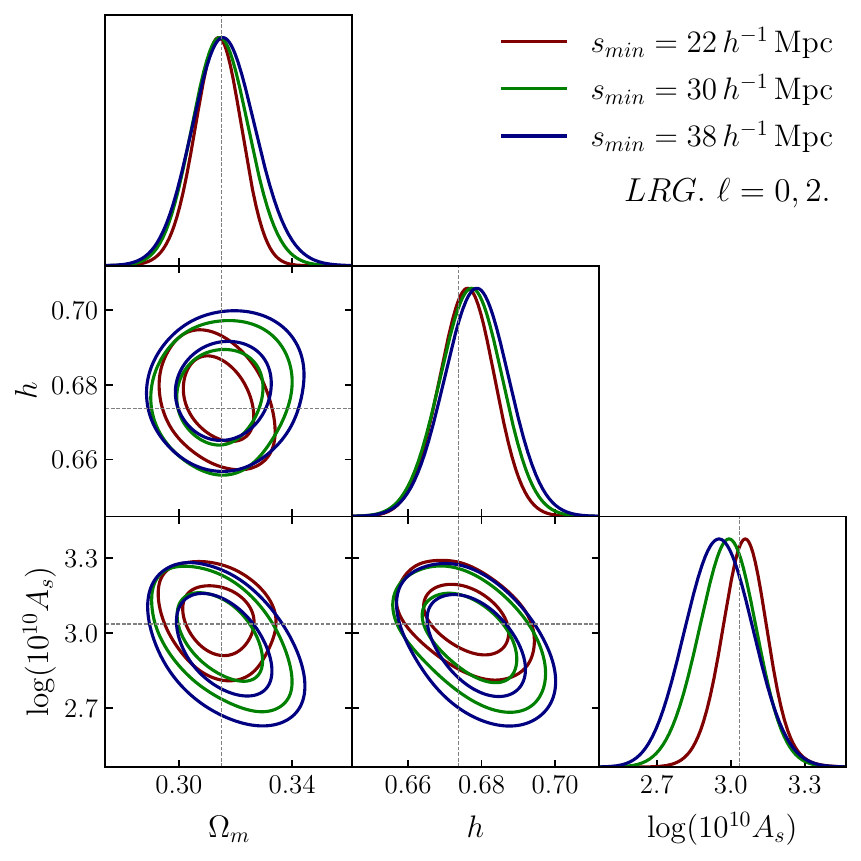}
\includegraphics[width=70mm,height=70mm]{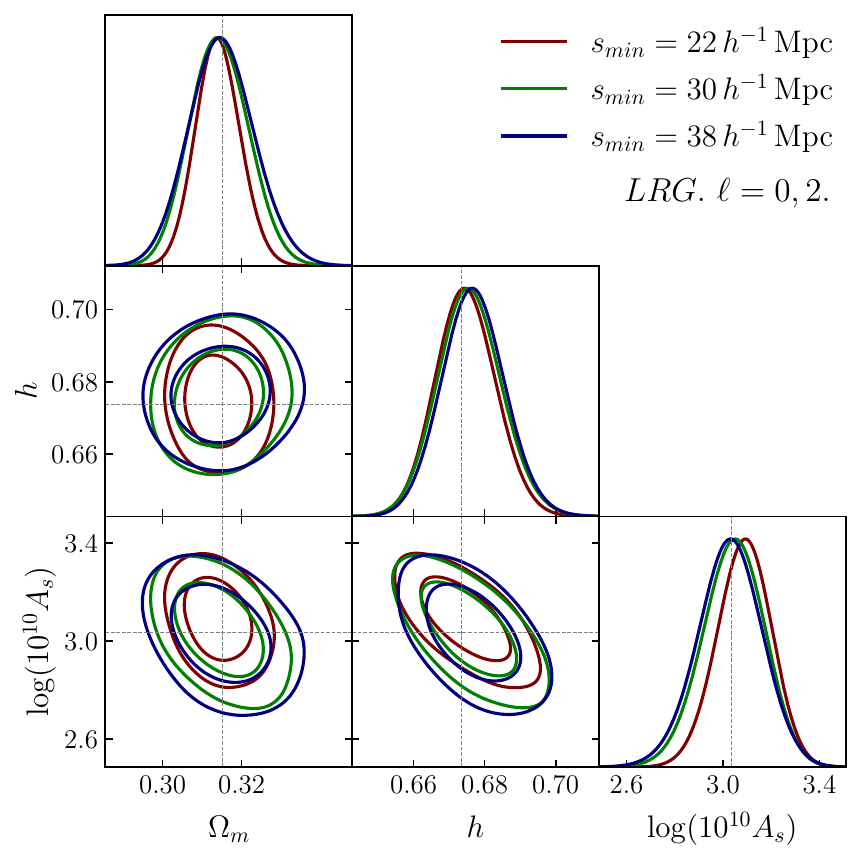}
\caption{The triangular plot shows the posterior distribution obtained from the mean of 25 \texttt{ABACUS-SUMMIT}  simulations using a non-rescaled covariance matrix for the LRG tracer following the {\it Full Modeling} methodology on the left side and the {\it ShapeFit} methodology on the right side. We plot three different ranges for the LRG tracers, a minimum scale  $s_{min}=22 \,h^{-1} \mathrm{Mpc}$ in red, $s_{min}=30\, h^{-1} \mathrm{Mpc}$ in green, and $s_{min}=38\, h^{-1} \mathrm{Mpc}$ in blue.}
\label{FM_traingular_plot_LRG_non_rescaled}
\end{figure*}

In \textit{Full Modeling} we define as our baseline analysis, 
four free cosmological parameters, $\{ \Omega_m$, $h$,  $\mathrm{ln}(10^{10}A_s)$, $\omega_b$ \}, however as we explained in the methodology we impose a Gaussian BBN prior on $\omega_b$, as well as flat priors in the nuisance parameters which are described in Table \ref{tab:gsm_priors}. 
In Figure \ref{FM_traingular_plot_LRG_non_rescaled} we plot the posterior distribution contours for our baseline analysis varying $s_{{min}}$ for three different values with a bin separation of $8 \,h^{-1} \mathrm{Mpc}$, on the right side we have a triangular plot for the {\it Full Modeling} methodology while on the left side, we have the {\it ShapeFit} methodology. We show the values in Table \ref{tab:smin_cosmo}. We notice that both methodologies are in good agreement with the true values of the cosmological parameters for all scales, within $ \leq 0.1 \sigma$ for $\Omega_m$, $\leq 0.6 \sigma$ for $h$ and, $\leq 0.7 \sigma$ for $A_s$ 
when $s_{min}=22\,h^{-1} \mathrm{Mpc}$ for Full modeling and ShapeFit. 
For ShapeFit we observe a similar trend, all deviations from the true value are $\leq 0.2\sigma$ for $\Omega_m$, $\leq 0.4 \sigma$ for $h$ and $\leq 0.5 \sigma$ for $\mathrm{ln}(10^{10}A_s)$. We conclude, we do not detect any significant shift with either of the methodologies employed. In Figure \ref{plot_corr_function}, we illustrate the shape of the correlation function. This depiction represents the mean of the values derived from the fit of the baseline analysis, with $s_{min}=30 ,h^{-1} \mathrm{Mpc}$, and a $\chi^2=60.57$ for 58 bins and 9 free parameter.

As we can notice from the results in this section, although we obtain results below $1\sigma$ for all the choices of minimum scales, the results with $s_{min}=22\,h^{-1} \mathrm{Mpc}$ provide the smallest error bars without any detectable bias, which is in agreement with what other works have concluded \cite{gsm2023arXiv231017834R}. Therefore, our optimal fitting range is $s_{\text{range}} = 22-130 \, h^{-1} \mathrm{Mpc}$. Several crucial results will be exclusively presented with this range, unless otherwise specified. Our baseline analysis is extended and compared in further sections.

\begin{center}
\begin{table*}
\ra{1.7}
\begin{center}

\begin{tabular} {cccccc}
\multicolumn{5}{c}{BASELINE } \\
\hline
\multicolumn{5}{c}
{\textit{ Full Modeling }(Max Freedom)} \\
\hline
$V$&$s_{\text{min}}$ &$\Omega_{m}$ &$h$ & $\mathrm{ln}(10^{10}A_s) $ \\
\hline
$V_1$& $22\, h^{-1} \mathrm{Mpc}$  & $0.3138\pm 0.0081$   & $0.6760\pm 0.0074$ &  $3.056\pm 0.093$       \\
$V_1$& $30\, h^{-1} \mathrm{Mpc}$  & $0.3147\pm 0.0099$   & $0.6771\pm 0.0083$ &  $2.98\pm 0.12$      \\
$V_1$&$38\, h^{-1} \mathrm{Mpc}$ &$0.316\pm 0.011$ &$0.6786\pm 0.0086$  & $2.95\pm 0.13$ \\
\hline
$V_5$& $22\, h^{-1} \mathrm{Mpc}$  & $0.3146\pm 0.0037$   & $0.6775\pm 0.0042$ &  $3.038\pm 0.044$       \\
$V_5$& $30\, h^{-1} \mathrm{Mpc}$  & $0.3146\pm 0.0046$   & $0.6768\pm 0.0052$ &  $3.003\pm 0.058$      \\
$V_5$&$38\, h^{-1} \mathrm{Mpc}$ &$0.3144\pm 0.0050$ &$0.6771\pm 0.0052$  & $2.984\pm 0.067$ \\
\hline
\multicolumn{5}{c}
{\textit{ ShapeFit } (Max Freedom)}  \\
\hline
$V$&$s_{\text{min}}$ &$\Omega_{m}$ &$h$ & $\mathrm{ln}(10^{10}A_s) $  \\
\hline
$V_1$& $22\, h^{-1} \mathrm{Mpc}$  & $0.3140\pm 0.0055$   & $0.6747\pm 0.0082$  &  $3.09\pm 0.11$ \\
$V_1$& $30\, h^{-1} \mathrm{Mpc}$  & $0.3144\pm 0.0072$   & $0.6758\pm 0.0088$  & $3.04\pm 0.12$ \\
$V_1$&$38\, h^{-1} \mathrm{Mpc}$ &$0.3149\pm 0.0081$      &$0.6768\pm 0.0086$   &$3.03\pm 0.13 $ \\
\hline
$V_5$& $22\, h^{-1} \mathrm{Mpc}$  & $0.3156\pm 0.0027$   & $0.6772\pm 0.0043$  &  $3.044\pm 0.053$ \\
$V_5$& $30\, h^{-1} \mathrm{Mpc}$  & $0.3153\pm 0.0036$   & $0.6760\pm 0.0054$  & $3.014\pm 0.061$ \\
$V_5$&$38\, h^{-1} \mathrm{Mpc}$ &$0.3151\pm 0.0039$      &$0.6771\pm 0.0055$   &$2.999\pm 0.069 $ \\
\hline
\hline
\end{tabular}
\caption{Constraints on our cosmological parameters when fitting our ABACUS LRG sample for three minimum scales when using different methodologies, {\it Full Modeling} methodology and {\it ShapeFit} methodology together with two different volumes $V_1=8 \, h^{-3} \mathrm{Mpc}^3$ and $V_5=40 \, h^{-3} \mathrm{Mpc}^3$.}\label{tab:smin_cosmo}

\end{center}
\end{table*}
\end{center}

\subsubsection{Maximal and minimal freedom}
We explore two different parameter space configurations of the galaxy bias terms, that we refer to as {\it maximal} and {\it minimal} freedom. In the first, we let all parameters vary simultaneously and it is considered as our baseline configuration. While the second configuration assumes a co-evolution for which non-local biases appear only through gravity evolution, and so are not present in Lagrangian coordinates scheme, corresponding to fix  $b_{s^2}=0$.

\begin{figure*}
\centering
\includegraphics[width=3 in]{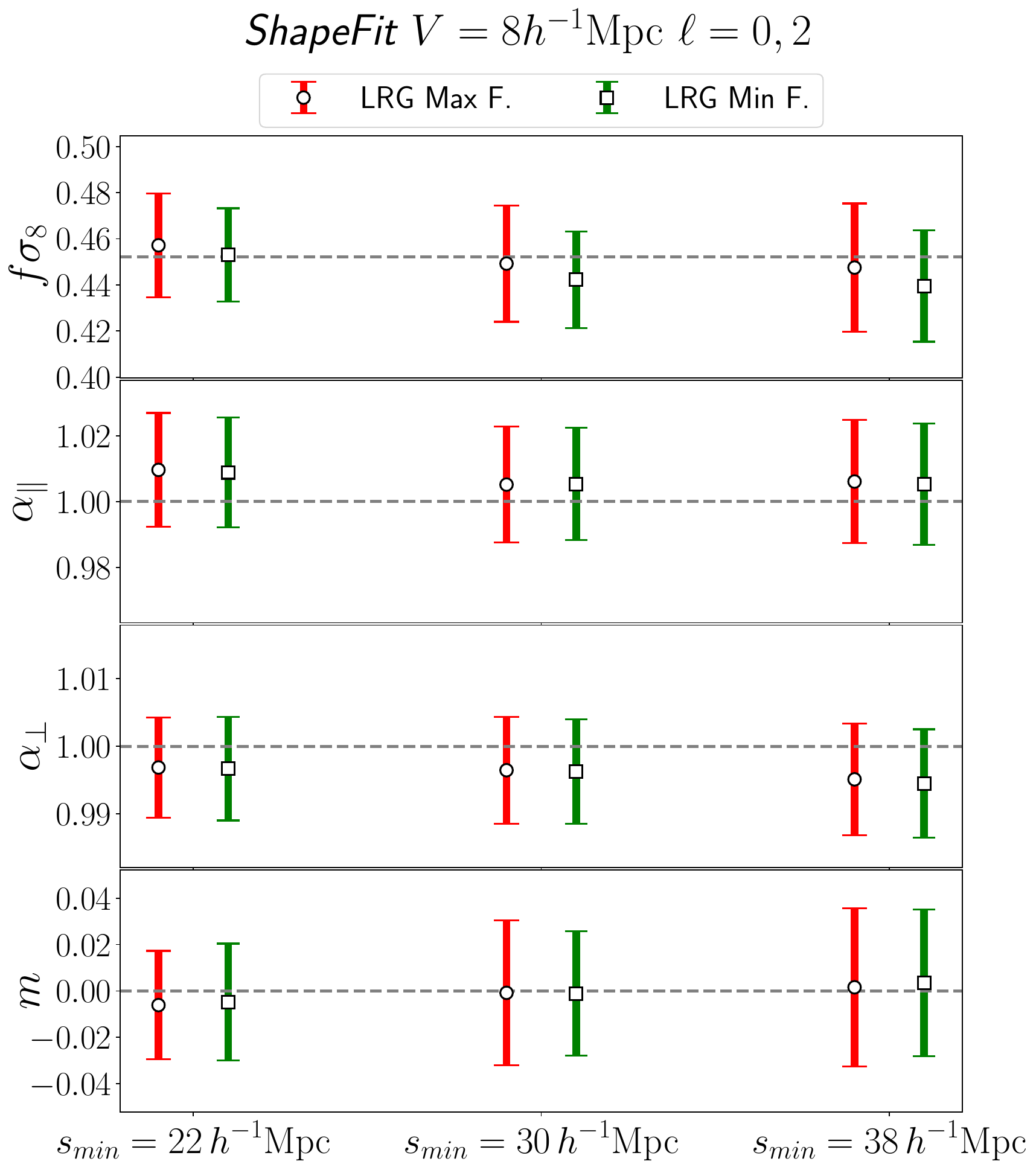}
\includegraphics[width=3 in]{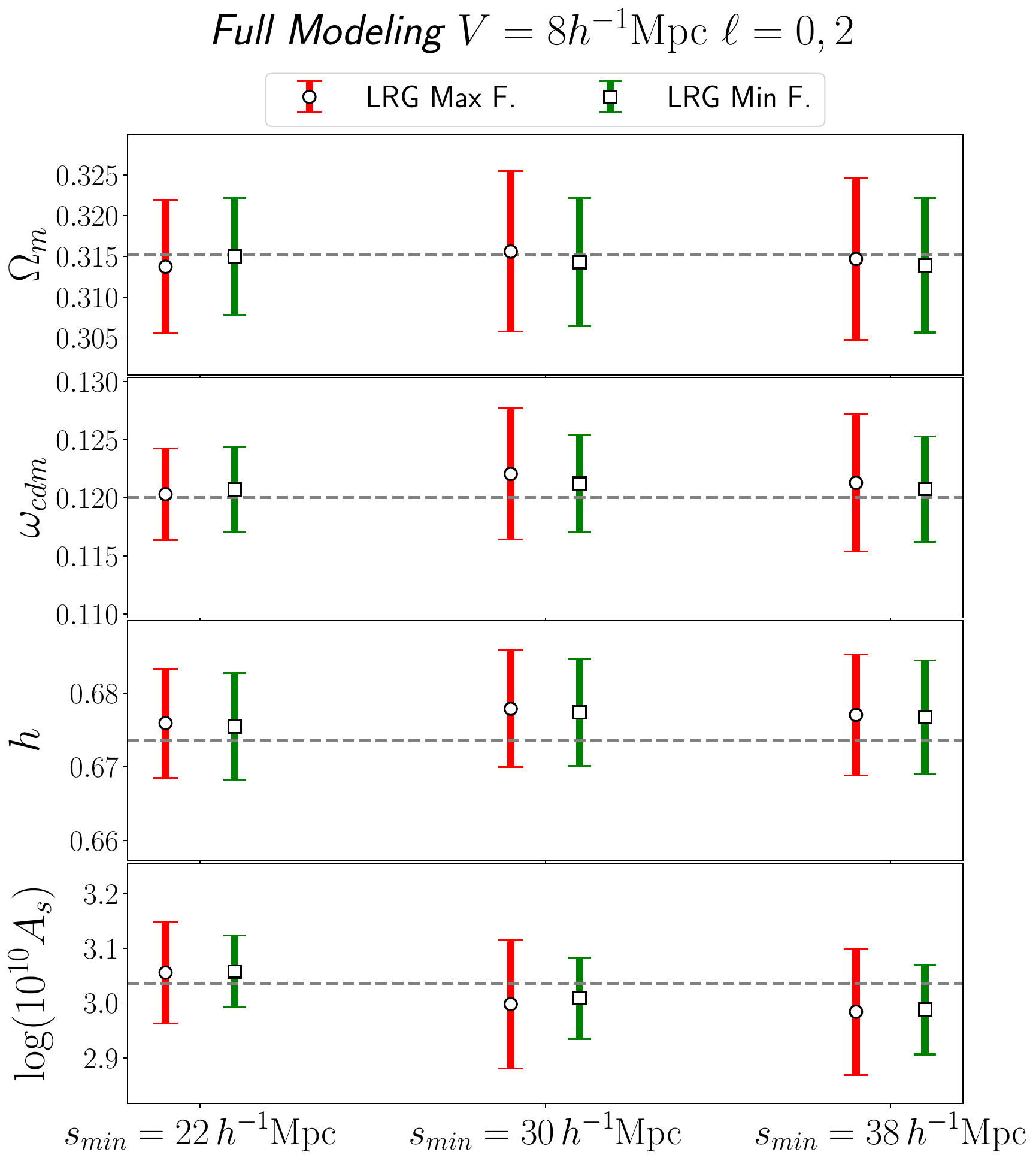}
\caption{Error bar comparison for LRG using a volume of $8\, h^{-3} \mathrm{Gpc}^3$. The red bars were obtained using maximal freedom, while the red ones were obtained using minimal freedom. We show the model's behavior at various minimum scales, each one separated by 1 bin of 8 $h^{-1} \mathrm{Mpc}$. On the left side, we have the {\it ShapeFit} fits, while on the right side, we have the {\it Full Modeling} fits.
}
\label{error_comparison_fm_cov5_min_vs_max}
\end{figure*}

Figure \ref{error_comparison_fm_cov5_min_vs_max} shows the mean value of the posterior distribution of the parameters found when running our methodology using both {\it maximal} and {\it minimal} freedom configurations and using both  {\it ShapeFit} and {\it Full Modeling} methods. The corresponding quantitative results are presented in Tables \ref{tab:smin_compress}, \ref{tab:smin_cosmo}, \ref{tab:minmaxFreedom_FM} and \ref{tab:minmaxFreedom_SF}.
We notice that in both our {it Full Modeling} and {it Shape Fit} results, the central values are fairly consistent with each other. However, the {\it Minimal} freedom configuration leads to $\sim 10\%$ smaller error bars, this trend is expected as we reduce the number of free parameters. 
We notice that the {it Full Modeling} scenario shows an underestimation of the $A_s$ parameters in this larger scale which is closer to its expected value at $22 \,h^{-1} \mathrm{Mpc}$.

On the other hand, if we explore the compressed parameter space for the $22 h^{-1} \mathrm{Mpc}$ scale specifically, we observe that the {\it Maximal} freedom mean value predictions are closer to the true value of the simulation for $f\sigma_8$, while the {\it Minimal} freedom case fares better with its predictions for $\alpha_{\parallel}$. Additionally, we show that the error bars are of similar size for all parameters except $m$, where the {\it Maximal} freedom case leads to tighter constraints.

\begin{center}
\begin{table*}
\ra{1.7}
\begin{center}

\begin{tabular} {  c c c c cc}
\multicolumn{4}{c}{ \textit{Full Modeling }(Min Freedom). }\\
\hline
$s_{\text{min}}$ &$\Omega_{m}$ &$h$ & $\mathrm{ln}(10^{10}A_s) $\\
\hline
 $22\, h^{-1} \mathrm{Mpc}$  & $0.3150\pm 0.0072$   & $0.6755\pm 0.0072$ &  $3.058\pm 0.065$        \\
 $30\, h^{-1} \mathrm{Mpc}$  & $0.3139\pm 0.0082$   & $0.6767\pm 0.0078$ &  $2.989\pm 0.082$        \\
$38\, h^{-1} \mathrm{Mpc}$ &$0.314\pm 0.010$ &$0.6786\pm 0.0080$  & $2.96\pm 0.11 $ \\
\hline
\end{tabular}
\caption{Constraints to our parameters when fitting our ABACUS LRG sample for three minimum scales when using different methodologies, {\it Full Modeling} methodology.}\label{tab:minmaxFreedom_FM}
\end{center}
\end{table*}
\end{center}

\begin{center}
\begin{table*}
\ra{1.7}
\begin{center}

\begin{tabular} {  c c c c cc}
\multicolumn{5}{c}{ {\it ShapeFit} (Min Freedom). }\\

\hline
$s_{\text{min}}$ &$f\sigma_{8}$ &$\alpha_{\parallel} $ & $\alpha_{\perp} $ & $m$ \\
\hline
 $22\, h^{-1} \mathrm{Mpc}$  & $0.453\pm 0.020$   & $1.009\pm 0.017 $ &  $0.9967\pm 0.0076$  &  $-0.004^{+0.028}_{-0.018}$ \\
 $30\, h^{-1} \mathrm{Mpc}$  & $0.442\pm 0.021$   & $1.005\pm 0.017$  & $0.9963\pm 0.0077$  & $-0.001\pm 0.027$  \\
$38\, h^{-1} \mathrm{Mpc}$ &$0.440\pm 0.024$      &$1.005\pm 0.018$   &$0.9945\pm 0.0080$   &$0.004\pm 0.032$ \\
\hline
\end{tabular}
\caption{Constraints to our parameters when fitting our ABACUS LRG sample for three minimum scales when using different methodologies for {\it ShapeFit} methodology.}\label{tab:minmaxFreedom_SF}

\end{center}
\end{table*}
\end{center}

In conclusion, our comparison between the predicted central values and the expected values of our parameters reveals similar performances between our {\it Minimal} and {\it Maximal} freedom configurations. This holds true for both the compressed parameter space and the cosmological parameter space. We observe no significant shift of the fitted parameters from their true value. However, our baseline analysis with $s_{min}=22 \, h^{-1} \mathrm{Mpc}$ indicates a improvement in most parameters of approximately $\sim 10\%$ in the size of the error bars when using the more restricted {\it Minimal} freedom configuration, this is true regardless of whether we employ the compressed or {\it Full Modeling} approach. However, throughout this work we opt for a conservative approach and designate the {\it Maximal} freedom scenario as our baseline configuration.
\subsection{Extensions to Baseline Analysis}\label{sec:extensions_base}

We will present three extensions to the main analysis we presented above. First we focus on the information we can extract from the hexadecapole of the correlation function, then we explore the possibility to constrain the spectral index $n_s$ and finally we allow the Dark Energy equation of state evolve with the time using the $(w_o, w_a)$ prescription.

\subsubsection{Effect of including hexadecapole}
In this subsection, we explore the effect of adding the hexadecapole in our baseline analysis. It is well known that the hexadecapole is considered to be the multipole where PT at first-loop starts to struggle to reproduce the small scales for both the 2-point correlation function and the power spectrum, particularly at small scales (below $30 \,h^{-1} \mathrm{Mpc}$ in configuration space), where non-linear effects start to be important.

\begin{figure*}
\centering
\includegraphics[width=75mm,height=80mm]{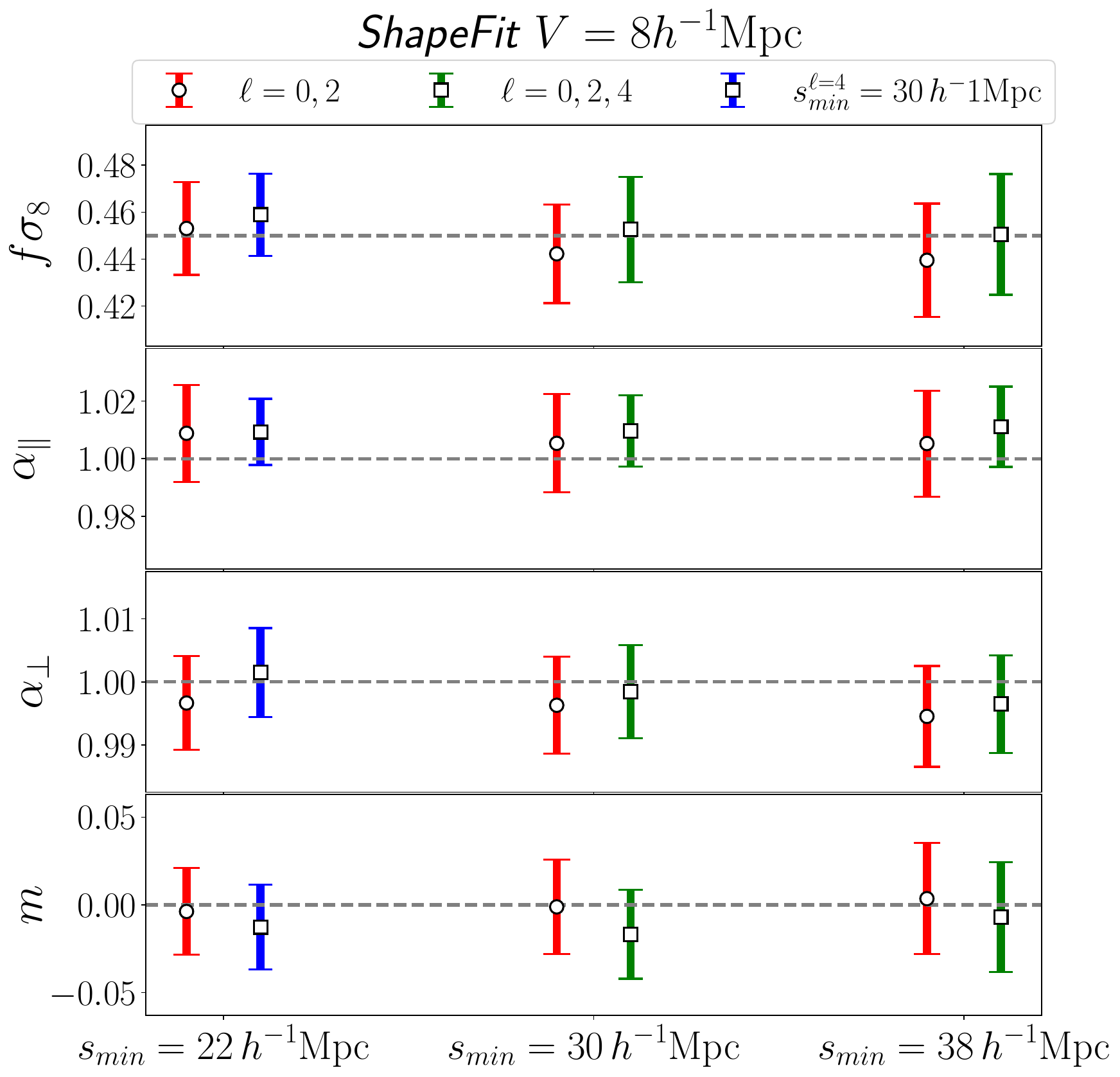}
\includegraphics[width=75mm,height=80mm]
{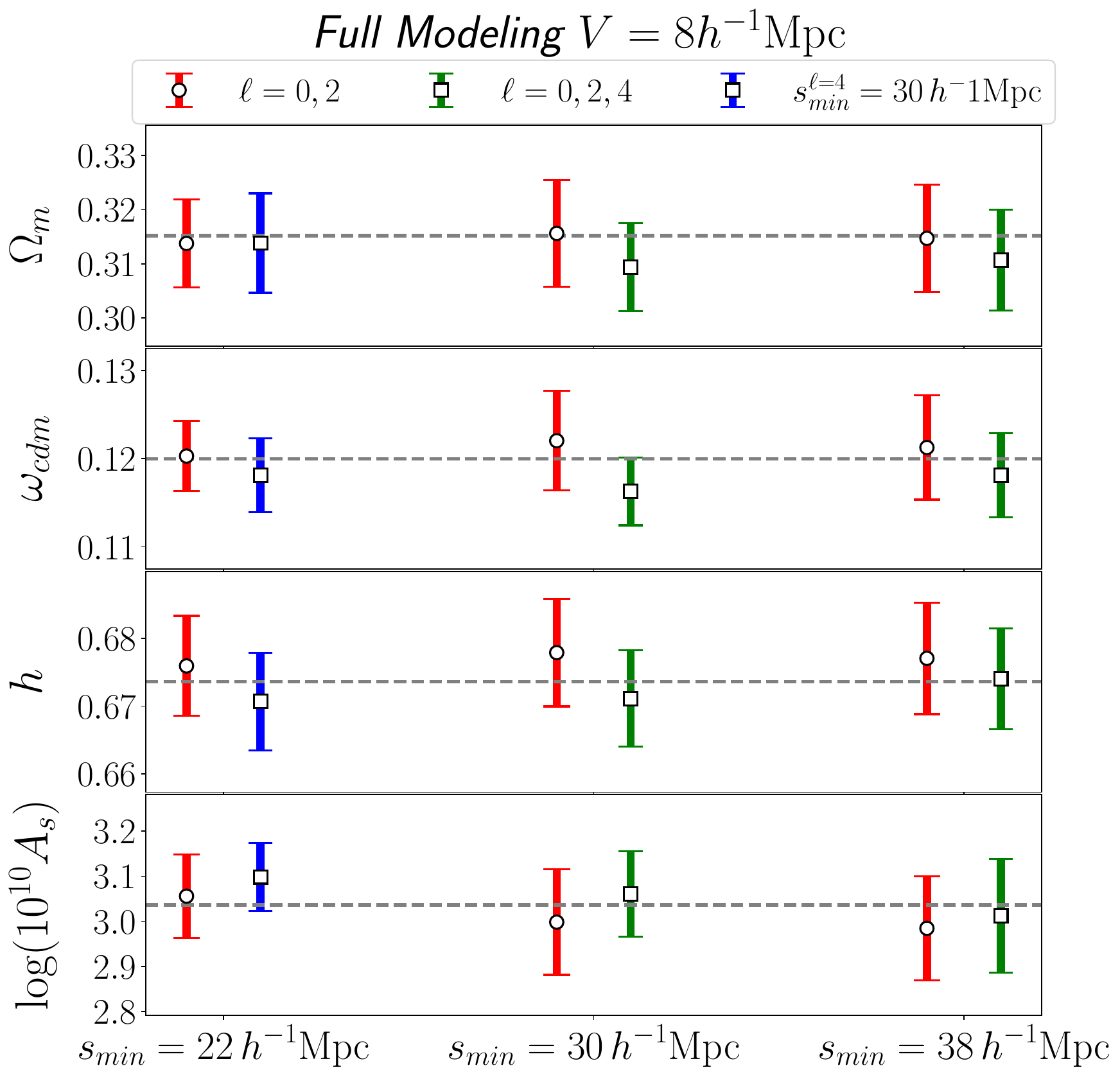}
\caption{1D constraints as a function of minimum scale for the LRGs using a volume of 8 $h^{-1} \mathrm{Mpc}$. The results for the baseline analysis without the hexadecapole are shown in red while the results including the hexadecapole are shown in green. On the left side, we have the {\it ShapeFit} fits, while on the right side, we have the {\it Full Modeling} fits. The lower bin, represented in blue, has the same settings, but the bin $s_{min}=22 \,h^{-1} \mathrm{Mpc}$ is excluded for the hexadecapole.
}

\label{error_comparison_fm_cov5_hexa_vs_max}
\end{figure*}

In Figure \ref{error_comparison_fm_cov5_hexa_vs_max} we can observe the effect of including the hexadecapole in the compressed parameter space and the cosmological parameter space for three different ranges, $s_{min}=22$, $30$, and $38\, h^{-1} \mathrm{Mpc}$. The fits following our baseline analysis are plotted in red, while fits adding hexadecapole are shown in green. Due to the difficulties in small scale model of the hexadecapole, we remove the bin $22 \, h^{-1} \mathrm{Mpc}$ and maintain the bin $s_{min}=22 \, h^{-1} \mathrm{Mpc}$ for the monopole and the quadrupole. This is done because when we include the smaller hexadecapole bin, we note that our posterior distribution becomes bimodal due to the appearance of a second region of the parameter space with high likelihood. However, this region has already been discarded by Planck to more than $3\sigma$. Given that we don't trust our model at these smaller $r$ scales, we opt to exclude this bin. The quantitative results are shown in Tables \ref{tab:smin_compress}, \ref{tab:smin_cosmo}, \ref{tab:hexa_fm} and \ref{tab:hexa_sf}. In the compressed parameters, we note slight variations within the 1$\sigma$ range for $f\sigma_8$, $\alpha_{\parallel}$, $\alpha_{\perp}$, and $m$. Similarly, the cosmological parameters exhibit consistent results, with $\Omega_m$, $h$, and $A_s$ all falling within the 1$\sigma$ range.

On the other side, the error bars when we include the hexadecapole for the three minimum scales have a small reduction in \textit{ShapeFit}. For $f\sigma_8$ the error bar is reduced by $0.26\%-0.07\%$ compared with the baseline analysis, for $\alpha_{\parallel}$ this change is $0.26\%-0.30\%$, while $\alpha_{\perp}$ remains in $0.06\%$, and finally for $m$ this change goes from an increase of $0.04\%$ to a reduction of $0.08\%$. For \textit{Full Modeling} we have a change in the error bars for $ \Omega_m$ that goes from an increment of $0.13\%$ to a decrease of $0.17\%$, for $h$ we have a diminution in the error bars of $0.03\%-0.14\%$, while $\mathrm{ln}(10^{10}A_s)$ has a reduction in the error bars of  $0\%-0.21\%$.

In conclusion, adding the hexadecapole does not significantly improve the results, even when removing specifically the lowest bin distance in order to prevent the problems on quasi-linear scale model. The main effect is slightly reducing the error on the \textit{Full Modeling} analysis. However, we can notice that all the constraints remain below 1$\sigma$ from the true value when we include the hexadecapole.

\begin{center}
\begin{table*}
\ra{1.7}
\begin{center}

\begin{tabular} {  c c c c }

\multicolumn{4}{c}{ {\it Full Modeling} (M+Q+H)} \\
\hline
$s_{\text{min}}$ &$\Omega_{m}$ &$h$ & $\mathrm{ln}(10^{10}A_s) $ \\
\hline
 $22\, h^{-1} \mathrm{Mpc}$  & $0.3138^{+0.0099}_{-0.0085}$   & $0.6707\pm 0.0072$ &  $3.098^{+0.066}_{-0.082}$        \\
 $30\, h^{-1} \mathrm{Mpc}$  & $0.3094\pm 0.0082$   & $0.6711\pm 0.0071$ &  $3.060\pm 0.095$        \\
$38\, h^{-1} \mathrm{Mpc}$ &$0.3107\pm 0.0093$ &$0.6741\pm 0.0075$  & $3.01\pm 0.13$ \\
\hline
\end{tabular}
\caption{1D constraints when fitting the ABACUS LRG mocks with (M+Q+H) and without (M+Q) the hexadecapole for three minimum scales for {\it Full Modeling}.
}\label{tab:hexa_fm}

\end{center}
\end{table*}
\end{center}

\begin{center}
\begin{table*}
\ra{1.7}
\begin{center}

\begin{tabular} {  c c c c c}

\multicolumn{5}{c}{ {\it ShapeFit} (M+Q+H)} \\
\hline
$s_{\text{min}}$ &$f\sigma_{8}$ &$\alpha_{\parallel} $ & $\alpha_{\perp} $ & $m$  \\
\hline
 $22\, h^{-1} \mathrm{Mpc}$  & $0.459\pm 0.017$   & $1.009^{+0.011}_{-0.012} $ &  $1.0015^{+0.0074}_{-0.0065}$  &  $-0.013^{+0.026}_{-0.023}$ \\
 $30\, h^{-1} \mathrm{Mpc}$  & $0.453\pm 0.022$   & $1.010^{+0.012}_{-0.013}$  & $0.9984\pm 0.0074$  & $-0.017\pm 0.025$  \\
$38\, h^{-1} \mathrm{Mpc}$ &$0.451\pm 0.026$      &$1.011\pm 0.014$   &$0.9965\pm 0.0078$   &$-0.007\pm 0.031$ \\
 $30\, h^{-1} \mathrm{Mpc}$  & $0.449\pm 0.025$   & $1.005\pm 0.018$  & $0.9965\pm 0.0079$ & $-0.001\pm 0.031           $\\
\hline
\end{tabular}
\caption{1D constraints when fitting the ABACUS LRG mocks with (M+Q+H) and without (M+Q) the hexadecapole for three minimum scales for {\it ShapeFit} methodology.
}\label{tab:hexa_sf}

\end{center}
\end{table*}
\end{center}

\subsubsection{Exploring the impact of $n_s$ free}
\label{ns_extension}

The scalar spectral index, denoted as $n_s$, serves as a key parameter that characterizes the primordial power spectrum of density perturbations in the early universe. As mentioned in Section \ref{Likelihood_and_Priors}, in our standard approach, we fix this parameter to its Planck 2018 reported value of $n_s=0.9649$ \cite{plank2018_cosmological}. However, in this section, we incorporate this parameter into our methodology as one of the extensions of our baseline analysis. This extension enables a more comprehensive and complex exploration of the parameter space. In this section, we impose the following prior on $n_s$:
 \begin{align}
   n_s &: \,  \mathcal{U}[0.5,1.5]   
 \end{align}

Given the new degree of freedom included, the models become more complex, in order to maintain the precision of our emulator, we expand the number of multipoles utilized in the neural network to 90,000. Among these, 80,000 are allocated for training the network, while 5,000 served as a validation set. The remaining 5,000 are our test set dedicated to generating statistics for evaluating the model's performance. 

\begin{figure*}
\centering
\includegraphics[width=140mm,height=140mm]{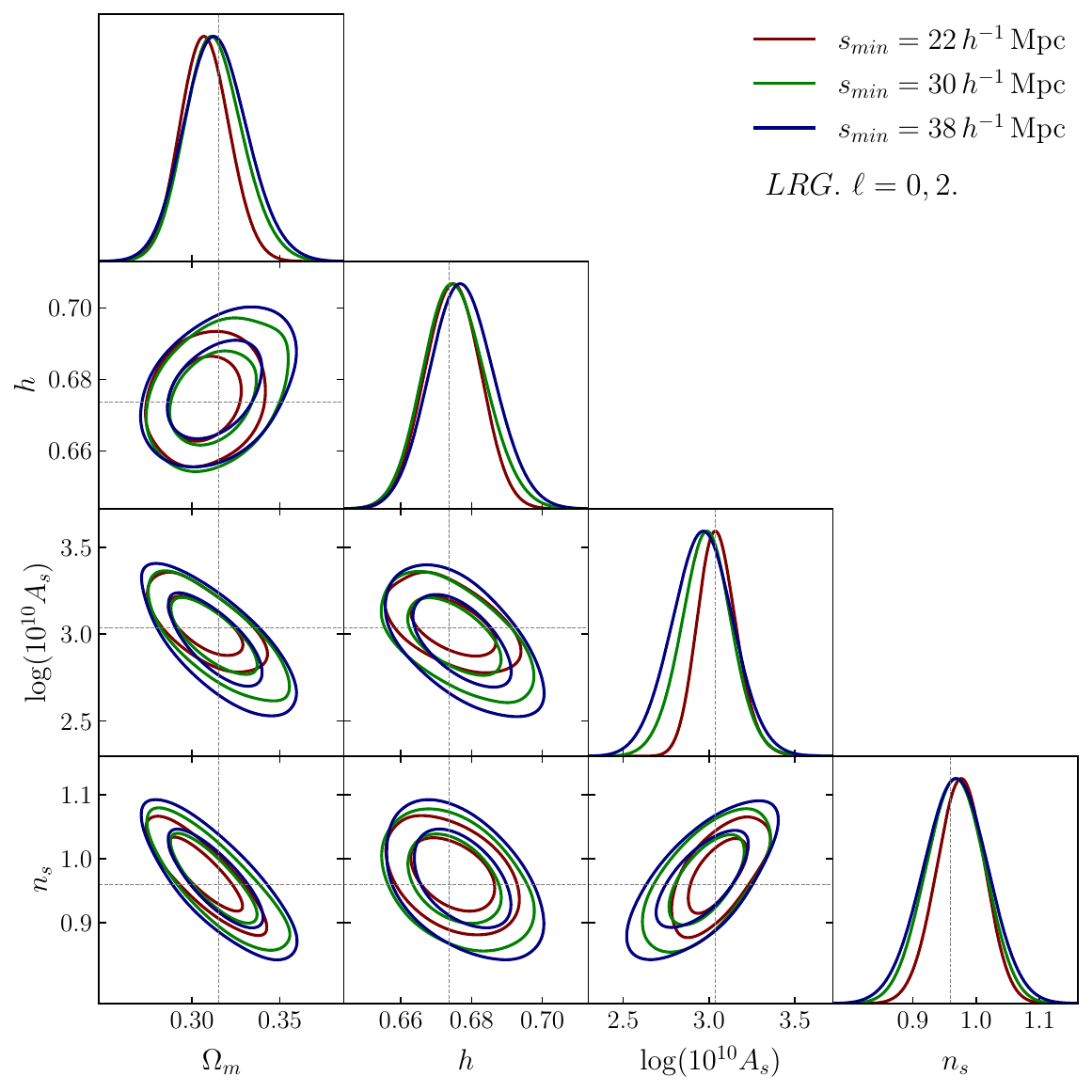}
\caption{The triangular plot displays the posterior one-sigma and two-sigma contours of our baseline analysis when $n_s$ is treated as a free parameter, using our {\it Full Modeling} methodology on the LRG sample. This plot illustrates the variation in predictions when adjusting the minimum scales. It is equivalent to the left panel of Figure \ref{FM_traingular_plot_LRG_non_rescaled} but with $n_s$ set as a free parameter.} 
\label{triangular_plot_ns}
\end{figure*}

Figure \ref{triangular_plot_ns} shows the results of our analysis when varying minimum scales, $s_{\text{min}}$, while treating $n_s$ as a free parameter for the LRG sample and using {\it Full Modeling}.  The results are presented for three distinct minimum scales,  and each one has a different performance in the precision at a parameter is recovered. Our analysis consistently captures the expected values of all parameters within the $1\sigma$ uncertainty contour, including the extra parameter $n_s$. Specifically, for $s_{\text{min}} = 22 \, h^{-1} \mathrm{Mpc}$, we achieve remarkable accuracy in determining $h$ and $\log(10^{10}A_s)$  with $0.13 \sigma$ and $0.13 \sigma$ deviations from the true value respectively. Similarly, when $s_{\text{min}} = 30 \,h^{-1} \mathrm{Mpc}$, our precision improves notably for $\Omega_m$, $h$, and $n_s$ with $0.13 \sigma$, $0.19 \sigma$ and $0.05 \sigma$ deviations respectively from the true value. Finally, when $s_{\text{min}} = 38 \,h^{-1} \mathrm{Mpc}$, our analysis aligns exceptionally well with $\Omega_m$ and $n_s$ within $0.07 \sigma$ and $0.04 \sigma$ deviations respectively. The resulting constraints on our parameters are summarized in Table \ref{tab:extended_ns} (to be compared with baseline analysis in Table \ref{tab:smin_cosmo}).

\begin{center}
\begin{table*}
\ra{1.7}
\begin{center}

\begin{tabular} {  c c c ccc}

\multicolumn{5}{c}{$n_s$ Free }\\
\hline
$s_{\text{min}}$ &$\Omega_{m}$ &$h$ & $\mathrm{ln}(10^{10}A_s) $ &  $ n_s$ \\
\hline
  $22 \,h^{-1} \mathrm{Mpc}$  & $0.307\pm 0.014 $   & $0.6746\pm 0.0076$ &  $3.05^{+0.10}_{-0.12}$        & $0.975\pm 0.037$ \\
  $30 \,h^{-1} \mathrm{Mpc}$  & $0.313^{+0.015}_{-0.017}$   & $0.6752\pm 0.0086$ &  $2.99\pm 0.15$        & $0.967\pm 0.045$ \\
 $38 \,h^{-1} \mathrm{Mpc}$ &$0.314\pm 0.018$&$0.6773\pm 0.0089$ &$2.96\pm 0.17$                          & $0.967\pm 0.050$\\
\hline
\end{tabular}
\caption{Summary of the constraints obtained when fitting the ABACUS LRG sample with our {\it Full Modeling} methodology for our three minimum scales and when we allow $n_s$ to vary as a free parameter
}.\label{tab:extended_ns}

\end{center}
\end{table*}
\end{center}

This shows how the sensitivity with which we can recover parameters varies with the minimum ($s_{\text{min}}$) scale when $n_s$ is free.The distinct performance levels observed across different $s_{\text{min}}$ values underscore the importance of carefully selecting this parameter in cosmological analyses. We see that, $s_{\text{min}} = 30 \,h^{-1} \mathrm{Mpc}$ is the scale exhibiting the most robust performance across all parameters, except $\A_s$ which is still predicted to within less than one sigma from the true value. Our results highlight the capability of our analysis to consistently capture parameters within one sigma uncertainty, highlighting its robustness in characterizing extended cosmological models.

The results presented in this section are built using our {\it Full Modeling} methodology only.

\subsubsection{Exploring the impact of  $w_0$ and $w_a$ as free parameters}
\label{w0wa_extension}
The parameters $w_0$ and $w_a$ 
characterize the behavior of dark energy within the Chevallier–Polarski–Linder model \cite{2001_Chevallier_Polarski, 2003Linder}. This model provides a framework to study an evolving equation of state for dark energy and is an extension of the cosmological constant $\Lambda$CDM model. The function $w(a) = w_0 + w_a(1 - a)$ describes the evolution of the equation of state parameter over cosmic times, offering insights into the dynamic nature of dark energy throughout the universe's history.

In this subsection, we analyze two different extensions of our methodology. In the first, we incorporate variations of the parameter $w_0$ only into our baseline analysis. Then, in the second extension, we vary both $w_0$ and $w_a$ and study their combined effect.

\begin{center}
\begin{table*}
\ra{1.7}
\begin{center}

\begin{tabular} {  c c c cccc}

\multicolumn{6}{c}{ $w_0$CDM Extension}\\
\hline
$s_{\text{min}}$ &$\Omega_{m}$ &$h$ & $\mathrm{ln}(10^{10}A_s) $ &  $ w_0$&  $ w_a$ \\

\hline
 $22 \,h^{-1} \mathrm{Mpc}$  & $0.309^{+0.015}_{-0.021}$   & $0.681^{+0.022}_{-0.019}$ &  $3.031\pm 0.090$        & $-1.031^{+0.077}_{-0.10}$& $0$ \\

 $30 \,h^{-1} \mathrm{Mpc}$  & $0.313^{+0.017}_{-0.023}$   & $0.679^{+0.025}_{-0.022}$ &  $2.99\pm 0.12$        & $-1.011^{+0.085}_{-0.11}$ & $0$ \\

$38 \,h^{-1} \mathrm{Mpc}$ &$0.314^{+0.018}_{-0.024}$ &$0.681^{+0.026}_{-0.022}$  & $2.94\pm 0.13$ & $-1.014^{+0.090}_{-0.12}   $& $0$\\
\hline

\hline
\multicolumn{6}{c}{ $w_0w_a$CDM Extension} \\
\hline
 $22\, h^{-1} \mathrm{Mpc}$  & $0.314^{+0.020}_{-0.026}$   &$0.674^{+0.027}_{-0.024}$ &  $3.13^{+0.11}_{-0.16}$ & $-0.88\pm 0.25$& $-0.67\pm 0.97$ \\
 $30\, h^{-1} \mathrm{Mpc}$  & $0.319^{+0.022}_{-0.027}$   & $0.673\pm 0.026$       & $3.07^{+0.14}_{-0.19}$   & $-0.86^{+0.29}_{-0.26}$ &$-0.7\pm 1.0$\\
$38\, h^{-1} \mathrm{Mpc}$ &$0.322^{+0.023}_{-0.029}$      &$0.673\pm 0.027$   &$3.07^{+0.16}_{-0.23}$  & $-0.83^{+0.30}_{-0.27}$& $-0.8\pm 1.1 $\\
\hline
\end{tabular}
\caption{Constraints when fitting the ABACUS LRG mock for three minimum scales when we allow $w_0$ an $w_a$ to vary as a free parameter and when using our {\it Full Modeling} methodology.}\label{tab:extended_wowa}

\end{center}
\end{table*}
\end{center}

We expand our standard setup, to incorporate $w_0$ and $w_a$ as additional parameters, and we use the resulting models to train a new neural network. To ensure the precision of our emulator, we increase the number of multipoles used to train and test our neural networks to 170,000 points. Of these, 150,000 are allocated for train the networks, with 10,000 designated for validation and the remaining 10,000 are our test set that evaluate the model's performance. We also apply the following priors to our parameters:
 \begin{align}
   w_0 &: \,  \mathcal{U}[-2,0]\\   
   w_a &: \,  \mathcal{U}[-3,3] 
 \end{align}
Here, we only consider the extension of our {\it Full Modeling} methodology.

\begin{figure*}
\centering
\includegraphics[width=140mm,height=140mm]{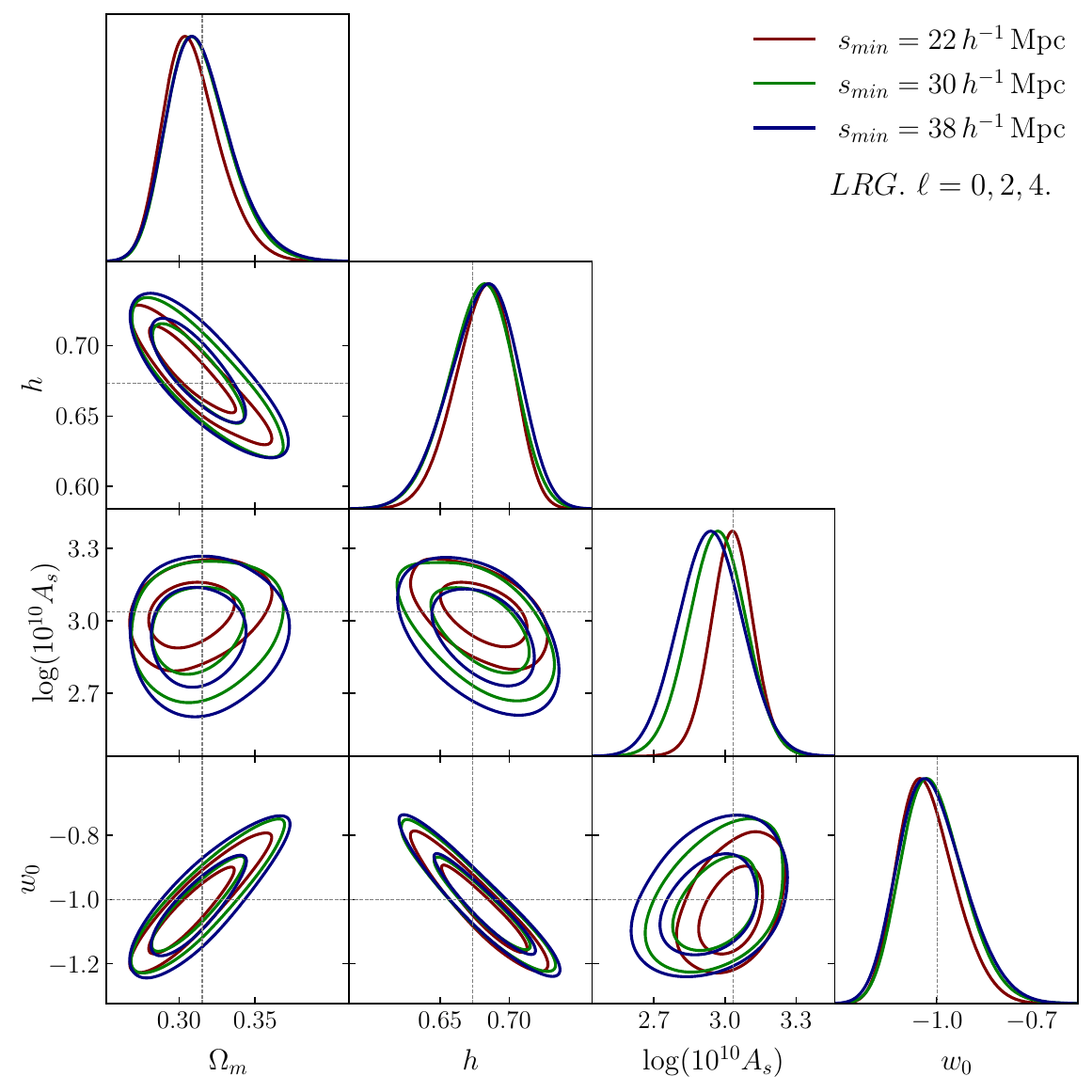}
\caption{The triangular plot displays the posterior one-sigma and two-sigma contours of our baseline analysis when $w_0$ is treated as a free parameter, using our {\it Full Modeling} methodology on the LRG sample. This plot illustrates the variation in predictions when adjusting the minimum scales.}
\label{triangular_plot_w0}
\end{figure*}

Let us first present our results for the extended methodology that only includes $w_0$. Figure \ref{triangular_plot_w0} shows the triangular plot of the constraints on the set of cosmological parameters, where we can see that our analysis yields parameter constraints that are all within 1$\sigma$ from their true value. The $s_{\text{min}} = 22 \,h^{-1} \mathrm{Mpc}$ case shows better performance when recovering cosmological parameters with smaller 1$\sigma$contours, while also being more accurate when predicting $A_s$. The model presents a $0.34 \,\sigma$ deviation for the true $\Omega_m$ value, $0.41 \sigma$ for $h$,  $0.05\, \sigma$ for $\log(10^{10}A_s)$, and $0.09 \,\sigma$ for $w_0$. Our extended model successfully recovers all parameters, including $w_0$, showcasing its robustness when recovering the cosmological parameters.

\begin{figure*}
\centering
\includegraphics[width=140mm,height=140mm]{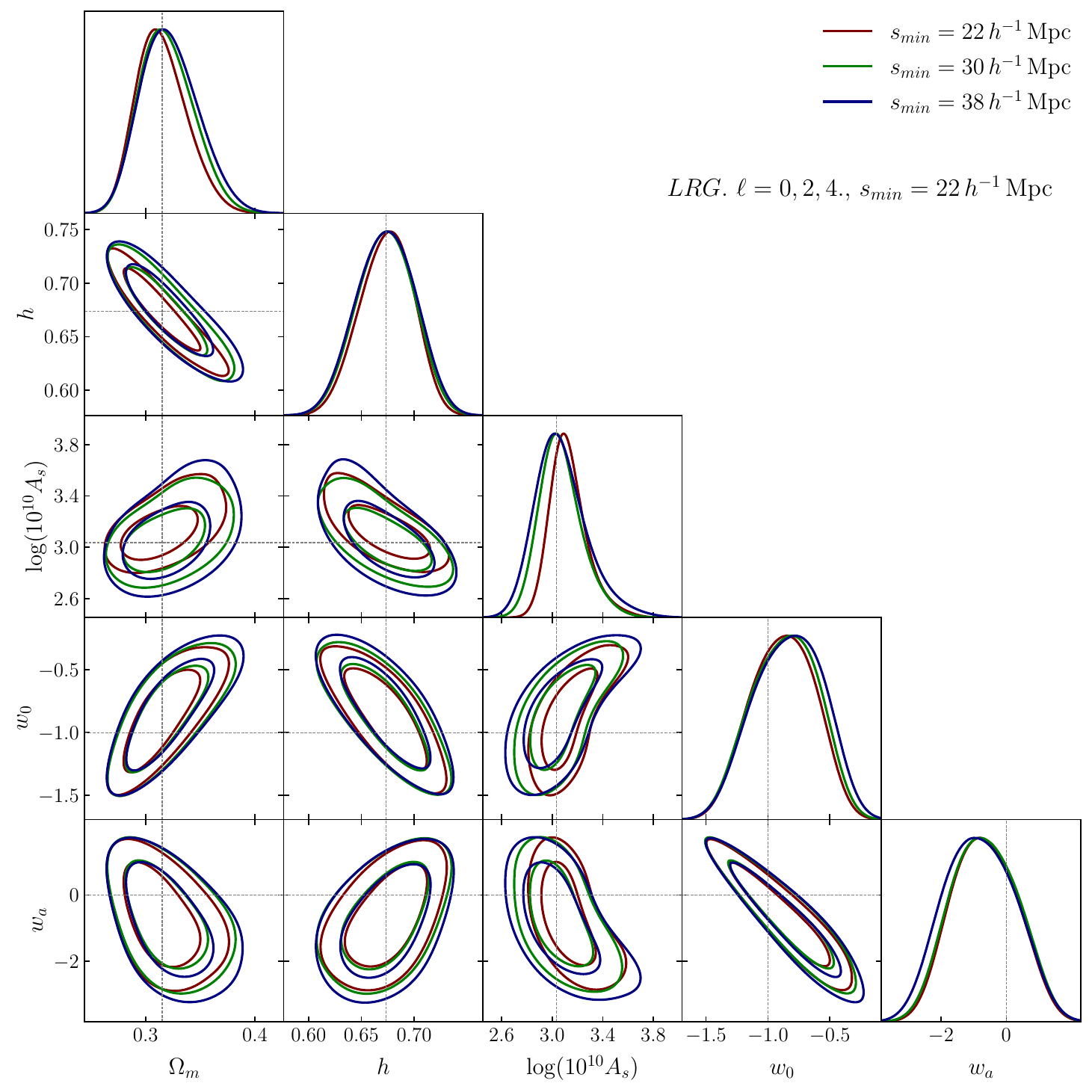}
\caption{This plot shows the error bar comparison for LRG using a volume of 8 $h^{-1} \mathrm{Mpc}$. The red bars were obtained using maximal freedom, while the red ones were obtained using minimal freedom. We show the model's behavior at various minimum scales, each one separated by 1 bin of 8 $h^{-1} \mathrm{Mpc}$. On the left side, we have the {\it ShapeFit} fits, while on the right side, we have the {\it Full Modeling} fits.}
\label{triangular_plot_w0wa}
\end{figure*}

Figure \ref{triangular_plot_w0wa} shows the results for our second extension when we also add $w_a$ as a free parameter and our extended model depends on both $w_0$ and $w_a$. As with the first extension, our analysis recovers all parameter constraints to within less than 1 $\sigma$ from the true value for all parameters. The three models perform similarly well showing similar accuracy and prediction in most parameters, we note that the model with a minimum scale of $s_{\text{min}} = 30\, h^{-1} \mathrm{Mpc}$ and $s_{\text{min}} = 38\, h^{-1} \mathrm{Mpc}$ slightly perform better than $s_{\text{min}} = 22\, h^{-1} \mathrm{Mpc}$ when recovering $A_s$ and $\Omega_m$, but it still has the smaller error bars.
The deviations for $s_{\text{min}} = 22\, h^{-1} \mathrm{Mpc}$ are within 1$\sigma$ for this model and are $0.05\, \sigma$ for $\Omega_m$, $0.01 \,\sigma$ for $h$,  $0.63\, \sigma$ for $\log(10^{10}A_s)$,  $0.47\, \sigma$ for $w_0$, and $0.69 \sigma$ for $w_a$. 

Table \ref{tab:extended_wowa} summarizes our final constraints for both extensions and all three scales. Our extended model successfully recovers all parameters, including $w_0$ and $w_a$, showing its ability to reliably recover the cosmological parameters.

\subsection{All DESI tracers: LRG, ELG, QSO mocks}
\label{sec:tracers}
In this section, we expand our analysis to include fits to the ELG and QSO mocks introduced in Section \ref{sec:data}. Additionally, we create a new dataset by fitting all three samples together, this is done by merging all three distinct data vectors. Each vector contributes to the total sum of the logarithm of the likelihood, which is subsequently fitted using our model.

As with our previous sections, we fit three iterations of our model, additionally, we perform all fits twice, once using our {\it ShapeFit} template and once using the {\it Full Modeling} approach.

\begin{figure*}
\includegraphics[width=75mm,height=80mm]{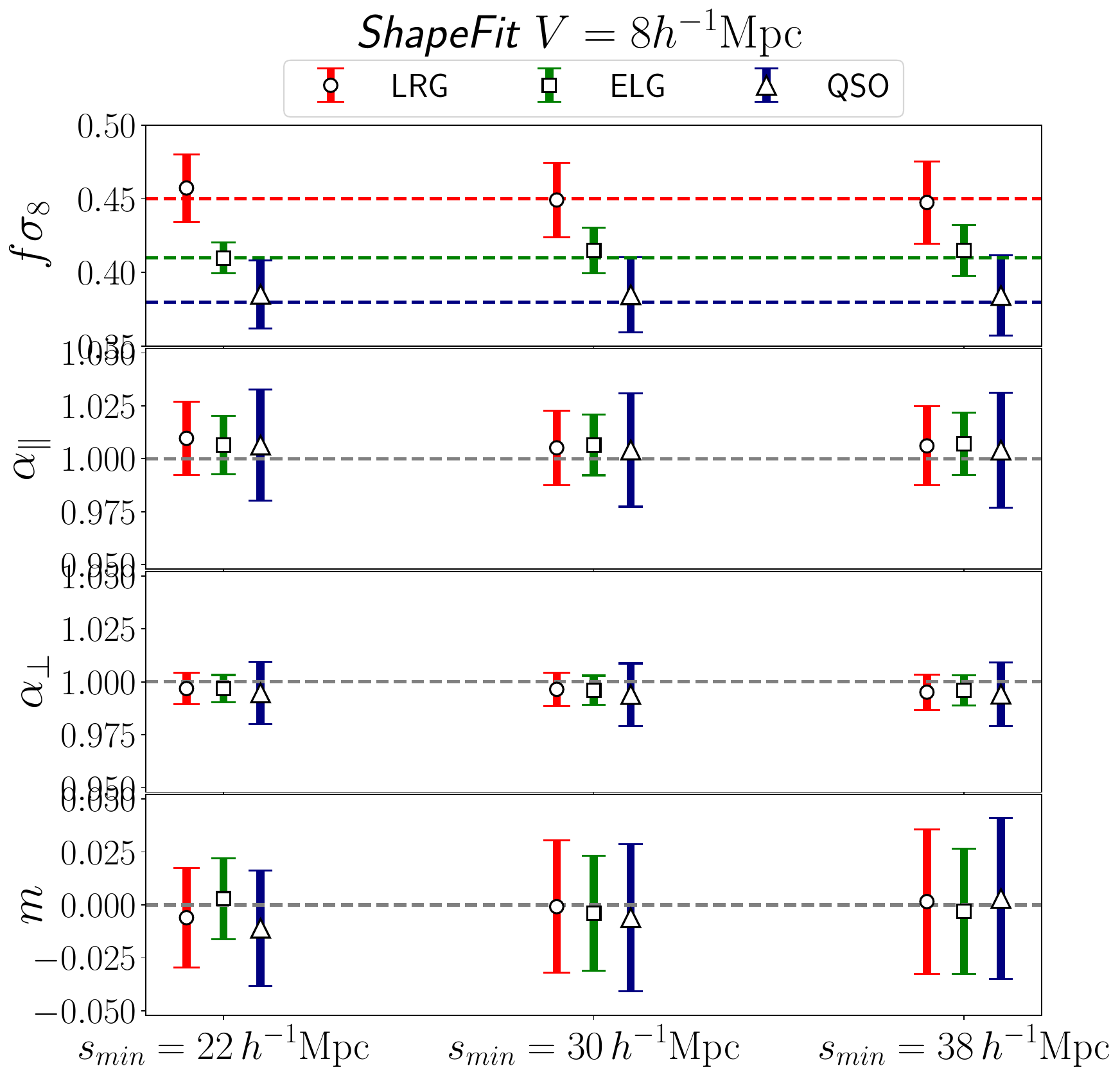}
\includegraphics[width=75mm,height=80mm]{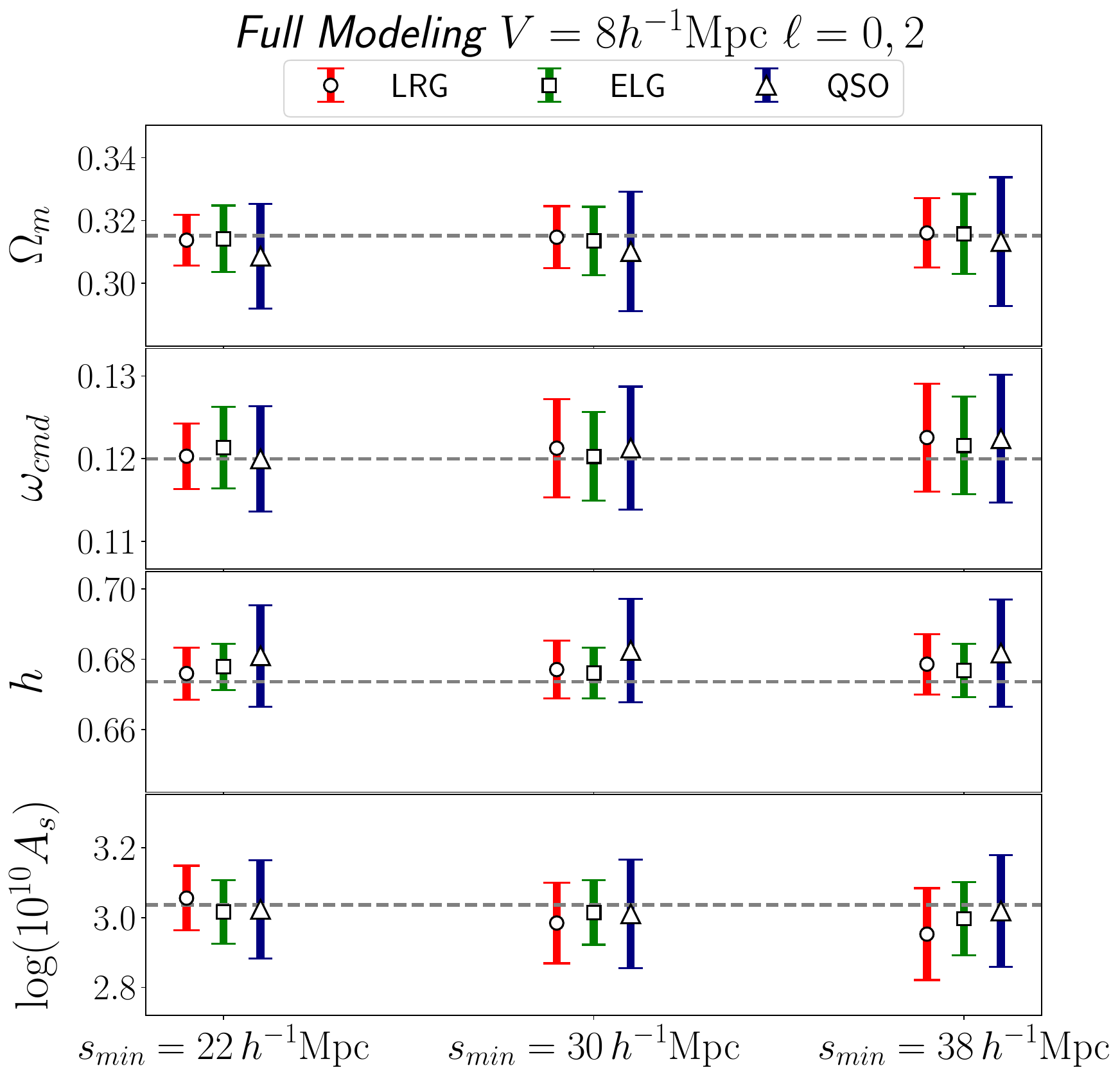}
\caption{Error bar plots on LRG (red), ELG (green), and QSO (blue) constraints for three minimum scales and for {\it Shape Fit} (left) and {\it Full Modeling} (right) parameters. The black dashed lines show the true value of the parameters (note that the true value of $f\sigma_8$ is dependent on redshift).}
\label{comparison_V1_all_tracers}
\end{figure*}

Tables \ref{tab:extended_Tracers_fm} and \ref{tab:extended_Tracers_sf} show the constraints on all \textit{ShapeFit} and \textit{Full Modeling} parameters for all fits mentioned above. The tables demonstrate that, in general, all constraints are tighter for our smaller minimum scale of $s_{\text{min}} = 22 ,h^{-1} \mathrm{Mpc}$. To visually compare these results more intuitively, Figure \ref{comparison_V1_all_tracers} summarizes them as a set of error bar plots that also include our previous LRG results from Section \ref{sec:baseline_lrg}. It is worth noting that all results in the figure are within one standard deviation from their expected value, indicating good accuracy. Given that all results are consistent with their expected value and that our measurements at the smaller minimum scale are slightly more precise, we select this as our main result.

\begin{figure*}
\centering
\includegraphics[width=110mm,height=110mm]{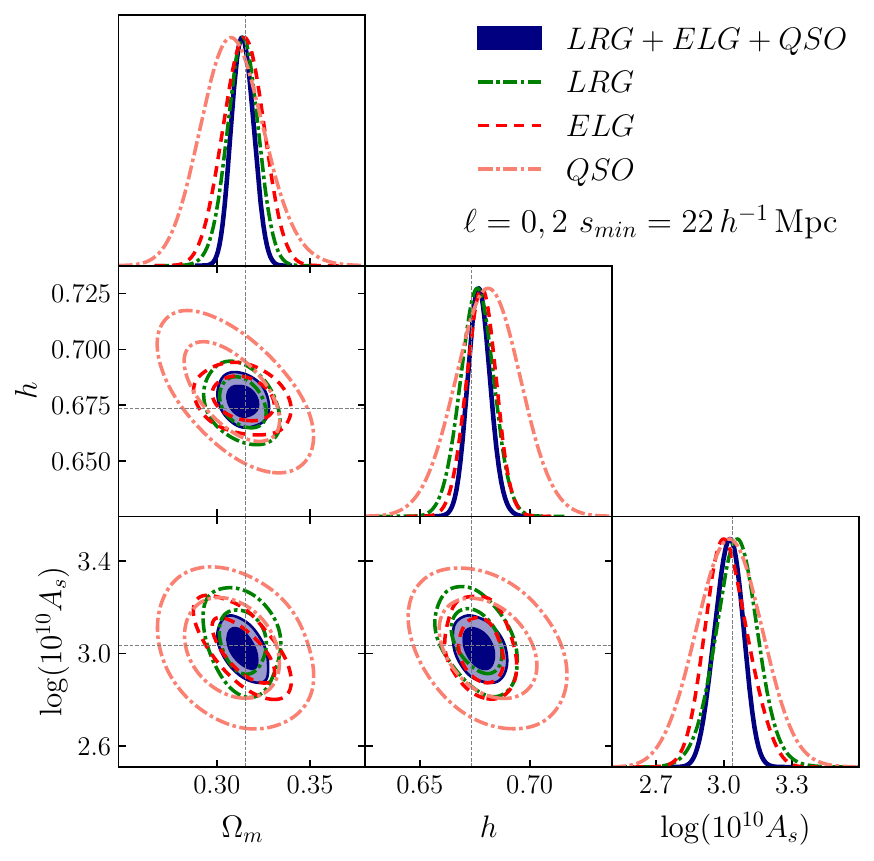}
\caption{We present results for our three tracers: LRG (green dashed lines), ELG (red dashed lines) and  QSO (salmon dashed lines). We also include the resulting contour of the combined $LRG+ELG+QSO$ (blue solid lines and contours), and the expected value of each parameter (gray faded lines).}
\label{LRG_ELG_QSO_smin22_Om}
\end{figure*}

Figure \ref{LRG_ELG_QSO_smin22_Om} depicts the triangular plot of our complete shape model alongside our findings for $s_{\text{min}} = 22, h^{-1} \mathrm{Mpc}$. We observe that, generally, the constraints from LRG and ELG exhibit similar precision, with ELG being marginally more precise, while QSO constraints are comparatively less precise overall. This aligns with our expectations based on the different effective volumes of the samples, with ELG being the largest DESI sample and QSO being the smallest. Additionally, we observe that our combined fit matches the expected value of all parameters within less than one sigma, with a $0.23 \, \sigma$ deviation for $\Omega_m$, a $0.68 \, \sigma$ deviation for $h$, and a $0.26 \,\sigma$ deviation for $\log(10^{10}A_s)$.

\begin{center}
\begin{table*}
\ra{1.7}
\begin{center}
\begin{tabular} {ccccc}
\multicolumn{4}{c}{DESI tracers}\\
\hline
$s_{\text{min}}$ &$\Omega_{m}$ &$h$ & $\mathrm{ln}(10^{10}A_s) $ \\
\hline
\multicolumn{4}{c}{ ELG \textit{Full Modeling}}\\
\hline
 $22 \,h^{-1} \mathrm{Mpc}$  & $0.314\pm 0.011$   & $0.6779\pm 0.0066$ &  $3.015^{+0.085}_{-0.10}$  \\
 $30\, h^{-1} \mathrm{Mpc}$  & $0.314\pm 0.011$   & $0.6761\pm 0.0072$ &  $3.014\pm 0.092$   \\
$38\, h^{-1} \mathrm{Mpc}$ &$0.316^{+0.012}_{-0.014}$ &$0.6769\pm 0.0076$  & $3.00^{+0.11}_{-0.10}$\\
\hline
\multicolumn{4}{c}{ QSO \textit{Full Modeling}}\\
\hline
 $22\, h^{-1} \mathrm{Mpc}$  & $0.309\pm 0.017 $   & $0.681\pm 0.014$ &  $3.02\pm 0.14$ \\
 $30\, h^{-1} \mathrm{Mpc}$  & $0.310\pm 0.019$   & $0.683\pm 0.015$  & $3.01\pm 0.16$\\
$38\, h^{-1} \mathrm{Mpc}$ &$0.3139\pm 0.0073$      &$0.6776\pm 0.0057$   &$2.983\pm 0.076$\\
\hline
\multicolumn{4}{c}{ COMBINED \textit{Full Modeling}} \\
\hline
 $22\, h^{-1} \mathrm{Mpc}$  & $0.3139\pm 0.0057$   & $0.6770\pm 0.0050$ &  $3.021\pm 0.060$  \\
 $30\, h^{-1} \mathrm{Mpc}$  & $0.3129\pm 0.0059$   & $0.6772\pm 0.0057$ &  $3.002\pm 0.062$  \\
 $38\, h^{-1} \mathrm{Mpc}$  & $0.309^{+0.015}_{-0.021}$   & $0.681^{+0.022}_{-0.019}$ &  $3.031\pm 0.090$ \\
\hline
\hline
\end{tabular}
\caption{Constraints to our {\it Full shape} (top) and {\it ShapeFit} (bottom) parameters when fitting our ABACUS ELG, QSO and Combined samples for all three minimum scales explored.}\label{tab:extended_Tracers_fm}

\end{center}
\end{table*}
\end{center}

\begin{center}
\begin{table*}
\ra{1.7}
\begin{center}
\begin{tabular} {ccccc}
\multicolumn{5}{c}{DESI tracers}\\
\hline
$s_{\text{min}}$ &$f\sigma_{8}$ &$\alpha_{\parallel}$ & $\alpha_{\perp}$& m \\
\hline
\multicolumn{5}{c}{ ELG \textit{ShapeFit}}\\
\hline
 $22\, h^{-1} \mathrm{Mpc}$  & $0.411\pm 0.019$   & $1.007\pm 0.014$ &  $0.9971\pm 0.0076$  &$-0.001^{+0.023}_{-0.020} $   \\
 $30\, h^{-1} \mathrm{Mpc}$  & $0.415^{+0.014}_{-0.016}$   & $1.007\pm 0.014$ &  $0.9960\pm 0.0070$   &  $-0.004\pm 0.027$\\
$38\, h^{-1} \mathrm{Mpc}$ &$0.415^{+0.016}_{-0.018}$ &$1.007\pm 0.015$  & $0.9959^{+0.0066}_{-0.0075}$& $-0.003\pm 0.030  $\\
\hline
\multicolumn{5}{c}{ QSO {\it ShapeFit}} \\
\hline
 $22\, h^{-1} \mathrm{Mpc}$  & $0.385\pm 0.023$   & $1.006\pm 0.026 $ &  $0.995\pm 0.015$ & $-0.011\pm 0.027  $ \\
 $30\, h^{-1} \mathrm{Mpc}$  & $0.385\pm 0.025$   & $1.004\pm 0.027$ &  $0.994\pm 0.015$   &  $-0.006\pm 0.035  $\\
$38\, h^{-1} \mathrm{Mpc}$ &$0.384^{+0.025}_{-0.029}$ &$1.004\pm 0.027$  & $0.994\pm 0.015 $& $0.003\pm 0.038  $\\
\hline
\end{tabular}
\caption{Constraints to our {\it Full shape} (top) and {\it ShapeFit} (bottom) parameters when fitting our ABACUS ELG, QSO, and Combined samples for all three minimum scales explored.}\label{tab:extended_Tracers_sf}

\end{center}
\end{table*}
\end{center}

\subsection{{\it ShapeFit} and {\it Full Modeling} methodology comparison}
\label{sec:comparison_sf_fm}
\begin{figure*}
\centering
\includegraphics[width=15cm]{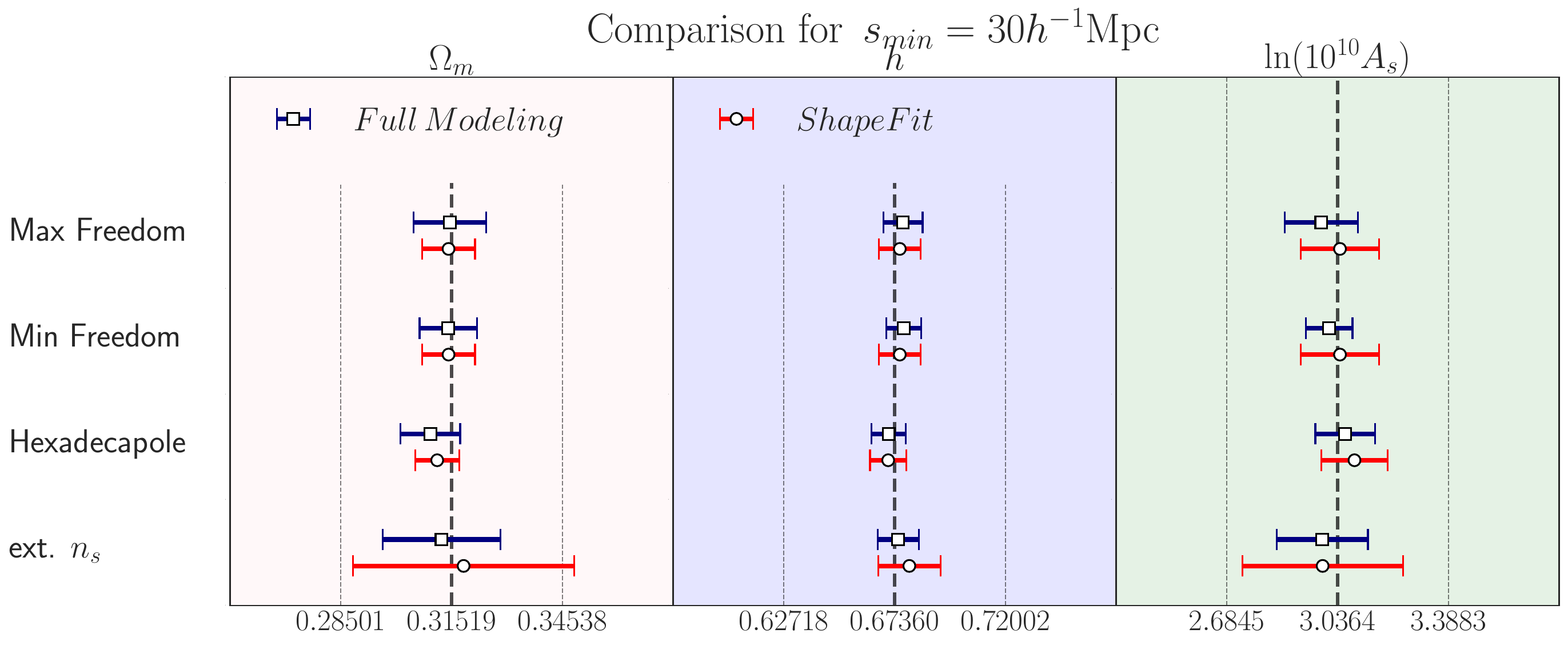}
\caption{Comparison of the {\it Full Modeling} and {\it ShapeFit} methodologies in the cosmological parameter space, the red bars and the blue bars represent respectively the \textit{ShapeFit} and  the \textit{Full Modeling} results obtained using the mean of the 25 realizations of the \texttt{ABACUS-SUMMIT} simulations with a non-rescaled covariance matrix when we fitted the monopole and quadrupole.}
\label{Shafefit_vs_fullmodeling_cosmo_params}
\end{figure*}

In this section, we transform {\it ShapeFit} from compressed parameters to cosmological parameters to be able to compare with {\it Full Modeling} in the same parameter space.
In Figure \ref{Shafefit_vs_fullmodeling_cosmo_params} we present a plot where we can observe small differences for {\it ShapeFit} and {\it Full Modeling} when we fit the correlation function using LRG for a volume of  $V=8 \,h^{-3} \mathrm{Gpc^3}$ for the scales, $s_{min}=30\, h^{-1} \mathrm{Mpc}$. In particular, we observe a good agreement for the Maximal Freedom, Minimal Freedom, the case when we include the hexadecapole and the $n_s$ extension having a similar constraining power.

First, we look at the baseline analysis, which shows good performance as we can notice from Table \ref{tab:smin_cosmo}, where for $\Omega_m$, we have nearly the same results in both methodologies, with a central value difference in the range of $10^{-3}$ and error bars of $0.68 \sigma_{FM}<\sigma_{SF}<0.74 \sigma_{FM}$ being {\it ShapeFit} more constrictive for the three minimum scales. For $h$, we have a similar result obtaining a difference in the central value of around $10^{-3}$ and error bars of $1.11 \sigma_{FM}   <  \sigma_{SF}  <1.0 \sigma_{FM}$ this time, shape fit shows less constrictive power. In the last parameter, $\mathrm{ln}(10^{10}A_s) $, the central value shows a difference of the order of $10^{-3}$ with error bars $1.0 \sigma_{FM}   <  \sigma_{SF}  <1.18 \sigma_{FM}$, {\it ShapeFit} again is less constrictive.

The Minimal Freedom case shows similar performance, the statistic is detailed in Table \ref{tab:cosmoresults}, where we obtain for $\Omega_m$ roughly the same results in the central value with a difference of the order of $10^{-3}$ and error bars $0.76 \sigma_{FM}   <  \sigma_{SF}  <0.80 \sigma_{FM}$ being {\it ShapeFit} more constrictive. For $h$, we have a central value which shows a difference around $10^{-3}$ and error bars of $1.05 \sigma_{FM}   <  \sigma_{SF}  <1.1 \sigma_{FM}$, with {\it ShapeFit} with less constrictive power. Finally, $\mathrm{ln}(10^{10}A_s) $ shows a central value difference of the order of $10^{-2}$ with error bars $1.09 \sigma_{FM}   <  \sigma_{SF}  <1.33 \sigma_{FM}$, again {\it ShapeFit} less constrictive.

When we include the hexadecapole in the baseline analysis, we obtain similar constraining power in both analyses. In Table \ref{tab:cosmoresults}, we can see that for $\Omega_m$, we have a central value with a difference around $10^{-3}$ and error bars of $0.71 \sigma_{FM}   <  \sigma_{SF}  <0.77 \sigma_{FM}$ being {\it ShapeFit} more constrictive for all the minimum scales. For the $h$ parameter, we have a central value with a difference around $10^{-3}$ and error bars of $1.00 \sigma_{FM}   <  \sigma_{SF}  <1.11 \sigma_{FM}$, with {\it ShapeFit} less constrictive. Lastly, $\mathrm{ln}(10^{10}A_s) $, shows a central value difference of the order of $10^{-2}$ with error bars $1.0 \sigma_{FM}   <  \sigma_{SF}  <1.09 \sigma_{FM}$, {\it ShapeFit} being less constrictive.

On the other hand, when we include $n_s$ as a model extension, we implement the extra parameter $n$, in the {\it ShapeFit} methodology, so we vary $\alpha_{\perp}$, $\alpha_{\parallel}$, $f\sigma_{8}$, $m$ and $n$, then, we convert these parameters into cosmological parameter using $n=n_s-n_{\texttt{ref}}$ \cite{brieden_2021}, these results are in Table \ref{tab:cosmoresults}. We notice that for $\Omega_m$, we have a central value of the order of $10^{-3}$, with a difference around $0.72 \sigma_{FM}   <  \sigma_{SF}  <0.93 \sigma_{FM}$ being {\it ShapeFit} more constrictive. In the case of $h$, we have a central value difference of the order $10^{-3}$ and error bars which show a difference of $0.99 \sigma_{FM}   <  \sigma_{SF}  <1.11 \sigma_{FM}$. Then, we have the $\mathrm{ln}(10^{10}A_s) $ parameter, which shows a central value difference of the order of $<10^{-2}$ with error bars $0.88 \sigma_{FM}   <  \sigma_{SF}  <1.18 \sigma_{FM}$. Finally, the $n_s$ parameter shows a difference in the central value of the order of $10^{-3}$ while the error bars are $1.32 \sigma_{FM}   <  \sigma_{SF}  <1.54 \sigma_{FM}$ with {\it ShapeFit} being less constrictive.

So far, we extensively explore two distinct methodologies for predicting the recovery of cosmological parameters: {\it ShapeFit} and {\it FullModeling}. We demonstrate the accuracy and efficacy of both approaches for the baseline analysis.

\begin{center}
\begin{table*}
\ra{1}
\begin{center}

\begin{tabular} {ccccccc}
\multicolumn{7}{c}{ Cosmological Parameter Constraints }\\
\hline
\multicolumn{7}{c}{ Min Freedom }\\
\hline
T& $s$ &$\Omega_{m}$ &$h$ & $\mathrm{ln}(10^{10}A_s) $& &  \\
\hline
FM& $22$  & $0.3150\pm 0.0072$   & $0.6755\pm 0.0072$ &  $3.058\pm 0.065$        &-&-\\
FM&$30$  & $0.3139\pm 0.0082$   & $0.6767\pm 0.0078$ &  $2.989\pm 0.082$        &-&-\\
FM&$38$ &$0.314\pm 0.010$ &$0.6786\pm 0.0080$  & $2.96\pm 0.11$ &-&-\\
\hline
 SF&$22$  & $0.3144\pm 0.0058$   & $0.6743\pm 0.0080$ &  $3.082\pm 0.087$       &-&-\\
 SF&$30$  & $0.3146\pm 0.0062$   & $0.6758\pm 0.0082$ &  $3.01\pm 0.10$    &-&-   \\
SF&$38$   &$0.3155\pm 0.0077$    &$0.6783\pm 0.0085$  & $22.97\pm 0.12$ &-&-\\
\hline
\multicolumn{7}{c}{ Hexadecapole }\\
\hline
FM& $22$  & $0.3138^{+0.0099}_{-0.0085}$   & $0.6707\pm 0.0072$ &  $3.098^{+0.066}_{-0.082}$        &-\\
FM& $30$  & $0.3094\pm 0.0082$   & $0.6711\pm 0.0071$ &  $3.060\pm 0.095$               &-&-\\
FM&$38$ &$0.3107\pm 0.0093$ &$0.6741\pm 0.0075$  & $3.01\pm 0.13$  &-&-\\
\hline
 SF&$22$  & $0.3142\pm 0.0065$   & $0.6684\pm 0.0072$ &  $3.117\pm 0.081$      &-&-\\
 SF&$30$  & $0.3113\pm 0.0060 $   & $0.6709\pm 0.0076$ &  $3.09\pm 0.11$     &-&-  \\
SF&$38$ &$0.3128\pm 0.0072$      &$0.6730\pm 0.0083$  & $3.07\pm 0.13$ &-&-\\
\hline
\multicolumn{7}{c}{$n_s$}\\

\hline
T& $s$ &$\Omega_{m}$ &$h$ & $\mathrm{ln}(10^{10}A_s) $& $n_s$&  \\
\hline

FM& $22$  & $0.307\pm 0.014 $   & $0.6746\pm 0.0076$ &  $3.05^{+0.10}_{-0.12}$        & $0.975\pm 0.037$         &-\\
FM& $30$  & $0.313^{+0.015}_{-0.017}$   & $0.6752\pm 0.0086$ &  $2.99\pm 0.15$        & $0.967\pm 0.045$         &-\\
FM&$38$ &$0.314\pm 0.018$&$0.6773\pm 0.0089$ &$2.96\pm 0.17$                          & $0.967\pm 0.050$ &-\\
\hline

 SF&$22$  & $0.326^{+0.033}_{-0.038}$   & $0.681\pm 0.013$ &  $2.97^{+0.25}_{-0.29}$       &$0.93\pm 0.14$&- \\
 SF&$30$  & $0.319^{+0.031}_{-0.037}$   & $0.680^{+0.013}_{-0.015}$ &  $2.99\pm 0.27$     &$0.96\pm 0.14$&-   \\
SF&$38$ &$0.317^{+0.031}_{-0.038}$ &$0.680^{+0.013}_{-0.015}$  & $3.01\pm 0.30$ &$0.97\pm 0.15$&-\\
\hline

\end{tabular}
\caption{Constraints on our parameters when fitting our ABACUS LRG sample for three minimum scales when using {\it Full Modeling} methodology and {\it ShapeFit} methodologies.}\label{tab:cosmoresults}

\end{center}
\end{table*}
\end{center}

\section{Comparisons with others Effective Field Theory models }
\label{sec:comparison_models}
As stated at the beginning of this manuscript, our \texttt{EFT-GSM} model is one of four EFT theories utilized in the analysis of galaxy Full-Shape within DESI year one data. In the companion paper \cite{KP5s1-Maus}, the three models in Fourier space (FS) (namely, the \texttt{Velocileptors} model of \cite{KP5s2-Maus}, the \texttt{Folps$\nu$} model of \cite{KP5s3-Noriega}, and the \texttt{PyBird} model of \cite{KP5s4-Lai}) are compared to each other, this section extends the comparison by showing how their results align with our configuration space (CS) counterpart presented in this work. 
We highlight that our \texttt{GSM-EFT} methodology is the only model fully developed in CS out of the four models considered here. 

We divide this comparison into two parts. The first section \ref{FS_comparison} shows two results where the analysis of the three models we are comparing with is done in FS. The first result is a direct comparison between our parameter constraints in CS and the original FS results of \texttt{Velocileptors}, \texttt{Folps$\nu$}, and \texttt{PyBird}. The second comparison involves using a different implementation of these three codes, where we calculate their results using the \texttt{desilike}\footnote{https://github.com/cosmodesi/desilike/} tool. This tool is a compendium of all the methodologies required for the likelihood analysis of DESI. It's designed to provide a common framework for all analyses. Therefore, it allows us to execute all the important parts of our methodologies using the same tool for all models. This includes the MCMC sampler, likelihood estimations, methods for fabricating the input linear power spectrum and so on. Therefore when using \texttt{desilike}, the only difference between the methodologies used to analyze each of the four models is the method used to compute the redshift space multipoles from the input linear power spectrum.

Working with \texttt{desilike} is not only useful because it allows us to standardize the methodology between codes, but it also provides a standardized tool to transform FS multipoles into the CS. Therefore, it permits us to reanalyze the \texttt{Velocileptors}, \texttt{Folps$\nu$}, and \texttt{PyBird} results, this time in CS. Section (\ref{CS_comparison}) shows how our EFT-GSM results compare with these new CS model constraints. 

Throughout this whole section, we present all analyses in both \textit{Full Modeling} and \textit{ShapeFit} configurations. Table \ref{table:eftcodes} provides a short description of the codes used in this comparison. For a full description of the codes and individual performance, we refer the reader to the companion papers. 
\begin{center}
\begin{table*}
\begin{center}
\begin{tabular} { ll}
Model               & Description  \\
\hline
\texttt{Folps-$\nu$}& EPT/EFT with beyond-EdS kernels \cite{KP5s3-Noriega}\\
\texttt{PyBird} &  EPT /EFT \cite{KP5s4-Lai}\\
\texttt{Velocileptors}&  LPT/EFT with the possibility to run EPT/EFT \cite{KP5s2-Maus}.\\
\hline
\end{tabular}
\caption{Brief description of the models compared in FS}\label{table:eftcodes}
\end{center}
\end{table*}
\end{center}
The four models utilize distinct conventions for their different galaxy-halo bias parameters. The translation between these conventions can sometimes be challenging to reproduce across all models. Therefore, in this study and its companion paper, we concentrate only on cases where this mapping is well-defined. While \cite{KP5s2-Maus} explores the two cases \textit{minimal} and \textit{maximal freedom}, all comparisons presented here are done using only the \textit{minimal freedom} configuration. This is done because our maximal freedom GSM-EFT model does not implement the tidal bias parameter of \cite{KP5s1-Maus} , and therefore our {\it maximal} freedom definition is not  equivalent to one of the other models.

The definitions of {\it minimal freedom} for FS models are presented in \cite{KP5s1-Maus}, equations 4.15 and 4.16. Our comparisons are conducted by fitting the mean of the Abacus mocks using our covariance matrix without any re-scaling, which corresponds to having a volume of V=8$h^{-1} \mathrm{Mpc}$.

\subsection{Comparison with EFT models in Fourier Space}
\label{FS_comparison}

As stated above, throughout this section, we present a set of results from analyses conducted in FS. All of these FS analyses select the same fiducial settings as those utilized in the companion paper. Therefore, we fit both the monopole and quadrupole up to $k_{\rm max}=0.25 h^{-1}$Mpc. We begin this section by introducing a comparison between our CS results and the original FS results of \texttt{Velocileptors}, \texttt{Folps$\nu$}, and \texttt{PyBird}. These comparisons are done for all three tracers: LRG, ELG, and QSO. We utilize the priors presented in Tables 1, 2, and 3 of the companion paper \cite{KP5s1-Maus}.
 
\begin{figure*}
\centering
\includegraphics[width=75mm]
{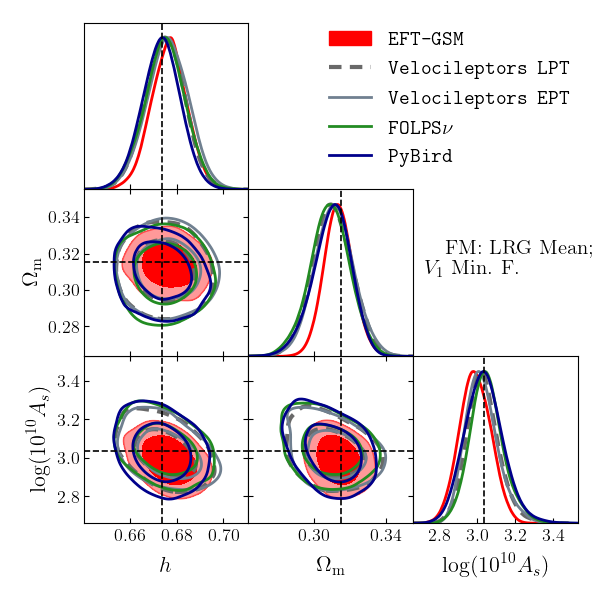}
\includegraphics[width=75mm]
{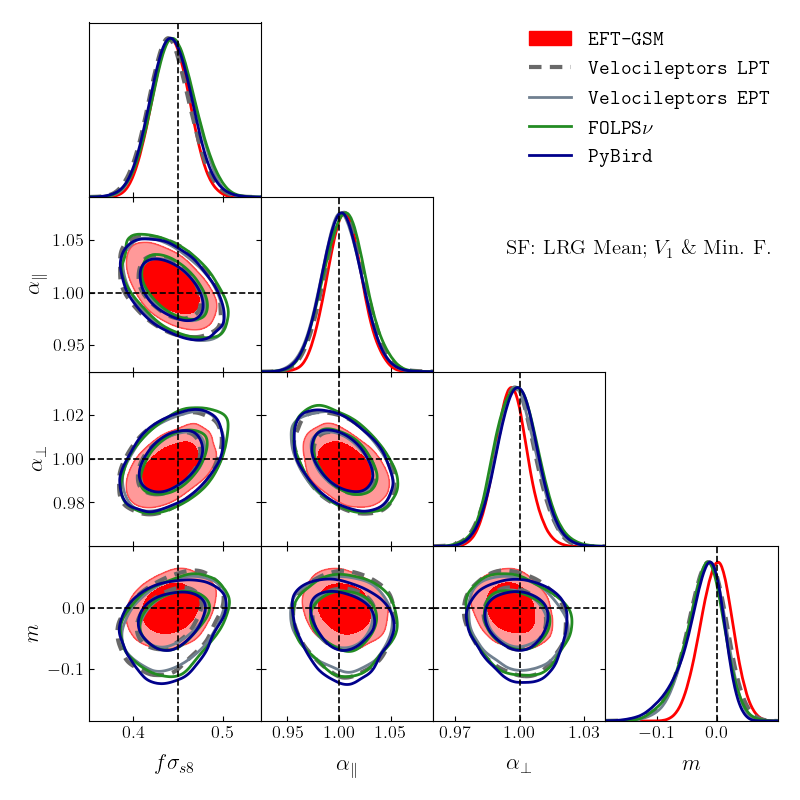}
\includegraphics[width=75mm]
{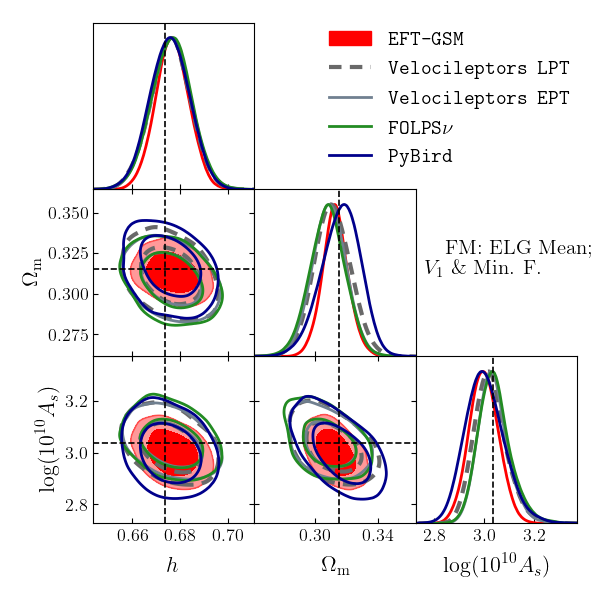}
\includegraphics[width=75mm]
{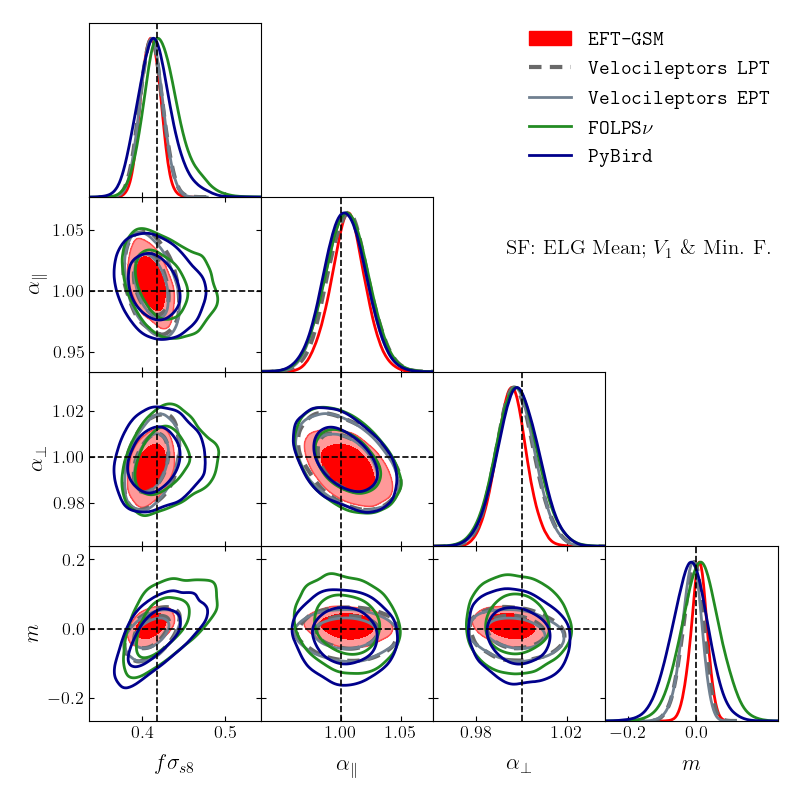}
\includegraphics[width=75mm]
{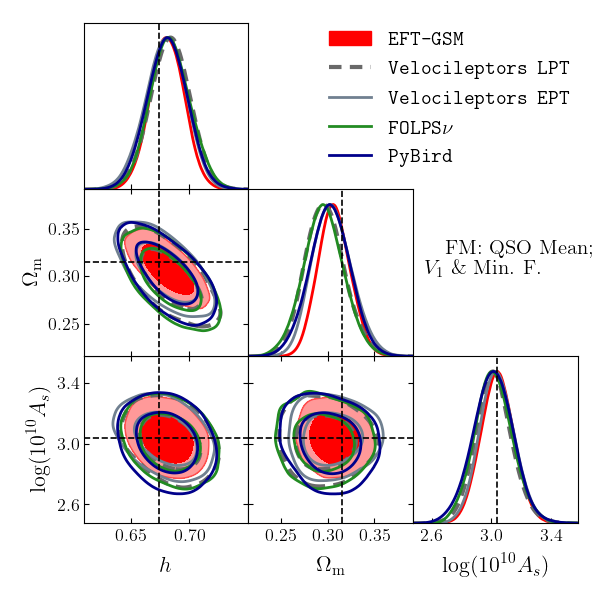}
\includegraphics[width=75mm]
{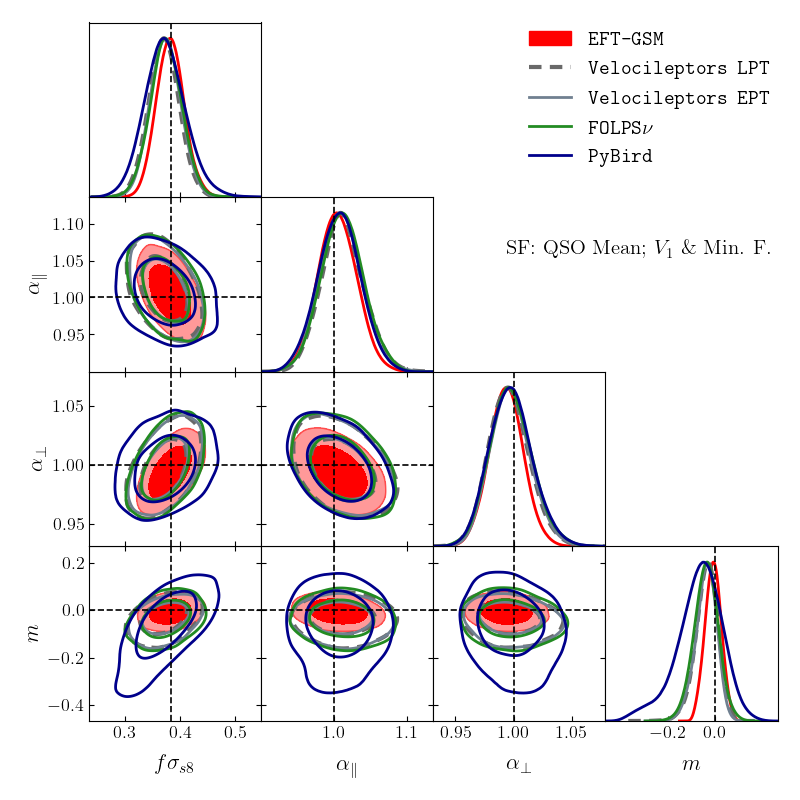}
\caption{All models using their baseline analysis for all tracers. Need to replace left plots with the {\it Full Modeling} cases, change the format of the plot, i.e put in solid GSM-EFT}   
\label{fig:modelsFS}
\end{figure*}

Figure \ref{fig:modelsFS} shows a set of triangular plots with the constraints we get using our {\it Full Modeling} (left panels) and {\it ShapeFit} (right panels) configurations, for our LRG (top panels), ELG (intermediate panels), and QSO (bottom panels) data respectively. It is important to notice that the posteriors of the FS results are generated with the
original codes and are the same as those analyzed in the companion paper. We observe that the 1-$\sigma$ contours of our \texttt{GSM-EFT} model align with those from the FS models showing very good agreement, while our model  gives slightly narrower constraints. The observed difference between results may not be unexpected, considering that not all analyses are conducted in the same space and are not done with one standardized framework.

\begin{figure*}
\centering
\includegraphics[width=75mm,height=80mm]{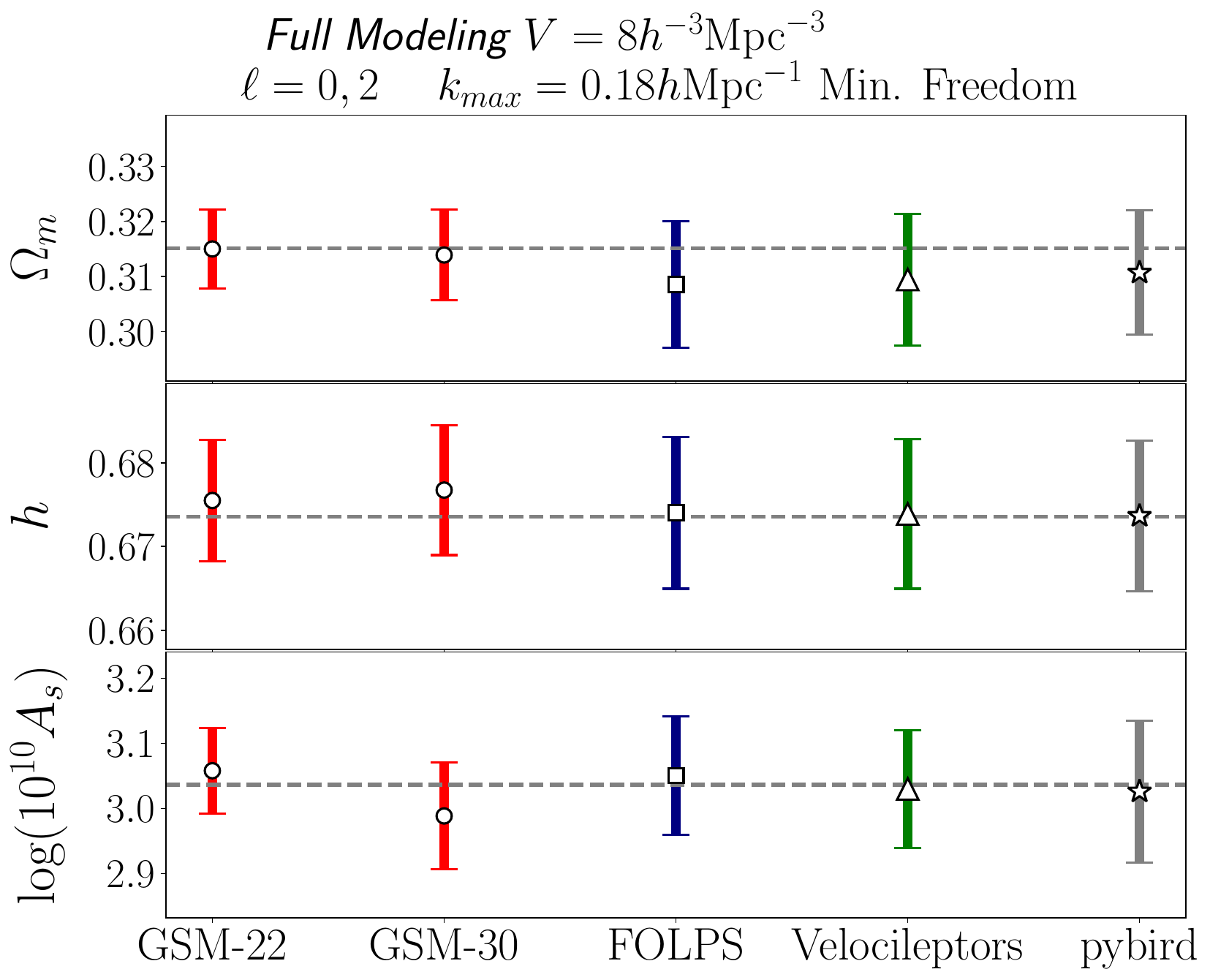}
\includegraphics[width=75mm,height=80mm]{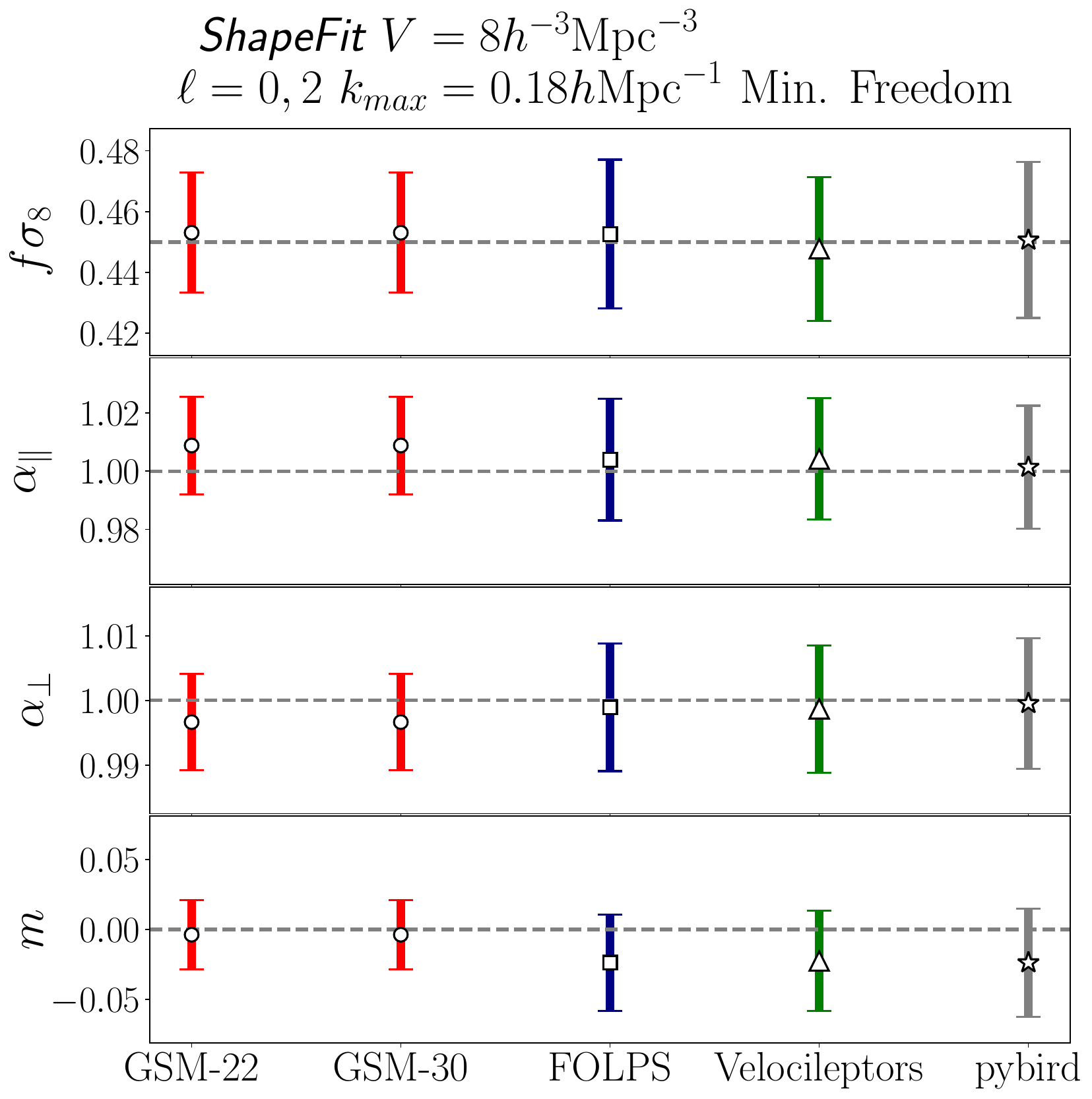}
\caption{All models using their baseline analysis for the minimal freedom case with desilike 
}   
\label{fig:modelsFS_desilike}
\end{figure*}

We observe that our CS constraints consistently yield smaller values compared to the three FS results obtained from other codes. Additionally, this trend holds true for all three tracers. As we mentioned, we have not standardized our framework in terms of MCMC sampler, likelihood evaluation and the fitting ranges between FS and CS are not exactly equivalent, which could explain the differences observed between methodologies. With this in mind, we test how our model compares with the rest when all the FS models are analyzed using the common framework from \texttt{desilike}. Figure \ref{fig:modelsFS_desilike} shows the results from this analysis and compares it with our GSM-EFT results. 
This comparison is complementary to the analysis of the next sub-section, where  we  use the \texttt{desilike} framework to redo all likelihood analysis in CS instead. These tests are intended as a verification that the differences observed so far are still present when we homogenize our analysis with \texttt{desilike}.
Tables \ref{tab:desilike_FM} and \ref{tab:desilike_SF}
present the quantitative results of this FS \texttt{desilike} analysis. Before analysing \texttt{desilike} results we verified that the results that the {\it Full Modeling} and {\it ShapeFit} constraints and errors were consistent between the original pipeline and their implementation within \texttt{desilike}. As we highlight in the next section, the difference found between FS and CS is also found in the CS implementation of the FS models, and is larger than the difference found between FS results with different pipelines.

\subsection{Comparison with EFT models in Configuration space with \texttt{DesiLike}}
\label{CS_comparison}
In this final section we discuss the performance of all four models in CS. It is important to note that the \texttt{Velocileptors}, \texttt{Folps$\nu$}, and \texttt{PyBird}, models work in FS by construction. The CS counterparts of the original FS code are also implemented in \texttt{desilike}, the transformation between one space and the other is done using the Hankel transform of the FS multipoles,
\begin{equation}
  \xi_{\ell}(s) = \frac{(-i)^{\ell}}{2 \pi^{2}} \int dk k^{2} P_{\ell}(k) j_{\ell}(ks)  
\end{equation}

The Hankel transform is implemented in the \texttt{cosmoprimo}\footnote{https://github.com/cosmodesi/cosmoprimo/} package using the \texttt{fftlog} algorithm. For the Hankel transform we used a $k_{min}=10^{-4}$ and $k_{max}=0.6$. Beyond 0.6, there is a Gaussian damping.

\begin{figure*}
\centering
\includegraphics[width=75mm,height=80mm]{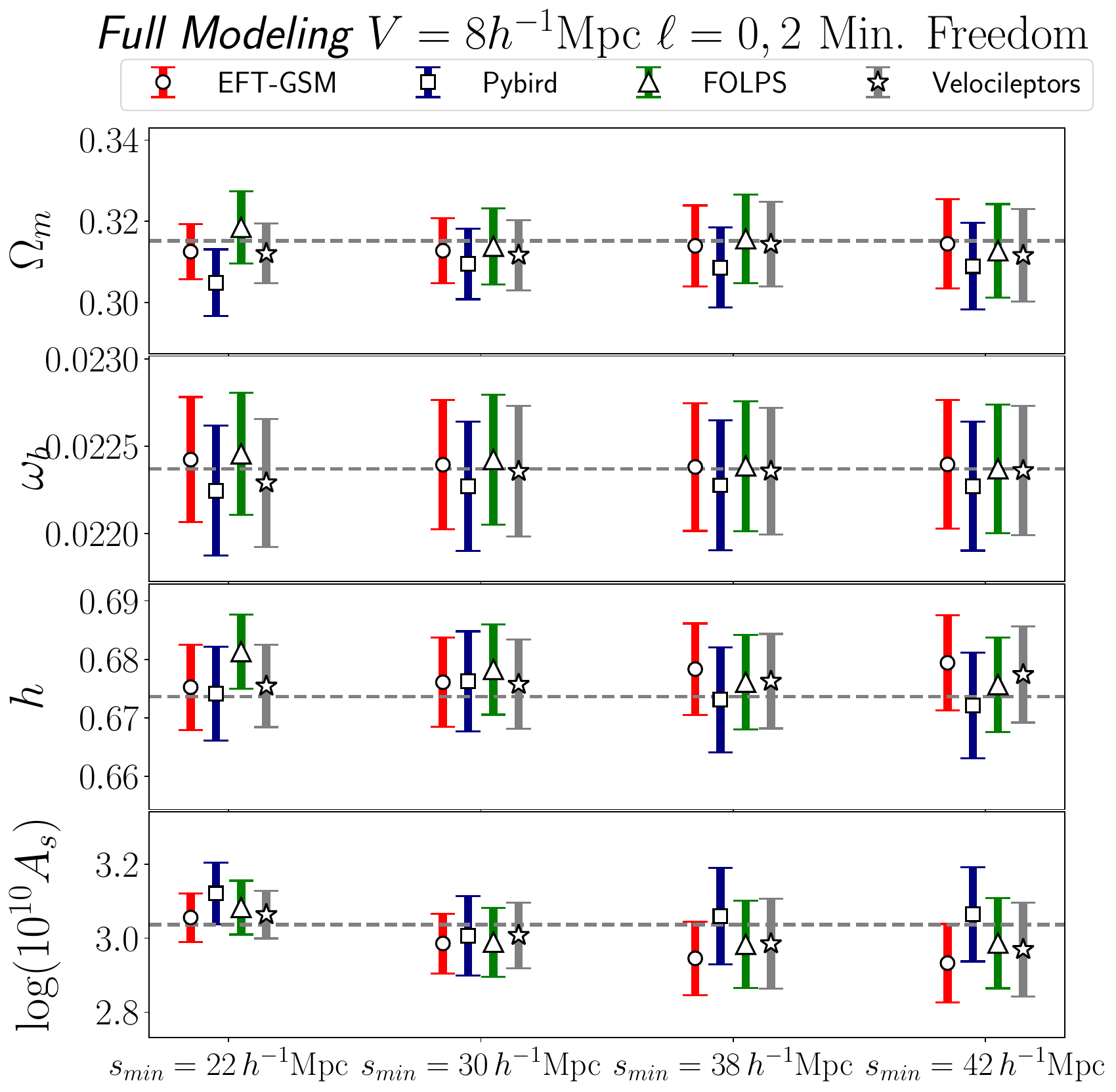}
\includegraphics[width=75mm,height=80mm]{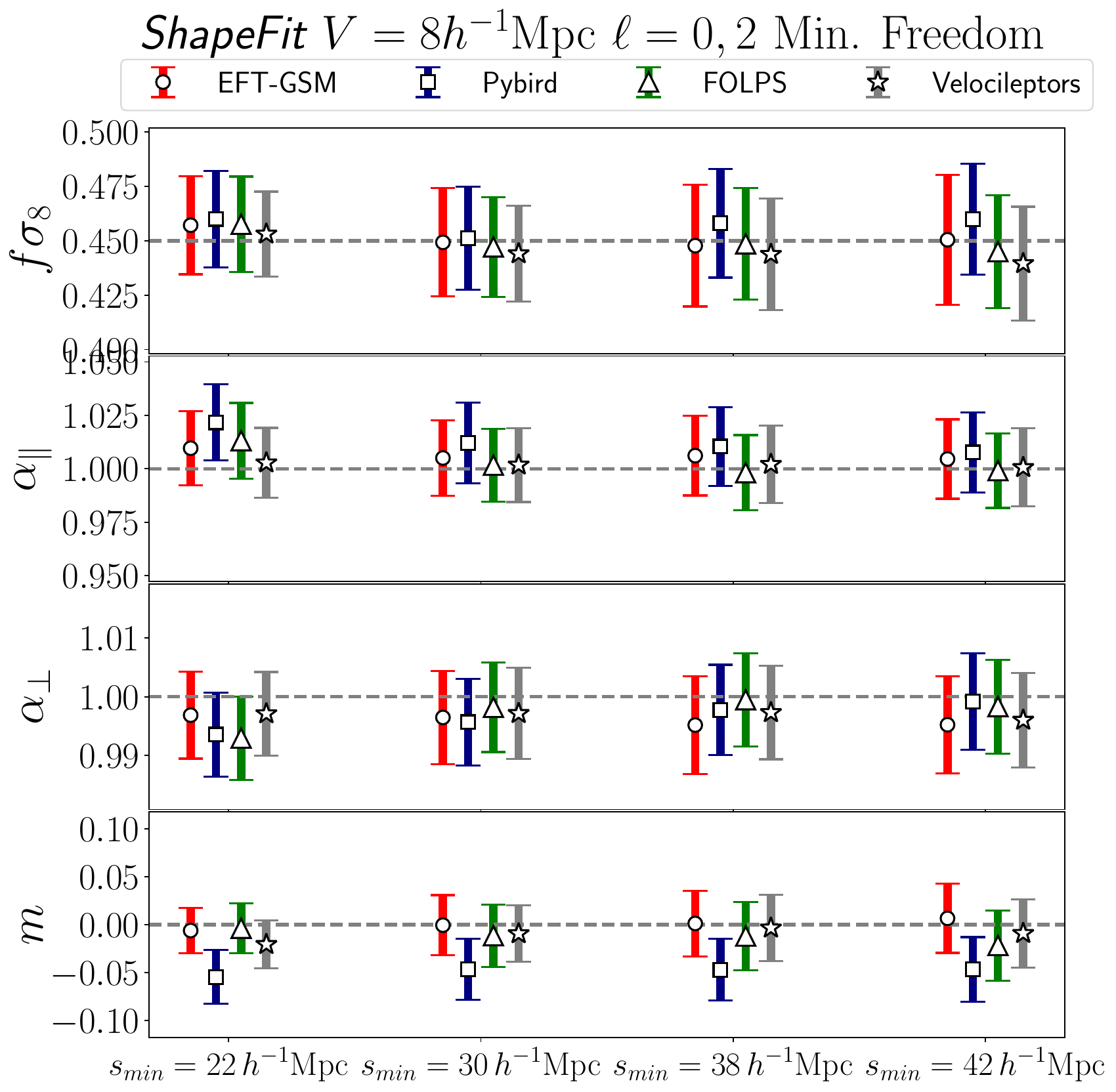}
\caption{All models using their baseline analysis for the minimal freedom case.
}   
\label{fig:modelsCS}
\end{figure*}

\begin{figure*}
\centering
\includegraphics[width=75mm,height=80mm]{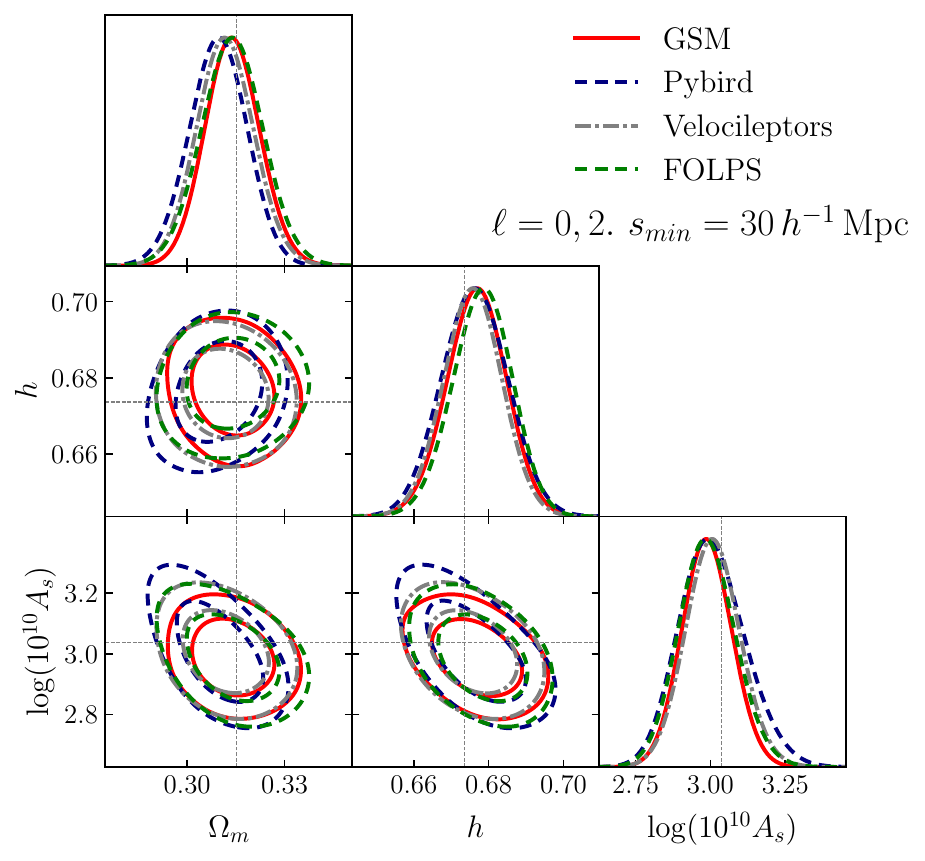}
\includegraphics[width=75mm,height=80mm]{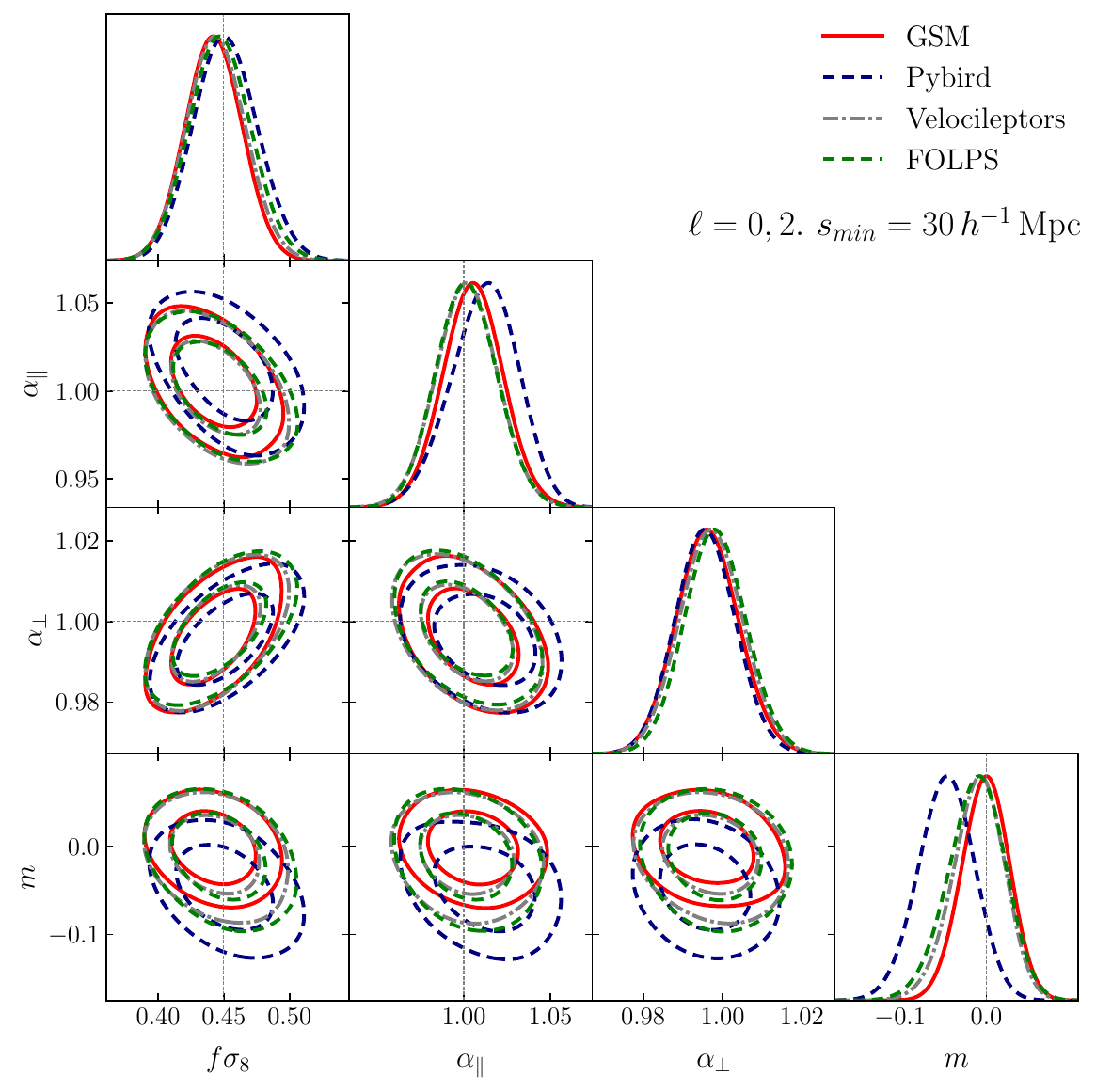}
\caption{All models using their baseline analysis for the minimal freedom case for the correlation function.
}   
\label{modelsCS_triangular}
\end{figure*}

Figure \ref{fig:modelsCS} shows the comparison between our  \texttt{GSM-EFT} results and the \texttt{desilike} configurations space versions of \texttt{Velocileptors}, \texttt{Folps$\nu$}, and \texttt{PyBird}. 
The left panel shows the \textit{Full Modeling}  results and the right panel the {\it ShapeFit} counterparts for the LRG sample and for different values of $s_{min}$. We observe good agreement between the models at all scales, we also note that there is slightly more dispersion than the one we observed in the FS analysis, which may be related to the sensitivity of the model to the parameters used in the Hankel transform. In order to getting a closer look, Figure \ref{modelsCS_triangular} shows the triangular plot for both methodologies for $s_{min}=30$. We observe that the contours are all well aligned with each other, and most of them are centered within less than $1\sigma$ from the true values. 
Tables \ref{tab:desilike_FM} and \ref{tab:desilike_SF}
present the quantitative results of the CS with \texttt{desilike}. We notice that the errors obtained from the FS models in CS are consistently lower than their counterparts in FS, confirming the conclusion of the past section that CS provides slightly smaller error bars. The error bars of \texttt{GSM-EFT} are smaller than the mean error by 5-10 percent depending on the parameter. We consider the level of agreement to be reasonable given that the only model fully developed in CS is \texttt{GSM-EFT} and the others are generated from transforming the multipoles which may add a layer of potential numeric errors that require to be explored by tuning the internal parameters of the Hankel transform some of the parameters, such as the binning and the value of $k_{max}$. A careful exploration of the dependence of the configuration space constraints on the numerical precision on the Hankel transform and the exploration of the differences between precision depending of the different spaces is left of a future job.

\begin{center}
\begin{table*}
\ra{1.7}
\begin{center}

\begin{tabular} {  l c c ccc}
\multicolumn{5}{c}{Comparison Models in CS \textit{Full Modeling}  with \texttt{desilike} }. \\
\hline
Model &  Space &$\Omega_{m}$ &$h$ & $\mathrm{ln}(10^{10}A_s) $ \\
\hline

\texttt{Velocileptors} LPT& FS  & $0.309\pm 0.012$   & $0.6739\pm 0.0089$ &  $3.030^{+0.083}_{-0.097}$  \\
\texttt{Folps$\nu$}   &FS   &$0.309\pm 0.011$ &$0.6740\pm 0.0091$  & $3.051^{+0.083}_{-0.097}$\\
\texttt{Pybird}   & FS  & $0.311^{+0.012}_{-0.010}$   & $0.6737\pm 0.0090$ &  $3.03\pm 0.11$  \\
\hline

\texttt{Velocileptors} LPT& CS  & $0.3117\pm 0.0086$   & $0.6757\pm 0.0076$ &  $3.008\pm 0.089$  \\
\texttt{Folps$\nu$}  &CS   &$0.3138\pm 0.0093$ &$0.6782\pm 0.0077$  & $2.989\pm 0.093$ \\
\texttt{Pybird}   & CS  & $0.3095\pm 0.0087$   & $0.6762\pm 0.0086$ &  $3.01^{+0.10}_{-0.12}$  \\

\texttt{GSM-EFT}  & CS  & $0.3139\pm 0.0082$   & $0.6767\pm 0.0078$ &  $2.989\pm 0.082$   \\
\texttt{GSM-EFT}-22  & CS  & $0.3150\pm 0.0072$   & $0.6755\pm 0.0072$ &  $3.058\pm 0.065$        \\
\hline

\hline
\end{tabular}
\caption{Constraints on {\it Full modeling} parameters when fitting the \texttt{ABACUS}-LRG for different models and different spaces, minimal freedom, using \texttt{desilike} compared with this work.}\label{tab:desilike_FM}

\end{center}
\end{table*}
\end{center}

\begin{center}
\begin{table*}
\ra{1.7}
\begin{center}

\begin{tabular} {  l c c ccc}

\multicolumn{6}{c}{Comparison Models in CS \textit{ShapeFit} with \texttt{desilike} }. \\
\hline
Model &  Space &$f\sigma_{8}$ &$\alpha_{\parallel}$ & $\alpha_{\perp} $& m\\
\hline

\texttt{Velocileptors} LPT& FS  & $0.448\pm 0.024$   & $1.004\pm 0.021$ &  $0.9987\pm 0.0099 $   & $-0.023^{+0.039}_{-0.032}  $ \\
\texttt{Folps$\nu$}   &FS   &$0.453\pm 0.024$ &$1.004\pm 0.021$  & $0.9989\pm 0.0099$& $-0.024^{+0.037}_{-0.031}  $\\
\texttt{Pybird}  & FS  & $0.451\pm 0.026$   & $1.001\pm 0.021$ &  $0.9995\pm 0.010$  & $-0.024^{+0.043}_{-0.031}  $  \\
\hline

\texttt{Velocileptors} LPT& CS  & $0.444\pm 0.022$   & $1.002\pm 0.017$ &  $0.9972\pm 0.0078 $   &  $-0.009^{+0.031}_{-0.028}  $\\
\texttt{Folps$\nu$}   &CS   &$0.447\pm 0.023$ &$1.002\pm 0.017$  & $0.9982\pm 0.0076$&  $-0.012^{+0.034}_{-0.031}  $\\
\texttt{Pybird}   & CS  & $0.451\pm 0.024 $   & $1.012^{+0.020}_{-0.018}$ &  $0.9957\pm 0.0074$  &  $-0.046\pm 0.032 $  \\

\texttt{GSM-EFT}-30  & CS  &  $0.442\pm 0.021$   & $1.005\pm 0.017$  & $0.9963\pm 0.0077$  & $-0.001\pm 0.027$   \\
\texttt{GSM-EFT}-22  & CS  &  $0.453\pm 0.020$   & $1.009\pm 0.017 $ &  $0.9966\pm 0.0074$  &  $-0.004^{+0.027}_{-0.021}$   \\

\hline
\hline
\end{tabular}
\caption{Constraints on {\it ShapeFit} parameters when fitting the \texttt{ABACUS}-LRG for different models and different spaces, minimal freedom using \texttt{desilike} compared with this work.}\label{tab:desilike_SF}

\end{center}
\end{table*}
\end{center}
\section{Conclusions}
\label{sec:conclusions}

In preparation for the analysis of the upcoming first batch of DESI data, it is imperative to analyze and test the tools that will be used to extract cosmological information. This includes testing different configurations and degrees of freedom that models can have when predicting the full-shape of galaxy two-point statistics. In particular, there is a lot of interest in understanding the differences in accuracy and precision between compressed methodologies like the {\it Standard} and {\it Shape-Fit} methods, which are more model agnostic, and direct methods like the {\it Full Modeling} methodology where a particular cosmological model is selected a priori. 

Currently, four independent methodologies that model the redshift space clustering statistics are being developed and tested within the collaboration in preparation for the first data release. These are the \textsc{PyBird}, \textsc{FOLPS}$\nu$, and \textsc{velocileptors} methodologies, all of which study the FS power spectrum of tracers. The fourth method is our own EFT-GSM methodology, which is the only approach working on CS.

Throughout this work, we have analyzed the performance of our GSM-EFT model, along with some extensions that allow us to explore a wider parameter range, in recovering the cosmological information from a new suite of 25 \texttt{ABACUS-SUMMIT} mocks. These mocks have been independently populated with different tracers of matter to generate samples resembling ELGs, LRGs, QSOs. Here, we have analyzed all three samples independently, and we also examine a combined sample. We have compared our constraints with those produced by the three other FS models when transforming their predictions into CS. The analysis presented here utilizes our optimized C-code tool introduced in this paper, which computes the CS multipoles of our GSM-EFT model. Furthermore, we have optimized our methodology by employing Neural Networks to construct surrogate models of our GSM-EFT models, accelerating their evaluation and thus reducing the convergence time required to obtain cosmological constraints.

We have analyzed the ability of our {\it Full Modeling} configuration to recover the {\tt ABACUS} {\tt -SUMMIT} LRG cosmology under variations in the volume of the simulation. This was achieved by running our analysis on volumes of $8,h^{-3},\textrm{Gpc}^{3}$, $40,h^{-3},\textrm{Gpc}^{3}$, and $200,h^{-3},\textrm{Gpc}^{3}$, respectively, which correspond to approximately the first and fifth years of DESI observations, and a much larger volume than DESI, respectively. We find that our methodology is relatively robust to variations in volume, with all parameters being recovered to within $1\sigma$ of their original values, with the exception of the values of $h$ and $\mathrm{ln}(10^{10}A_s)$ for the larger volume, which are recovered to within $1.14\sigma$ and $1.4\sigma$, respectively. Throughout this work, we choose to report our results based on an $8,h^{-3},\textrm{Gpc}^{3}$ box, as this approximates the size of the DESI year one volume. 

We have also tested how our results vary in our baseline analysis of the LRG sample under variations of the minimum scale utilized in our analysis, employing both \textit{Full Shape} and \textit{Shape fit} methodologies. We explore three possible values of the minimum scale, namely $s_{\text{min}} = 22,h^{-1} \mathrm{Mpc}$, $s_{\text{min}} = 30,h^{-1} \mathrm{Mpc}$, and $s_{\text{min}} = 38,h^{-1} \mathrm{Mpc}$. We found that both methodologies are in good agreement with each other and with the true values of the simulation for all minimum scales explored and for all parameters within $1\sigma$. Given that the smaller scale corresponds to tighter constraints, we select $s_{\text{min}} = 22,h^{-1} \mathrm{Mpc}$ as the baseline value utilized throughout the rest of this work. We have also tested how our methodology is affected by switching between the minimal and maximal freedom scenarios, where $b_{s^2}$ is either set free or fixed to zero, and we define maximal freedom as our baseline configuration. We found that while all predictions are consistent within $1\sigma$ for $V_{1}$, the minimal freedom case yields constraints approximately $10\%$ tighter. This improvement in constraining power is expected when one reduces the parameter space explored.

In this work, we have also analyzed several extensions to our standard configuration. The first of these extensions is the inclusion of the hexadecapole of the correlation function into our analysis, which is considered as the multipole where first-loop PT starts to find difficulties modeling our scales of interest. In this extension, we include a model of the hexadecapole but only down to a minimum scale of $s_{\text{min}} = 30\, h^{-1} \mathrm{Mpc}$ as we do not trust our model at smaller scales. We note that this extension does not significantly improve our results, with its only advantage being a small improvement in the precision of the \textit{Full Modeling} estimates. We also highlight that all predictions recover the true value of all parameters within $1\sigma$.

We have explored the inclusion of the spectral index $n_s$ into our methodology as another extension to our model and analyzed how this affects the fits on the LRG sample using the \textit{Full Shape} configuration of our model. We found that our model predicts all parameters, including $n_s$, to within $1\sigma$. Additionally, we observe that the ability to recover the original parameters improves over their baseline counterparts at all scales. It is noteworthy that under this extension, our minimum scale of $s_{\text{min}} = 30 \, h^{-1} \mathrm{Mpc}$ shows the best performance overall.

As another extension to our baseline analysis we have expanded our standard setup to include the $w_0$ and $w_a$ parameters of the Chevallier–Polarski–Linder model of dark energy. We run our fitting methodology on our LRG sample and for two distinct configurations of these extensions, first we allow for only $w_0$ to vary while keeping $w_a$ fixed at its true values. Then we allow both parameters to vary simultaneously. We find that all constrains using our {\it Full-Modeling} configuration are capable of recovering the {\tt ABACUS-SUMMIT} parameters, including $w_0$ and $w_a$ well bellow the $1\sigma$ range, we also note that in both cases the $s_{\text{min}} = 22 \, h^{-1} \mathrm{Mpc}$ shows the overall best performance. We believe the accuracy of our fits is in part a result of working in CS, where volume effects play a smaller role on these new parameters. 

All the tests and extensions discussed so far have been analyzed using our LRG sample. As a separate test, we have analyzed the accuracy of our methodology when fitting the QSO and ELG samples from {\tt ABACUS-SUMMIT}. We have conducted fits for these two new samples using our standard configuration for three different minimum scales. In general, we observe that the ELG fits are slightly more accurate and precise than the LRG fits, while the QSO fits are generally less precise. This discrepancy is expected due to the differences in the number of tracers between the samples. Nevertheless, our methodology successfully recovers all parameters within $1\sigma$ for all three samples. Additionally, we find that using a minimum scale of $s_{\text{min}} = 22, h^{-1} \mathrm{Mpc}$ provides tighter constraints on our parameter space while also predicting all parameters precisely.

We have also conducted tests comparing the performance of \textit{Full Modeling} and \textit{ShapeFit} within cosmological parameter space. They involved examining the impact of variations in certain settings, including bias settings and minimum scale ranges. Additionally, we explored how these configurations were affected by our three model extensions: hexadecapole and $n_s$ free. We observed that the various configurations of our vanilla model generally exhibit agreement in their estimates of the central values between both methodologies. There are slight variations in the size of their error estimates, ranging from a few to 30 percent depending on the case studied and the specific parameter. On the other hand, when considering extensions beyond the baseline that involve varying additional cosmological parameters, we observed larger disparities in both the central values and the size of the error bars. Specifically, when allowing $n_s$ to vary freely, we found that the {\it Full Modeling} approach imposes greater constraints, particularly on $n_s$ compared to {\it ShapeFit}. 

The last tests we have presented analyze the performance of our CS methodology, \texttt{GSM-EFT}, when compared to three FS codes: \texttt{Velocileptors}, \texttt{Folps$\nu$}, and \texttt{PyBird}. This analysis serves as a complement to the main comparison between RSD methodologies presented in a companion paper dedicated to the consistency of all FS models. We found that our \texttt{GSM-EFT} results are in good agreement with those from all FS models, with our results showing smaller error bars for both \textit{Full Modeling} and \textit{ShapeFit} configurations. We have also investigated the performance of models in CS through the common framework \texttt{desilike}, confirming the aforementioned consistency of models in both spaces and the reduction of error bars in CS.

\begin{section}{Data Availability}
\label{appendix:DataAvailability}

The data will be made publicly available once the paper is accepted.

\end{section}
\acknowledgments

SR, HN, SF, FR and MV are supported by Investigacion in Ciencia Basica CONAHCYT grant No. A1-S-13051 and PAPIIT IN108321 and IN116024, and Proyecto PIFF. SF and FR is supported by  PAPIIT IN115424. HN and AA are supported by Ciencia de Frontera grant No. 319359 and PAPIIT IG102123. 

MIL recognizes the support of the National Research Foundation of Korea (NRF) grant funded by the Korea government (MIST), Grant No. 2021R1A2C1013024

This material is based upon work supported by the U.S. Department of Energy (DOE), Office of Science, Office of High-Energy Physics, under Contract No. DE–AC02–05CH11231, and by the National Energy Research Scientific Computing Center, a DOE Office of Science User Facility under the same contract. Additional support for DESI was provided by the U.S. National Science Foundation (NSF), Division of Astronomical Sciences under Contract No. AST-0950945 to the NSF’s National Optical-Infrared Astronomy Research Laboratory; the Science and Technology Facilities Council of the United Kingdom; the Gordon and Betty Moore Foundation; the Heising-Simons Foundation; the French Alternative Energies and Atomic Energy Commission (CEA); the National Council of Humanities, Science and Technology of Mexico (CONAHCYT); the Ministry of Science and Innovation of Spain (MICINN), and by the DESI Member Institutions: \url{https://www.desi.lbl.gov/collaborating-institutions}. Any opinions, findings, and conclusions or recommendations expressed in this material are those of the author(s) and do not necessarily reflect the views of the U. S. National Science Foundation, the U. S. Department of Energy, or any of the listed funding agencies.

The authors are honored to be permitted to conduct scientific research on Iolkam Du’ag (Kitt Peak), a mountain with particular significance to the Tohono O’odham Nation.

\appendix

\section{Author Affiliations}
\label{sec:affiliations}

\begin{hangparas}{.5cm}{1}

$^{1}${Instituto de F\'{\i}sica, Universidad Nacional Aut\'{o}noma de M\'{e}xico,  Cd. de M\'{e}xico  C.P. 04510,  M\'{e}xico}

$^{2}${Korea Astronomy and Space Science Institute, 776, Daedeokdae-ro, Yuseong-gu, Daejeon 34055, Republic of Korea}

$^{3}${Instituto de Ciencias F\'{\i}sicas, Universidad Aut\'onoma de M\'exico, Cuernavaca, Morelos, 62210, (M\'exico)}

$^{4}${Lawrence Berkeley National Laboratory, 1 Cyclotron Road, Berkeley, CA 94720, USA}

$^{5}${Physics Dept., Boston University, 590 Commonwealth Avenue, Boston, MA 02215, USA}

$^{6}${University of Michigan, Ann Arbor, MI 48109, USA}

$^{7}${Institute for Astronomy, University of Edinburgh, Royal Observatory, Blackford Hill, Edinburgh EH9 3HJ, UK}

$^{8}${Department of Physics \& Astronomy, University College London, Gower Street, London, WC1E 6BT, UK}

$^{9}${Institute for Computational Cosmology, Department of Physics, Durham University, South Road, Durham DH1 3LE, UK}

$^{10}${NSF NOIRLab, 950 N. Cherry Ave., Tucson, AZ 85719, USA}

$^{11}${Department of Physics \& Astronomy and Pittsburgh Particle Physics, Astrophysics, and Cosmology Center (PITT PACC), University of Pittsburgh, 3941 O'Hara Street, Pittsburgh, PA 15260, USA}

$^{12}${Kavli Institute for Particle Astrophysics and Cosmology, Stanford University, Menlo Park, CA 94305, USA}

$^{13}${SLAC National Accelerator Laboratory, Menlo Park, CA 94305, USA}

$^{14}${Departamento de F\'isica, Universidad de los Andes, Cra. 1 No. 18A-10, Edificio Ip, CP 111711, Bogot\'a, Colombia}

$^{15}${Observatorio Astron\'omico, Universidad de los Andes, Cra. 1 No. 18A-10, Edificio H, CP 111711 Bogot\'a, Colombia}

$^{16}${Institut d'Estudis Espacials de Catalunya (IEEC), 08034 Barcelona, Spain}

$^{17}${Institute of Cosmology and Gravitation, University of Portsmouth, Dennis Sciama Building, Portsmouth, PO1 3FX, UK}

$^{18}${Institute of Space Sciences, ICE-CSIC, Campus UAB, Carrer de Can Magrans s/n, 08913 Bellaterra, Barcelona, Spain}

$^{19}${Departament de F\'{\i}sica Qu\`{a}ntica i Astrof\'{\i}sica, Universitat de Barcelona, Mart\'{\i} i Franqu\`{e}s 1, E08028 Barcelona, Spain}

$^{20}${Institut de Ci\`encies del Cosmos (ICCUB), Universitat de Barcelona (UB), c. Mart\'i i Franqu\`es, 1, 08028 Barcelona, Spain.}

$^{21}${Center for Cosmology and AstroParticle Physics, The Ohio State University, 191 West Woodruff Avenue, Columbus, OH 43210, USA}

$^{22}${Department of Physics, The Ohio State University, 191 West Woodruff Avenue, Columbus, OH 43210, USA}

$^{23}${The Ohio State University, Columbus, 43210 OH, USA}

$^{24}${School of Mathematics and Physics, University of Queensland, 4072, Australia}

$^{25}${Departament de F\'{i}sica, Serra H\'{u}nter, Universitat Aut\`{o}noma de Barcelona, 08193 Bellaterra (Barcelona), Spain}

$^{26}${Institut de F\'{i}sica d’Altes Energies (IFAE), The Barcelona Institute of Science and Technology, Campus UAB, 08193 Bellaterra Barcelona, Spain}

$^{27}${University of California, Berkeley, 110 Sproul Hall \#5800 Berkeley, CA 94720, USA}

$^{28}${Instituci\'{o} Catalana de Recerca i Estudis Avan\c{c}ats, Passeig de Llu\'{\i}s Companys, 23, 08010 Barcelona, Spain}

$^{29}${Department of Physics and Astronomy, University of Sussex, Brighton BN1 9QH, U.K}

$^{30}${Department of Physics \& Astronomy, University  of Wyoming, 1000 E. University, Dept.~3905, Laramie, WY 82071, USA}

$^{31}${National Astronomical Observatories, Chinese Academy of Sciences, A20 Datun Rd., Chaoyang District, Beijing, 100012, P.R. China}

$^{32}${Department of Physics and Astronomy, University of Waterloo, 200 University Ave W, Waterloo, ON N2L 3G1, Canada}

$^{33}${Perimeter Institute for Theoretical Physics, 31 Caroline St. North, Waterloo, ON N2L 2Y5, Canada}

$^{34}${Waterloo Centre for Astrophysics, University of Waterloo, 200 University Ave W, Waterloo, ON N2L 3G1, Canada}

$^{35}${Space Sciences Laboratory, University of California, Berkeley, 7 Gauss Way, Berkeley, CA  94720, USA}

$^{36}${Department of Physics, Kansas State University, 116 Cardwell Hall, Manhattan, KS 66506, USA}

$^{37}${Department of Physics and Astronomy, Sejong University, Seoul, 143-747, Korea}

$^{38}${CIEMAT, Avenida Complutense 40, E-28040 Madrid, Spain}

$^{39}${Department of Physics, University of Michigan, Ann Arbor, MI 48109, USA}

$^{40}${Department of Physics \& Astronomy, Ohio University, Athens, OH 45701, USA}

$^{41}${Physics Department, Stanford University, Stanford, CA 93405, USA}

$^{42}${Sorbonne Universit\'{e}, CNRS/IN2P3, Laboratoire de Physique Nucl\'{e}aire et de Hautes Energies (LPNHE), FR-75005 Paris, France}

\end{hangparas}


\bibliographystyle{JHEP}
\bibliography{Bib,DESI2024_bib}

\end{document}